\documentclass[11pt,a4paper]{article}

\usepackage[T1]{fontenc}
\usepackage[utf8]{inputenc} 
\usepackage[english]{babel}

\usepackage{geometry}
\usepackage{lmodern} 

\usepackage{amsmath,amssymb,amsthm,mathtools,bm,dsfont} 

\usepackage{graphicx}
\usepackage{float}
\usepackage{booktabs,array,longtable,multirow}
\usepackage{subcaption} 
\usepackage{caption}    
\usepackage{dsfont}
\usepackage{mathrsfs}
\usepackage{xcolor}
 \usepackage{booktabs}
 \usepackage{threeparttable}
\usepackage{algorithm}
\usepackage{algpseudocode}
\usepackage{listings}

\usepackage{enumitem}
\newenvironment{assumptionlist}[1]{%
  \begin{enumerate}[
    label=\textnormal{(\textbf{#1\arabic*})},
    ref=\textnormal{(\textbf{#1\arabic*})},
    leftmargin=3.2em, labelsep=0.5em, start=1]
}{\end{enumerate}}
\usepackage{titlesec}
\usepackage{titletoc}
\usepackage{fancyhdr}
\usepackage{bbm}

\usepackage{booktabs}
\usepackage{longtable}
\usepackage{dcolumn}
\usepackage{xcolor}
\usepackage{colortbl}
\usepackage{array}

\usepackage{setspace}
\usepackage{needspace}
\usepackage{tikz}
\usetikzlibrary{automata,positioning,arrows.meta}

\DeclareCaptionFormat{myformat}{\hspace{-4.40in}#1#2\\#3}
\DeclareCaptionFormat{myformatT}{\hspace{-5.0in}#1#2\\#3}
\usepackage[round]{natbib}

\usepackage[colorlinks=true,linkcolor=blue!80!black,citecolor=red!70!black,urlcolor=blue!80!black]{hyperref}

\usepackage{etoolbox,xpatch}

\theoremstyle{definition}

\newtheorem{remark}{Remark}

\theoremstyle{plain}
\newtheorem{theorem}{Theorem}
\newtheorem{lemma}{Lemma}

\newtheorem{corollary}{Corollary}
\newtheorem{proposition}{Proposition}

\BeforeBeginEnvironment{theorem}{\Needspace{5\baselineskip}}
\BeforeBeginEnvironment{lemma}{\Needspace{5\baselineskip}}
\BeforeBeginEnvironment{tlemma}{\Needspace{5\baselineskip}}
\BeforeBeginEnvironment{proposition}{\Needspace{5\baselineskip}}
\BeforeBeginEnvironment{corollary}{\Needspace{5\baselineskip}}
\BeforeBeginEnvironment{definition}{\Needspace{5\baselineskip}}
\BeforeBeginEnvironment{example}{\Needspace{5\baselineskip}}
\BeforeBeginEnvironment{remark}{\Needspace{5\baselineskip}}

\makeatletter
\AtEndEnvironment{theorem}{\@endparpenalty=10000}
\AtEndEnvironment{lemma}{\@endparpenalty=10000}
\AtEndEnvironment{tlemma}{\@endparpenalty=10000}
\AtEndEnvironment{proposition}{\@endparpenalty=10000}
\AtEndEnvironment{corollary}{\@endparpenalty=10000}
\AtBeginEnvironment{corollary}{\postdisplaypenalty=10000}
\xpatchcmd{\proof}{\topsep6\p@\@plus6\p@\relax}
  {\topsep0pt\partopsep0pt\@beginparpenalty=10000\relax}
  {}{\PackageError{article_v6}{Could not adjust proof spacing}{}}
\makeatother

\AfterEndEnvironment{proof}{\par\addvspace{0.5\baselineskip}}

\DeclareMathOperator*{\argmin}{argmin}

\def\E{\mathds{E}}
\def\P{\mathds{P}}
\def\V{\mathds{V}}

\def\bx{\mathbf{x}}

\def\E{\mathds{E}}
\def\P{\mathds{P}}
\def\V{\mathds{V}}

\def\cB{\mathcal{B}}

\def\cL{\mathcal{L}}

\title{
	\Large\textbf{A Debiased Machine Learning Single-Imputation Framework for Item Nonresponse in Surveys}\\
	\vspace{0.5cm}
}

\author{
	\textbf{Mehdi Dagdoug}\\
	\small{Department of Mathematics and Statistics}\\
	\small{McGill University}\\
	\small{Email: mehdi.dagdoug@mcgill.ca}
	\and
	\textbf{David Haziza}\\
	\small{Department of Mathematics and Statistics}\\
	\small{University of Ottawa}\\
	\small{Email: dhaziza@uottawa.ca}
}

\date{\today}

\begin{document}
	
	\maketitle
	
	\begin{abstract}
		\noindent	Machine-learning methods offer flexible alternatives to parametric
	models for imputing missing survey values. However, using their
	predictions in the customary imputed estimator does not automatically
	yield valid inference. We explain why the error from estimating the
	regression function can prevent root-$n$ consistency and develop
	inference for a finite population mean using a cross-fitted augmented
	inverse-probability-weighted (AIPW) estimator. The analysis combines
	an imputation model, a sampling design, and a missing-at-random response
	mechanism, while allowing dependence among sampling indicators.
	We establish double robustness and show that, under suitable
	regularity conditions, consistency of both nuisance estimators
	together with a product-rate condition yields oracle equivalence
	and asymptotic normality. We also derive consistent variance
	estimators accounting for sampling and nonresponse. To obtain a
	completed survey variable, we construct adjusted imputed values
	that preserve observed responses and whose weighted mean reproduces
	the AIPW estimate exactly. We further discuss aggregation of nuisance
	learners and the conditions under which it preserves the prediction
	rates needed for inference. Simulations show that cross-fitting can
	substantially reduce undercoverage, but may increase mean squared
	error when nuisance predictions are inaccurate.

		\vspace{0.3cm}
		\vspace{0.3cm}
		
		\noindent

	\end{abstract}
	
	\thispagestyle{empty} 
	\newpage
	\setcounter{page}{1}
	

\section{Introduction}

Machine learning (ML) methods are increasingly used in national statistical offices. They are useful because they can handle many covariates and capture nonlinear relationships without requiring the analyst to specify a simple parametric model. One important area where these methods may help is the treatment of missing survey data. This is the topic of this paper.\\

We consider the problem of estimating the finite population mean of a survey variable $Y$ from a sample of size $n$ when some sampled units do not respond to item $Y$. A common approach is to impute the missing values. In practice, this means fitting a model to estimate the relationship between survey variables and available covariates, represented by
$$
m(\bx)=\E_m[Y\mid \bx]
$$
using the respondents, and then replacing the missing $Y$-values by predictions from the fitted model. The resulting estimator is the customary imputed estimator, that is, the weighted mean of the observed and imputed values.\\

If the imputation model is parametric, for example a linear regression model, the customary imputed estimator can be root-$n$ consistent and asymptotically normal when the conditional mean model is correctly specified and standard regularity conditions hold. However, if this model is misspecified, the estimator may be asymptotically biased. ML methods provide a natural alternative because they are more flexible and do not impose a simple parametric form on $m(\cdot)$. However, inserting ML predictions into the customary imputed estimator does not automatically lead to valid inference. The difficulty does not come from ML alone. It also comes from the estimator in which the predictions are used. In the customary imputed estimator, the error from estimating $m(\cdot)$ enters directly through a weighted average of prediction errors over the nonrespondents. Equivalently, its estimating equation is not Neyman-orthogonal with respect to $m(\cdot)$. As a result, if $\widehat m(\cdot)$ converges slowly, the resulting imputed estimator may fail to be root-$n$ consistent.\\

Consequently, standard inference can be difficult for the customary ML imputed estimator. Even if $\widehat m(\cdot)$ is consistent, its convergence rate may be slower than $n^{-1/2}$. The error from estimating $m(\cdot)$ may then affect the bias and the first-order distribution of the estimator, as well as the construction of a variance estimator. Data reuse may create an additional difficulty, because the same observations are used both to fit the prediction rule and to construct the estimator of the finite population mean. The literature on finite-population survey inference after ML imputation remains limited; see, for example, \cite{dagdoug2025statistical} for imputation based on regression trees and random forests.\\

The solution studied in this paper is to change the point estimator. Instead of using the customary imputed estimator, we use an augmented inverse-probability-weighted (AIPW) estimator. The AIPW estimator combines two nuisance functions: the outcome regression function $m(\cdot)$, and the response propensity
$$
p(\bx)=\P(r=1\mid \bx).
$$
Here, $r$ is the response indicator, equal to one when $Y$ is observed and zero otherwise. Under standard regularity conditions, the AIPW estimator is doubly robust: it remains consistent if either the imputation model or the response propensity model is correctly specified; see, for example, \cite{robins1994estimation}, \cite{haziza2006nonresponse}, \cite{schafer2008average}, \cite{Cao2009}, and \cite{kim2014doubly}.\\

This change of estimator is especially important when ML methods are used. Unlike the customary imputed estimator, the AIPW estimator is based on a Neyman-orthogonal score. This means that the leading bias due to nuisance estimation depends on the product of the errors in $\widehat m$ and $\widehat p$, rather than on either error separately. The nuisance estimators therefore do not both need to converge at the parametric rate. A typical sufficient set of conditions is that both nuisance estimators be consistent and that, in suitable $L_2$ norms,
$$
\|\widehat m-m\|_2
\|\widehat p-p\|_2
=
o_{\P}(n^{-1/2}).
$$
This product-rate condition is central in double machine learning; see \cite{chernozhukov2018Double,chernozhukov2022locally,chernozhukov2022automatic,kennedy2023review}. Double robustness and the product-rate condition play different roles: the former protects consistency when one of the two nuisance models is correct, whereas the latter supports oracle equivalence and root-$n$ inference when both nuisance functions are estimated flexibly.\\

For flexible learners, orthogonality does not by itself remove the dependence created by using the same observations both to fit the nuisance functions and to evaluate the AIPW score. We therefore use cross-fitting. The sample is divided into folds, and the nuisance functions used for a given unit are estimated without using the fold containing that unit. This produces out-of-sample nuisance predictions in the AIPW score and breaks the most direct form of data reuse; see \cite{newey2018cross,zheng2011crossvalidated,zivich2021machine}.\\

The paper makes four main contributions. First, we explain why customary ML imputation need not yield root-$n$ consistency. Second, we establish oracle equivalence for a cross-fitted AIPW estimator in a framework that explicitly allows dependent sampling indicators. Under the stated model, design, and nuisance rate conditions, ordinary random-fold cross-fitting suffices. Third, we derive a consistent variance estimator accounting for sampling and nonresponse, supporting asymptotically valid inference. Fourth, we construct adjusted imputed values that preserve observed responses and reproduce the AIPW estimate. We also discuss learner aggregation and assess finite-sample performance through simulation.\\

The remainder of the paper is organized as follows. Section~\ref{Sec:Setup} lays out the theoretical framework and introduces the notation used throughout the paper. Section~\ref{sec:literature} revisits the customary imputed estimator and explains why combining this estimator with ML-based imputation may lead to difficulties for statistical inference. Section~\ref{sec:aipw1} introduces the oracle AIPW estimator and establishes its properties. Sections~\ref{sec:aipw2} and~\ref{sec:aipw3} develop its feasible cross-fitted counterpart and establish the corresponding theoretical results. Section~\ref{sec:aggregation} discusses methods for aggregating the estimated nuisance functions, and Section~\ref{sec:simulation} reports the results of the simulation study. Section~\ref{sec:finalRemarks} concludes the paper. The proofs of all theoretical results, together with the more technical arguments, are deferred to the appendices.

\section{Theoretical setup}\label{Sec:Setup}

\subsection{Population and observed data}

Consider a finite population
$
U=\{1,...,N\}
$
of size $N$. We are interested in estimating the finite population mean of a survey variable $Y$,
$$
\mu=\frac{1}{N}\sum_{k\in U}y_k,
$$
where $y_k$ denotes the outcome for unit $k$. Write $\mathbf y=(y_1,\ldots,y_N)^\top$ for the population outcome vector. A sample $S$, of size $n$, is selected according to a sampling design
$
\mathcal F(\mathbf I\mid \mathbf Z),
$
where $\mathbf Z$ denotes the matrix of design information available prior to sampling for all population units, and
$
\mathbf I=(I_1,\ldots,I_k,\ldots,I_N)^\top
$
is the vector of sample membership indicators. Thus, $I_k=1$ if $k\in S$, and $I_k=0$ otherwise. The first- and second-order inclusion probabilities are defined as
$$
\pi_k=\E_p[I_k\mid \mathbf Z],
\qquad
\pi_{k\ell}=\E_p[I_kI_\ell\mid \mathbf Z],
$$
where $\E_p$ denotes expectation with respect to the sampling design. We assume that both first and second-order inclusion probabilities are strictly positive, i.e., $\pi_{k\ell}>0$ for all $k,\ell \in U$. We write
$
w_k=\pi_k^{-1}
$
for the design weight attached to unit $k$.\\

In the absence of missing values to item $Y$, the usual Horvitz--Thompson estimator of $\mu$ is given by
\begin{equation*}
	\widehat\mu_F
	=
	\frac{1}{N}\sum_{k\in S}w_ky_k.
\end{equation*}
The Horvitz--Thompson  estimator $\widehat\mu_F$ is design-unbiased for $\mu$, i.e., $\E_p[	\widehat\mu_F]=\mu$.\\

In practice, $Y$ may be missing for some sampled units. Let $r_k$ be the potential response indicator for unit $k$, with
$r_k=1$ if $y_k$ would be observed were unit $k$ sampled, and $r_k=0$ otherwise. These indicators are defined for all population units; only those for sampled units are observed.
Let
$
\mathbf r=(r_1,\ldots,r_N)^\top
$
denote the vector of response indicators. The set of respondents to item $Y$ is defined by
$
S_r=\{k\in S: r_k=1\},
$
and the set of nonrespondents is defined by
$
S_m=\{k\in S: r_k=0\}.
$
At the imputation stage, we assume that a vector of fully observed covariates $\bx_k$ is available for each sampled unit $k\in S$. We denote by $\mathbf X$ the population covariate matrix, whose $k$-th row is $\bx_k^\top$. The design variables $\mathbf Z$ are a subset of the covariates in $\mathbf X$.

\subsection{Probabilistic assumptions}

We impose the following assumptions on sampling, the population, and item nonresponse.

\begin{assumptionlist}{A}
	
	\item \label{A1}
	The sampling design satisfies
	$$
	\mathbf I\perp(\mathbf y,\mathbf r)\mid\mathbf X.
	$$
	
	\item \label{A2}
The population triples $(\bx_k,y_k,r_k)$, $k\in U$, are independent and identically distributed with common distribution $P$.
	
	\item \label{A3}
The item nonresponse mechanism is missing at random (MAR):
$$
y_k\perp r_k\mid\bx_k,
\qquad k\in U.
$$

	\item \label{A4}
	There exists $\delta>0$ such that
	$$
	\P(r_k=1\mid \bx_k)
	=
	p(\bx_k)
	\ge \delta,
	\qquad\text{almost surely,}\qquad k\in U.
	$$
	
\end{assumptionlist}

Assumptions~\ref{A1}--\ref{A3} imply that $\mathbf y$, $\mathbf I$, and $\mathbf r$ are mutually independent conditionally on $\mathbf X$; see Lemma~\ref{lem:conditional_moments}. Dependence among the sampling indicators themselves is allowed. Assumption~\ref{A4} is a classical positivity assumption.\\

Unless stated otherwise, expectations, variances, and probabilities are conditional on $\mathbf X$. Subscripts $m$, $p$, and $q$ refer to the imputation model, sampling design, and response mechanism, respectively. Thus, $\E_m$, $\E_p$, and $\E_q$ average over $\mathbf y$, $\mathbf I$, and $\mathbf r$, respectively, holding the other two vectors fixed; $\V_m$, $\V_p$, and $\V_q$ follow the same convention. The operator $\E_P$ averages under the joint superpopulation law $P$ in~\ref{A2}, including covariate randomness.\\

For the imputation model, write
\begin{equation*}\label{imp_mod}
	y_k=m(\bx_k)+\epsilon_k,
	\qquad k\in U,
\end{equation*}
where $m(\bx_k)=\E_m[y_k]$. The errors are independent conditionally on $\mathbf X$, with
$$
\E_m[\epsilon_k]=0,
\qquad
\V_m(\epsilon_k)=\sigma^2(\bx_k)=:\sigma_k^2.
$$

\subsection{Asymptotic framework}

Following \cite{isaki1982survey}, consider nested finite populations $U_v\subset U_{v+1}$ with sizes $N_v\xrightarrow[v \to \infty]{}\infty$, generated from a common infinite i.i.d. sequence with law $P$ in~\ref{A2}. A sample $S_v$ of size $n_v$ is selected from each $U_v$ under the sampling design defined above. For simplicity of notation, the population index is displayed only in $U_v$, $S_v$, $n_v$, and $N_v$.\\

We consider the following assumptions on the sampling design.

\begin{assumptionlist}{D}
	
	\item \label{D1}
	The sampling fraction satisfies
	$$
	\frac{n_v}{N_v}\xrightarrow[v \to \infty]{}\pi_*>0.
	$$
	
	\item \label{D2}
	There exist constants $\lambda_1,\lambda_2>0$ such that, for all $v\in\mathbb N$,
	$$
	\min_{k\in U_v}\pi_{k}>\lambda_1,
	\qquad
	\min_{\substack{k,\ell\in U_v\\ k\neq \ell}}\pi_{k\ell}\geq\lambda_2.
	$$
	
	\item \label{D3}
	The sampling covariances satisfy
	$$
	\limsup_{v \to \infty}
	n_v
	\max_{\substack{k,\ell\in U_v\\ k\neq \ell}}
	\left|
	\pi_{k\ell}-\pi_{k}\pi_{\ell}
	\right|
	<\infty.
	$$
	
	\item \label{D4}
	Let $D_{t}$ denote the set of ordered $t$-tuples of pairwise distinct units in $U_v$. The sample membership indicators satisfy
	$$
	\limsup_{v \to \infty}
	\max_{(i,j,k,\ell)\in D_{4}}
	\left|
	\E_p\left[
	\left(I_{i}I_{j}-\pi_{i}\pi_{j}\right)
	\left(I_kI_\ell-\pi_{k}\pi_{\ell}\right)
	\right]
	\right|
	=
	0.
	$$
	
\end{assumptionlist}

Assumption~\ref{D1} requires the sample and population sizes to grow at the same rate. Assumption~\ref{D2} bounds the inclusion probabilities away from zero, while~\ref{D3} controls the dependence between sample membership indicators. Assumption~\ref{D4} controls joint inclusion moments involving four distinct units and is used to prove consistency of the Horvitz--Thompson variance estimator. These are classical assumptions in the literature, see e.g., \cite{breidt2000local}. \\

We impose the following moment and variance assumptions on the superpopulation distribution $P$.

\begin{assumptionlist}{M}
	\item \label{M1}
	The outcome has a finite second moment:
	$$
	\E_P[y_1^2]<\infty.
	$$

	\item \label{M2}
	The outcome has a finite fourth moment:
	$$
	\E_P[y_1^4]<\infty.
	$$

	\item \label{M3}
	There exists a finite constant $\bar\sigma^2$ such that
	$$
	\sigma^2(\bx_1)\leq\bar\sigma^2,
	\qquad P\text{-almost surely}.
	$$
\end{assumptionlist}

Assumptions~\ref{M1} and~\ref{M2} distinguish the moment requirements for consistency of the point and variance estimators. Assumption~\ref{M3} bounds the conditional error variance and controls the error from estimating the response probabilities. Conditional asymptotic statements are understood to hold for almost every realized covariate sequence under $P$; the required population moment bounds follow from~\ref{A2} and the corresponding moment assumptions.

\section{Challenges with the customary imputed estimator}
\label{sec:literature}

The imputation task consists of selecting a regression function estimator $\widehat{m}$ (e.g.,  weighted linear regression, a machine-learning method, etc.), fitting it to the respondents to estimate
$m(\bx)=\E_m[Y\mid\bx]$, and using the predictions
$\widehat m(\bx_k)$ to replace the missing values $y_k$. The usual imputed estimator of $\mu$ is given by
\begin{equation}\label{imp_est}
	\widehat{\mu}_I(\widehat{m})
	=
	\frac{1}{N}\sum_{k \in S} w_k\,\widetilde{y}_k,
\end{equation}
where
$$
\widetilde{y}_k
=
r_k y_k+(1-r_k)\widehat{m}(\bx_k).
$$
The formulation \eqref{imp_est} is convenient for data users because they can simply apply the weighting system $\{w_k:k\in S\}$ to the completed variable $\widetilde Y$ to obtain weighted estimates.\\

\noindent In this setup, the regression function $m(\cdot)$ plays the role of a nuisance function: it is not the main object of interest, but it must be estimated in order to construct the imputed values for the nonrespondents. More precisely, its fitted version, $\widehat m(\cdot)$, is used to replace the missing $y_k$'s. The quantity of interest remains the finite population mean $\mu$, and $\widehat m(\cdot)$ enters the procedure only through the imputed estimator $\widehat\mu_I(\widehat m)$.\\

\noindent To isolate the effect of estimating $m$, consider the oracle imputed estimator
$$
\widehat{\mu}_{I}(m)
:=
\frac{1}{N}\sum_{k\in S} w_k
\{r_k y_k+(1-r_k)m(\bx_k)\},
$$
which replaces each missing $y_k$ by the true conditional mean $m(\bx_k)$. The additional error from using fitted predictions can be written as
\begin{equation}
	\label{eq:customary_remainder}
	B_I(\widehat m)
	:=
	\widehat{\mu}_I(\widehat{m})-\widehat{\mu}_{I}(m)
	=
	\frac{1}{N}\sum_{k\in S} w_k(1-r_k)
	\{\widehat m(\bx_k)-m(\bx_k)\}.
\end{equation}
With this notation, the total error decomposes as
\begin{equation}\label{tot_err}
	\widehat{\mu}_I(\widehat{m})-\mu
	=
	\big(\widehat{\mu}_F-\mu\big)
	+
	\{\widehat{\mu}_{I}(m)-\widehat{\mu}_{F}\}
	+
	B_I(\widehat m).
\end{equation}

\noindent The first term on the right-hand side of \eqref{tot_err} is the sampling error. The second term is the oracle nonresponse error, since
$$
\widehat{\mu}_{I}(m)-\widehat{\mu}_{F}
=
-\frac{1}{N}\sum_{k\in S} w_k(1-r_k)\{y_k-m(\bx_k)\}.
$$
The first two terms on the right-hand side of \eqref{tot_err} do not depend on how $m$ is estimated; they depend only on the sampling design, the response mechanism, and the residuals $y_k-m(\bx_k)$. Under \ref{A1}--\ref{A3}, \ref{D1}--\ref{D3}, and \ref{M1},
$$
\widehat{\mu}_F-\mu = \mathcal{O}_{\P}(n_v^{-1/2}),
\qquad
 \widehat{\mu}_{I}(m)-\widehat{\mu}_{F} = \mathcal{O}_{\P}(n_v^{-1/2}).
$$
The key issue is therefore the behavior of the remainder $B_I(\widehat m)$. The latter represents the additional error introduced by replacing the true regression function ($m$) with its estimate ($\widehat m$).  Its asymptotic behavior depends on how ($m$) is estimated and can differ substantially between parametric and nonparametric methods. We examine this distinction in Sections \ref{Sec:3.1} and \ref{Sec:3.2}, which discuss parametric imputation and nonparametric and machine-learning imputation, respectively.

\subsection{Parametric imputation}\label{Sec:3.1}

Assume that the regression function is correctly specified by a parametric model,
$m(\bx)=g(\bx;\boldsymbol{\beta}_0)$, where $\boldsymbol{\beta}_0$ is a finite-dimensional parameter. For this illustrative discussion, assume that $\boldsymbol{\beta} \mapsto g(\bx;\boldsymbol{\beta})$ is twice continuously differentiable on a neighborhood of $\boldsymbol{\beta}_0$, with its Hessian uniformly bounded over this neighborhood and the covariate support. Write
$$
\dot g(\bx;\boldsymbol{\beta}_0)
=
\frac{\partial g(\bx;\boldsymbol{\beta})}{\partial \boldsymbol{\beta}}
\bigg|_{\boldsymbol{\beta}=\boldsymbol{\beta}_0}.
$$
We also assume that $\E_P[\|\dot g(\bx_1;\boldsymbol{\beta}_0)\|_2^2]<\infty$.
Let $\widehat m(\bx)=g(\bx;\widehat{\boldsymbol{\beta}})$, where $\widehat{\boldsymbol{\beta}}$ is estimated from the respondents and is $\sqrt{n_v}$-consistent for $\boldsymbol{\beta}_0$, that is,
$$
\|\widehat{\boldsymbol{\beta}}-\boldsymbol{\beta}_0\|_2
=
\mathcal{O}_{\P}(n_v^{-1/2}).
$$

\noindent In the parametric case, the remainder $B_I(\widehat m)$ in \eqref{eq:customary_remainder} reduces to
$$
B_I(\widehat m)
=
\frac{1}{N}\sum_{k\in S} w_k(1-r_k)
\{g(\bx_k;\widehat{\boldsymbol{\beta}})
-
g(\bx_k;\boldsymbol{\beta}_0)\}.
$$
By Taylor's theorem, the bounded Hessian and the bounded design weights in~\ref{D2} give an averaged remainder $\mathcal{O}_{\P}(n_v^{-1})$. Thus,
\begin{equation*}
	B_I(\widehat m)
	=
	\mathbf A_N^\top
	(\widehat{\boldsymbol{\beta}}-\boldsymbol{\beta}_0)
	+
	\mathcal{O}_{\P}(n_v^{-1}),
\end{equation*}
where
$$
\mathbf A_N
:=
\frac{1}{N}\sum_{k\in S} w_k(1-r_k)
\dot g(\bx_k;\boldsymbol{\beta}_0).
$$
To find the limit of $\mathbf A_N$, write
\begin{align*}
\mathbf A_N-\frac{1}{N}\sum_{k\in U}
\{1-p(\bx_k)\}\dot g(\bx_k;\boldsymbol\beta_0)
&=\frac{1}{N}\sum_{k\in U}
\left(\frac{I_k}{\pi_k}-1\right)(1-r_k)
\dot g(\bx_k;\boldsymbol\beta_0)\\
&\quad+\frac{1}{N}\sum_{k\in U}
\{p(\bx_k)-r_k\}\dot g(\bx_k;\boldsymbol\beta_0).
\end{align*}
The first sum on the right is a Horvitz--Thompson mean error in each
coordinate. Under \ref{A1}--\ref{A3} and \ref{D1}--\ref{D3},
Lemma~\ref{lem:ht_mean_bound} bounds it by
$\mathcal O_{\P}(n_v^{-1/2})$; its moment condition follows from the
assumed second moment of $\dot g$.
The second sum has independent, centered response factors and is
$\mathcal O_{\P}(N_v^{-1/2})$ by the same moment condition.
The derivative moment condition and the law of large numbers therefore
give $\mathbf A_N\xrightarrow[v\to\infty]{\P}\mathbf a$, where
\begin{equation*}
	\label{eq:def_a_parametric}
	\mathbf a
	:=
	\E_P\left[
	\{1-p(\bx_1)\}
	\dot g(\bx_1;\boldsymbol{\beta}_0)
	\right].
\end{equation*}
Hence,
\begin{equation}
	\label{eq:parametric_expansion_limit}
	B_I(\widehat m)
	=
	\mathbf a^\top
	(\widehat{\boldsymbol{\beta}}-\boldsymbol{\beta}_0)
	+
	o_{\P}(n_v^{-1/2})
	=
	\mathcal{O}_{\P}(n_v^{-1/2});
\end{equation}
see \cite{kim2009unified}. Equation~\eqref{eq:parametric_expansion_limit} shows that the error from estimating $m(\cdot)$ is controlled at the parametric rate. Consequently, the customary imputed estimator $\widehat{\mu}_I(\widehat m)$ remains $\sqrt{n_v}$-consistent for $\mu$.\\

\noindent However, $\sqrt{n_v}$-consistency does not imply asymptotic equivalence to the oracle imputed estimator: the first-order term in~\eqref{eq:parametric_expansion_limit} need not be negligible. To make this explicit, suppose that the following joint central limit theorem holds:
$$
\sqrt{n_v}
\begin{pmatrix}
	\widehat{\mu}_I(m)-\mu \\
	\widehat{\boldsymbol\beta}-\boldsymbol\beta_0
\end{pmatrix}
\xrightarrow[v \to \infty]{\cL}
N\left[
\mathbf 0,
\begin{pmatrix}
	V_{\mathrm{or}} & \mathbf c^\top \\
	\mathbf c & \mathbf V_{\beta}
\end{pmatrix}
\right].
$$
Here $V_{\mathrm{or}}$, $\mathbf V_{\beta}$, and $\mathbf c$ denote the covariance blocks of the limiting distribution of these normalized estimation errors. If $\mathbf a^\top\mathbf V_{\beta}\mathbf a>0$, then~\eqref{eq:parametric_expansion_limit} implies that $\sqrt{n_v}B_I(\widehat m)$ has a nondegenerate normal limit. In this case, $B_I(\widehat m)$ is not $o_{\P}(n_v^{-1/2})$, and the feasible and oracle imputed estimators are not asymptotically equivalent. Combining the joint central limit theorem with~\eqref{eq:parametric_expansion_limit} gives
$$
\sqrt{n_v}\{\widehat{\mu}_I(\widehat m)-\mu\}
\xrightarrow[v \to \infty]{\cL}
N(0,V_{\mathrm{par}}),
$$
where
\begin{equation*}
	\label{eq:Vpar}
	V_{\mathrm{par}}
	=
	V_{\mathrm{or}}
	+
	2\mathbf a^\top \mathbf c
	+
	\mathbf a^\top \mathbf V_{\beta}\mathbf a.
\end{equation*}
Thus, the feasible parametric imputed estimator is asymptotically normal, but with a variance that accounts for the estimation of $\boldsymbol{\beta}_0$. A central limit theorem of this type for linear regression imputation was established by \cite{an2026variable}. Variance estimation procedures for parametric imputation have been studied extensively; see, for example, \cite{sarndal1992methods}, \cite{shao1999variance}, \cite{kim2009unified}, and \cite{haziza2020variance}.\\

\noindent If $V_{\mathrm{par}}>0$ and $\widehat V_{\mathrm{par}}\xrightarrow[v \to \infty]{\P}V_{\mathrm{par}}$, an asymptotic $(1-\alpha)$-level confidence interval for $\mu$ is given by
$$
\widehat{\mu}_I(\widehat m)
\pm
z_{1-\alpha/2}
\sqrt{\frac{\widehat V_{\mathrm{par}}}{n_v}}.
$$

\begin{remark}
	The first-order effect of estimating $m(\cdot)$ can be understood through
	Neyman orthogonality. Let $\theta$ be a generic scalar parameter and let
	$\varphi(\cdot)$ be a generic regression function. Consider the estimating
	equation
	\begin{equation*}
		\label{eq:customary_score}
		\widehat\Psi_N(\theta,\varphi)
		=
		\frac{1}{N}
		\sum_{k\in S}
		w_k
		\left\{
		r_k y_k+(1-r_k)\varphi(\bx_k)
		\right\}
		-\theta.
	\end{equation*}
	For a fixed function $\varphi$, solving $\widehat\Psi_N(\theta,\varphi)=0$
	gives $\widehat\mu_I(\varphi)$. Under assumptions~\ref{A1}--\ref{A3},
	the corresponding population moment is defined by
	$$
	\Psi_N(\varphi)
	:=
	\E_{mpq}
	\left[
	\widehat\Psi_N\{\mu,\varphi\}
	\right]
	=
	\frac{1}{N}\sum_{k\in U}
	\{1-p(\bx_k)\}\{\varphi(\bx_k)-m(\bx_k)\}.
	$$
	In particular, $\Psi_N(m)=0$. Neyman orthogonality with respect to the
	regression function would require the directional derivative at $m$ to
	vanish for every admissible perturbation $h$. Here,
	$$
	\left.
	\frac{\partial}{\partial t}
	\Psi_N(m+t h)
	\right|_{t=0}
	=
	\frac{1}{N}
	\sum_{k\in U}
	\{1-p(\bx_k)\}
	h(\bx_k),
	$$
	which is generally nonzero. Thus, the customary imputation estimating
	equation is not Neyman orthogonal with respect to $m(\cdot)$. The exact identity
	$$
	\widehat\Psi_N\{\mu,\widehat m\}
	-
	\widehat\Psi_N(\mu,m)
	=
	\widehat\mu_I(\widehat m)-\widehat\mu_I(m)
	=
	B_I(\widehat m)
	$$
	shows how regression estimation error enters the final estimator.
	In the parametric case, expansion~\eqref{eq:parametric_expansion_limit}
	identifies its first-order contribution as
	$\mathbf a^\top(\widehat{\boldsymbol\beta}-\boldsymbol\beta_0)$.
\end{remark}
\subsection{Nonparametric and machine-learning imputation}\label{Sec:3.2}

Nonparametric regression can yield consistency under weak conditions.
For example, in the i.i.d. setting, suitably tuned nearest-neighbor and
partitioning estimators are universally $L_2$-consistent: for every
distribution $P$ with $\E_P[y_1^2]<\infty$,
$$
\E_P\left[\{\widehat m(\bx)-m(\bx)\}^2\right]
\xrightarrow[v \to \infty]{}0,
$$
where the expectation is over the i.i.d. training sample and an independent test
covariate $\bx$. No smoothness assumption on $m$ is needed;
see \cite{stone1977consistent,gyorfi2002distribution}.\\

\noindent For the customary imputed estimator, consistency follows from a
corresponding prediction-consistency condition over the sampled
nonrespondents. Since the sampling and oracle nonresponse errors in
\eqref{tot_err} are $\mathcal{O}_{\P}(n_v^{-1/2})$, we have
\begin{equation}
	\label{eq:plugin_rate_decomposition}
	\widehat{\mu}_I(\widehat m)-\mu
	=
	B_I(\widehat m)+ \mathcal{O}_{\P}(n_v^{-1/2}).
\end{equation}
It is therefore enough that $B_I(\widehat m)=o_{\P}(1)$.
Cauchy--Schwarz and \ref{D2} give
\begin{align*}
|B_I(\widehat m)|
&\leq\left\{\frac{1}{N}\sum_{k\in S}w_k(1-r_k)\right\}^{1/2}
\left\{\frac{1}{N}\sum_{k\in S}w_k(1-r_k)
\{\widehat m(\bx_k)-m(\bx_k)\}^2\right\}^{1/2}\\
&\leq\lambda_1^{-1/2}
\left\{\frac{1}{N}\sum_{k\in S}w_k(1-r_k)
\{\widehat m(\bx_k)-m(\bx_k)\}^2\right\}^{1/2}.
\end{align*}
Therefore, weighted $L_2$-consistency of $\widehat m$
over the sampled nonrespondents is sufficient for
$B_I(\widehat m)=o_{\P}(1)$ and hence
$$
\widehat\mu_I(\widehat m)-\mu
\xrightarrow[v \to \infty]{\P}0.
$$
No particular convergence rate is needed. Under MAR, respondents share the
population regression function, and positivity ensures covariate overlap.
Applying a classical universal-consistency result here requires verifying
this weighted form of prediction consistency under the sampling design.
For Poisson sampling with fixed $\pi_k=\pi(\bx_k)$, this is a classical
covariate-shift problem \citep{shimodaira2000improving}, and positivity
transfers universal $L_2$-consistency of respondent fits to imputation
consistency under the joint law including covariate randomness.\\

\noindent Consistency alone gives no control of $B_I(\widehat m)$ at the
$n_v^{-1/2}$ scale needed for first-order inference. Oracle equivalence
requires the stronger condition
$$
B_I(\widehat m)=o_{\P}(n_v^{-1/2}).
$$
If instead
$B_I(\widehat m)=\mathcal{O}_{\P}(n_v^{-1/2})$, but not
$o_{\P}(n_v^{-1/2})$, then the plug-in estimator remains
$\sqrt{n_v}$-consistent, but its first-order distribution may differ from the
oracle distribution and must account for the estimation of $m$. If
$B_I(\widehat m)$ is not $\mathcal{O}_{\P}(n_v^{-1/2})$, then the estimator may remain
consistent, but it is no longer $\sqrt{n_v}$-consistent. In that case, a
classical central limit theorem centered at $\mu$ with the usual
$n_v^{1/2}$ scaling cannot hold.\\

\noindent To illustrate how slow prediction rates can affect inference,
consider classical nonparametric regression. Under \ref{A1}--\ref{A4} and
\ref{D1}, the respondent fraction is bounded away from zero with probability
tending to one as $v\to\infty$, so we express the rates in terms of $n_v$.
To make the discussion concrete, suppose that $m$ belongs to an $s$-smooth
Hölder class over $d$ continuous covariates. Here $d$ is the number of
predictors and $s$ measures the smoothness of the regression function. For
example, when $0<s\leq 1$, this means that
$|m(\bx)-m(\bx')|\leq C\|\bx-\bx'\|^s$ for some
constant $C>0$; the case $s=1$ corresponds to a Lipschitz function. More
generally, if $s=\ell+\alpha$, with $\ell$ a nonnegative integer and
$0<\alpha\leq 1$, then $m$ has derivatives up to order $\ell$, and the
derivatives of order $\ell$ are $\alpha$-Hölder continuous.\\

\noindent As an i.i.d. benchmark, classical minimax results
\citep{stone1982optimal,tsybakov2009introduction,gyorfi2002distribution} show
that, under standard design and noise regularity conditions, the optimal
root mean squared error for estimating $m$ over such classes is
$\mathcal{O}(n_v^{-s/(2s+d)})$, with matching minimax lower bounds.
This rate improves when the function is
smoother, that is, when $s$ is larger, but it deteriorates as the dimension
$d$ increases. Importantly,
$$
\frac{s}{2s+d}<\frac{1}{2}
\qquad\text{whenever } d>0.
$$
Thus, the usual nonparametric prediction rate is slower than the parametric
rate $n_v^{-1/2}$.\\

\noindent Recall that the plug-in remainder $B_I(\widehat m)$ is controlled by
the prediction error of $\widehat m$ over the sampled nonrespondents. Indeed,
by the Cauchy--Schwarz argument above, if
$$
\left\{
\frac{1}{N}
\sum_{k\in S}
w_k(1-r_k)
\{\widehat m(\bx_k)-m(\bx_k)\}^2
\right\}^{1/2}
=
\mathcal{O}_{\P}\left(n_v^{-s/(2s+d)}\right),
$$
then
$$
B_I(\widehat m)
=
\mathcal{O}_{\P}\left(n_v^{-s/(2s+d)}\right).
$$
This argument provides an upper bound on the plug-in remainder, rather than its
exact rate. In principle, cancellations among the prediction errors could make
$B_I(\widehat m)$ smaller than the corresponding prediction root mean squared error. However,
because the customary imputed estimator is not based on a Neyman-orthogonal
estimating equation, such cancellations cannot be guaranteed in general.
The bound $B_I(\widehat m)=\mathcal{O}_{\P}(n_v^{-s/(2s+d)})$ ensures consistency.
If, in addition, $B_I(\widehat m)$ is not $o_{\P}(n_v^{-s/(2s+d)})$, then
\eqref{eq:plugin_rate_decomposition} and $s/(2s+d)<1/2$ imply that the
estimator is not $\sqrt{n_v}$-consistent, so a standard central limit theorem
centered at $\mu$ with $n_v^{1/2}$ scaling cannot hold.\\

\noindent The preceding calculation provides a useful benchmark, but it does
not cover all modern machine-learning procedures. Some specific methods admit
convergence rates under structural assumptions such as sparsity, additivity,
honesty, or compositional smoothness. However, there is no single convergence
rate that applies to arbitrary random forests, boosted trees, neural networks,
or ensemble learners. For any given procedure, the available rate depends on
the assumptions imposed on the regression function, the dimension and empirical
configuration of the covariates, and the way in which the method is tuned.
Moreover, many such algorithms involve thresholding, tree splits,
regularization, model selection, or early stopping. These operations may
introduce non-negligible regularization bias and may make the map from the data
to $\widehat m$ non-smooth. Consequently, the condition
$B_I(\widehat m)=o_{\P}(n_v^{-1/2})$ generally has to be studied for
the particular learner and data-generating structure under consideration.\\

\noindent These facts also make variance estimation difficult. In the
parametric case, the effect of estimating $m$ can often be linearized through
$\widehat{\boldsymbol\beta}-\boldsymbol\beta_0$, leading to a tractable
first-order variance formula. With a flexible nonparametric or
machine-learning learner, there is generally no comparable finite-dimensional
expansion of $\widehat m-m$. As a result, there is no generic variance
estimator obtained by simply adding a standard ``imputation variance'' term to
the oracle variance. The first-order distribution of
$\widehat{\mu}_I(\widehat m)$, when it exists, depends on the learning
algorithm and on the behavior of $B_I(\widehat m)$.\\

\noindent One might be tempted to rely on resampling methods such as the
bootstrap. To capture the uncertainty due to $\widehat m$, however, the
imputation model would need to be refitted in each replicate. Even then,
validity is not automatic. The bootstrap is known to fail for some non-smooth
or non-regular estimators; see, for example, \cite{abadie2008failure} for
matching estimators and \cite{fang2019inference} for directionally
differentiable functionals. These examples are not identical to the present
imputation problem, but they illustrate the broader point that resampling does
not provide a generic solution when the estimator depends on a complex,
non-smooth learning procedure.\\

\subsection{Rate of convergence: An empirical investigation}

We conducted a limited simulation study to examine how the total error of the
customary plug-in imputed estimator changes with sample size under parametric
and nonparametric imputation. We considered both the raw error and the error
multiplied by $\sqrt N$.\\

We followed an $mpq$-type simulation scheme with three independent covariates
$x_1,x_2,x_3$, each generated from a uniform distribution on $[0,1]$ and
kept fixed throughout the Monte Carlo experiment. For each sample size
$
n\in\{200,400,800, 1600\},
$
we set
$
N={n}/{0.05},
$
so that the sampling fraction was fixed at $n/N=0.05$. Thus, the corresponding population sizes were given by
$
N\in\{4000,8000,16000, 32000\}.
$
For each value of $N$, the finite population covariates were taken as the first
$N$ rows of the fixed covariate matrix. Since $n/N$ is fixed, scaling by
$\sqrt N$ or $\sqrt n$ differs only by a constant and gives the same assessment
of convergence rates.\\

We then repeated $B=25,\!000$ Monte Carlo iterations. At each iteration, new survey variables were generated conditionally on the fixed covariates. We considered two imputation models. The first one was linear:
$$
m_1(\bx)
=
1
+
2(x_1-0.5)
-
1.5(x_2-0.5)
+
(x_3-0.5).
$$
The second one was nonlinear:
$$
\begin{aligned}
	m_2(\bx)
	=
	1
	&+
	2.5\sin(2\pi x_1)
	+
	2\cos(2\pi x_2)
	+
	2.5(x_3-0.5)^2  \\
	&+
	1.5x_1x_2
	+
	2\exp\{-12(x_3-0.65)^2\}.
\end{aligned}
$$
The finite population values were generated as
$$
y_{jk}=m_j(\bx_k)+\epsilon_{jk},
\qquad j=1,2,\quad k\in U,
$$
where the errors $\epsilon_{jk}$ were independent normal random variables with mean zero. For each imputation model, the error variance was calibrated so that the coefficient of determination was approximately equal to $R^2=0.5$. The finite population means
$$
\mu_j=\frac{1}{N}\sum_{k\in U}y_{jk},
\qquad j=1,2,
$$
were then computed from the generated finite population.\\

From each finite population, a sample $S$ of size $n$ was selected by simple random sampling without replacement. The corresponding design weight was therefore $w_k=N/n$ for all $k\in S$.

Item response indicators were then generated independently in the sample according to a MAR mechanism. The response probability was of the form
$$
p(\bx)
=
0.15+0.70\,\mathrm{expit}\{a+\eta(\bx)\},
$$
where $\mathrm{expit}(u)=\{1+\exp(-u)\}^{-1}$, and
$$
\eta(\bx)
=
-0.85 z_2(\bx)
+
0.35(x_1-0.5)
-
0.25(x_2-0.5)
+
0.25\sin(2\pi x_3).
$$
Here $z_2(\bx)$ denotes the standardized version of the nonlinear mean function $m_2(\bx)$ over the finite population. The intercept $a$ was chosen numerically so that the average response probability over each finite population was approximately equal to $50\%$.\\

For each sample and each outcome, missing $Y$-values were imputed using two methods. The first method was linear regression imputation, based on the model
$$
Y \sim x_1+x_2+x_3
$$
fitted to the respondents. The second method was $K$-nearest-neighbor imputation, also using  $x_1, x_2, x_3$, with the number of nearest neighbors allowed to increase with the number of respondents. Specifically, we used
$$
K=\min\left\{\max\left(10,\lfloor n_r^{0.55}\rfloor\right),n_r\right\},
$$
where $n_r$ denotes the number of respondents in the sample. This choice ensures
that $K$ increases with the sample size, while $K/n_r$ decreases. \\

For each imputation method and Monte Carlo iteration, we recorded the total
error $\widehat\mu_I(\widehat m)-\mu$.\\

\begin{figure}[!t] \centering \includegraphics[width=0.95\textwidth]{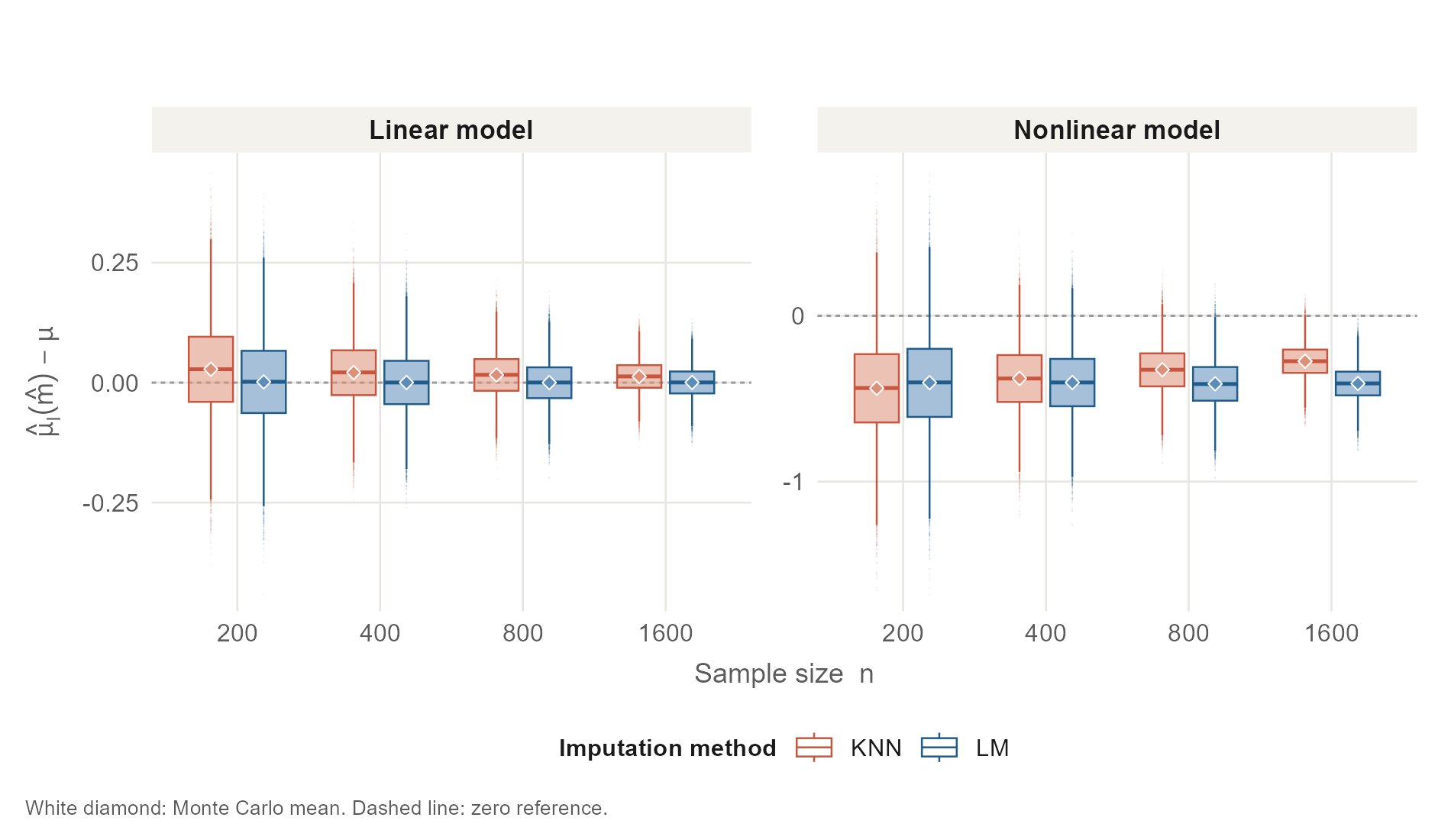} \vspace{0.5em} \refstepcounter{figure} \label{fig:customary_raw_error} {\normalfont\small\noindent \textbf{Figure}~\thefigure: Monte Carlo distribution of the raw error $\widehat\mu_I(\widehat m)-\mu$ of the customary plug-in imputed estimator.} 
\end{figure}

\begin{figure}[!t] \centering \includegraphics[width=0.95\textwidth]{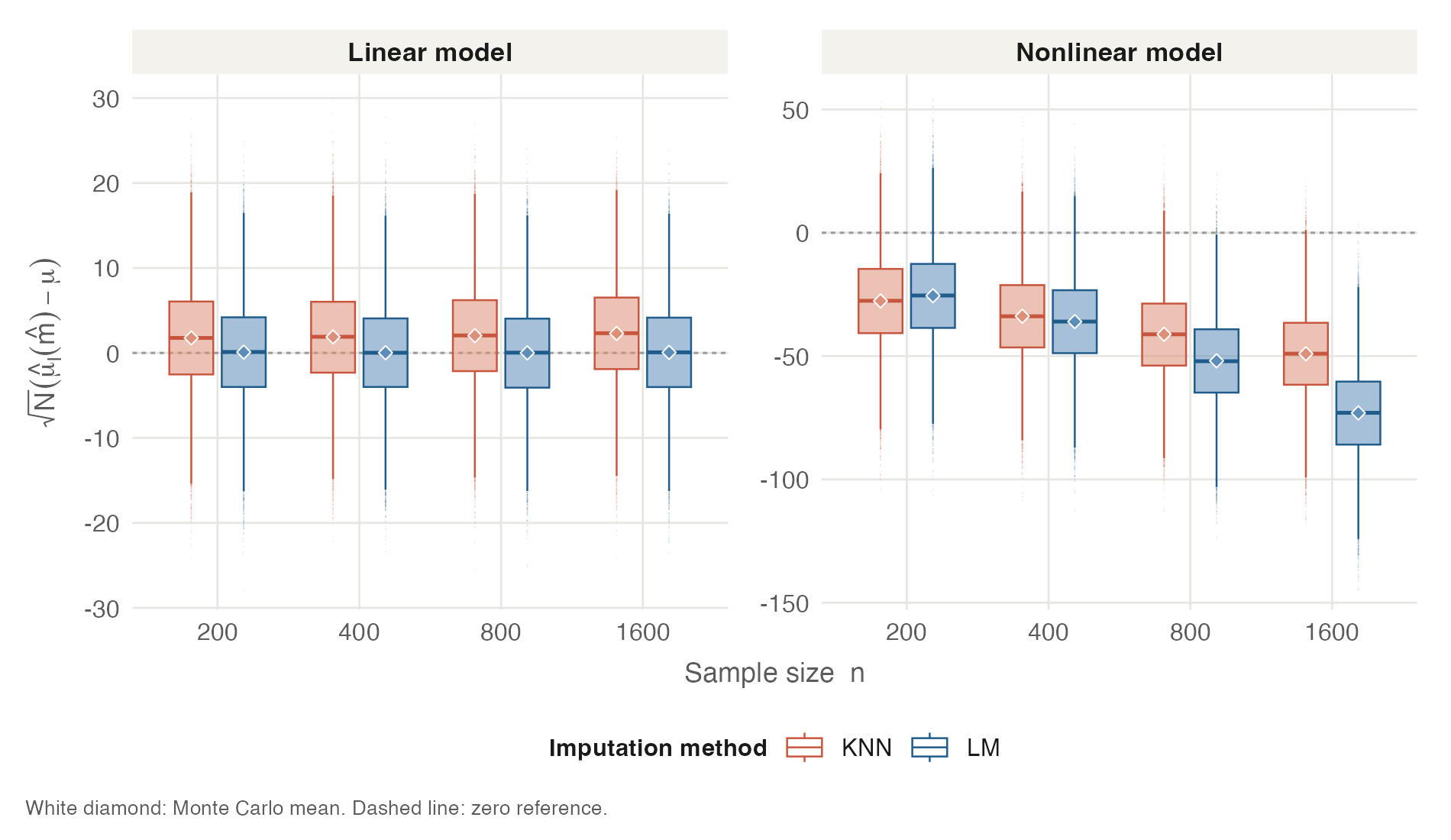} \vspace{0.5em} \refstepcounter{figure} \label{fig:customary_scaled_error} {\normalfont\small\noindent \textbf{Figure}~\thefigure: Monte Carlo distribution of the root-$N$-scaled error $\sqrt N\{\widehat\mu_I(\widehat m)-\mu\}$ of the customary plug-in imputed estimator.} 
\end{figure}

Figure~\ref{fig:customary_raw_error} displays the Monte Carlo distribution of the raw error
$
\widehat\mu_I(\widehat m)-\mu.
$
In the linear imputation model, linear regression imputation is nearly unbiased,
whereas KNN imputation has a small positive bias. The dispersion decreases
with sample size for both methods. In the nonlinear imputation model, linear
regression imputation has a persistent negative bias, reflecting model
misspecification. For KNN, both the magnitude of the bias and the dispersion
decrease with sample size, although the bias remains substantial.\\

Figure~\ref{fig:customary_scaled_error} shows the corresponding root-$N$-scaled error
$
\sqrt N\{\widehat\mu_I(\widehat m)-\mu\}.
$
Under a standard root-$n$ central limit theorem, the scaled distribution would
stabilize as $n$ and $N$ increase together. In the linear imputation model, linear
regression imputation remains centered close to zero with a stable spread.
KNN also has a fairly stable spread, but its small positive bias persists
after scaling.\\

In the nonlinear imputation model, although the KNN bias decreases with sample
size, its magnitude increases after multiplication by $\sqrt N$, while the
scaled dispersion remains approximately stable. This suggests that bias is
the main obstacle to root-$n$ inference over the sample sizes considered.
For linear regression imputation, the persistent misspecification bias becomes
increasingly pronounced after scaling.\\

\section{Oracle AIPW estimation}
\label{sec:aipw1}

The previous section showed that the customary plug-in imputed estimator
$\widehat\mu_I(\widehat m)$ is difficult to use as a general inferential tool
when $m(\cdot)$ is estimated by flexible machine-learning methods. The reason
is not only that ML estimators may converge slowly. The deeper issue is that the
customary imputed estimator is not based on a Neyman-orthogonal estimating
equation. As a result, errors in the fitted regression function may enter the
estimator at first order.\\

To recover root-$n_v$ inference under weaker conditions on the nuisance
estimators, we change the point estimator. We consider an augmented
inverse-probability-weighted estimator. The AIPW estimator uses both the imputation
regression function $m(\cdot)$ and the response propensity $p(\cdot)$.
For fixed working functions, the estimator is unbiased if either the
imputation regression or the response propensity is correct. Its estimating
equation is also Neyman-orthogonal. This section first studies the oracle version of the
AIPW estimator, where $m(\cdot)$ and $p(\cdot)$ are treated as known. The
feasible and cross-fitted versions are studied in the following sections.

\subsection{Point estimation}

The oracle AIPW estimator of $\mu$ is given by
\begin{align}
	\label{DR_OR}
	\widehat{\mu}_{\mathrm{aipw}}(m,p)
	&=
	\frac{1}{N}
	\left\{
	\sum_{k\in S}w_k m(\bx_k)
	+
	\sum_{k\in S_r}
	w_k
	\frac{y_k-m(\bx_k)}{p(\bx_k)}
	\right\} \nonumber\\
	&=
	\frac{1}{N}
	\sum_{k\in S}w_k\eta_k,
\end{align}
where the oracle pseudo-value is defined by
\begin{equation*}
	\label{pseudo-value_DR_OR}
	\eta_k
	=
	m(\bx_k)
	+
	\frac{r_k}{p(\bx_k)}
	\{y_k-m(\bx_k)\},
	\qquad k\in U.
\end{equation*}
Under \ref{A1}--\ref{A3}, the population-level response indicators make these pseudo-values well defined before sample selection; see Lemma~\ref{lem:conditional_moments}.
The residual correction uses respondents to estimate the prediction errors over the full sample. Under MAR, it recovers the full-sample estimator on average:
$$
\E_q
\left[
\widehat{\mu}_{\mathrm{aipw}}(m,p)
\right]
=
\frac{1}{N}\sum_{k\in S}w_k y_k
=
\widehat\mu_F.
$$
Taking the design expectation gives
$$
\E_{pq}\left[\widehat{\mu}_{\mathrm{aipw}}(m,p)-\mu\right]=0.
$$

The following bound shows that this correction preserves the usual $n_v^{-1/2}$ estimation rate, providing the benchmark for the fitted estimators studied later.

\begin{proposition}\label{prop:oracle_consistency}
	Assume \ref{A1}--\ref{A4}, \ref{D1}--\ref{D3}, and \ref{M1}. Then, there exists a constant $C>0$, independent of $v$, such that, for all sufficiently large $v$, $$\E_{mpq} \left[ \left(\widehat{\mu}_{aipw}(m,p)-\mu\right)^2\right]\leqslant\dfrac{C}{n_v}.$$
\end{proposition}
\begin{proof}
	See Appendix \ref{proof:oracle_consistency}.
\end{proof}

\subsection{Variance and variance estimation}
In this section, we study the total variance of $\widehat\mu_{\mathrm{aipw}}(m,p)$. Following the reverse approach of \cite{shao1999variance}, we decompose the total error as
\begin{align*}
	\widehat{\mu}_{\mathrm{aipw}}(m,p)-\mu
	&=
	\frac{1}{N}\sum_{k\in U}
	\left(\frac{I_k}{\pi_k}-1\right)\eta_k
	+
	\frac{1}{N}\sum_{k\in U}
	\left\{\frac{r_k}{p(\bx_k)}-1\right\}
	\{y_k-m(\bx_k)\}.
\end{align*}
Since the first term has zero sampling expectation and the second does not depend on sample selection, they are uncorrelated, leading to
\begin{equation}
	\label{Vtot_OR}
	V_{\mathrm{tot}}(\widehat{\mu}_{\mathrm{aipw}}(m,p))
	=
	V_1(\widehat{\mu}_{\mathrm{aipw}}(m,p))
	+
	V_2(\widehat{\mu}_{\mathrm{aipw}}(m,p)).
\end{equation}
The first component can be written as
\begin{equation}\label{V1_or}
	V_1(\widehat{\mu}_{\mathrm{aipw}}(m,p))
	=
	\E_{mq}
	\left[
	\frac{1}{N^2}
	\sum_{k\in U}
	\sum_{\ell\in U}
	\Delta_{k\ell}
	\frac{\eta_k}{\pi_k}
	\frac{\eta_\ell}{\pi_\ell}
	\right],
\end{equation}
where $\Delta_{k\ell}=\pi_{k\ell}-\pi_k\pi_\ell$. The second component is given by
\begin{equation}
	\label{V2_or}
	V_2(\widehat{\mu}_{\mathrm{aipw}}(m,p))
	=
	\frac{1}{N^2}
	\sum_{k\in U}
	\frac{1-p(\bx_k)}
	{p(\bx_k)}
	\sigma_k^2.
\end{equation}


If $V_1(\widehat{\mu}_{\mathrm{aipw}}(m,p))>0$, $\max_{k\in U_v}\pi_k=\mathcal{O}(n_v/N_v)$, and $\limsup_{v\to\infty}\max_{k\in U_v}\pi_k<1$, comparison with \eqref{V2_or} yields
\begin{equation}
	\label{eq:oracle_variance_ratio_bound}
	\frac{V_2(\widehat{\mu}_{\mathrm{aipw}}(m,p))}
	{V_1(\widehat{\mu}_{\mathrm{aipw}}(m,p))}
	\leq\frac{\max_{k\in U_v}\pi_k}{1-\max_{k\in U_v}\pi_k}
	=\mathcal{O}\left(\frac{n_v}{N_v}\right).
\end{equation}
For equal inclusion probabilities, the bound is $n_v/(N_v-n_v)=\mathcal{O}(n_v/N_v)$ when the sampling fraction is bounded away from one. The bound also applies when $n_v/N_v\xrightarrow[v\to\infty]{}0$, beyond our standing asymptotic regime, and then gives $$\frac{V_2(\widehat{\mu}_{\mathrm{aipw}}(m,p))}{V_1(\widehat{\mu}_{\mathrm{aipw}}(m,p))} \xrightarrow[v\to\infty]{} 0.$$
We now turn to variance estimation, starting with the first component
$V_1(\widehat{\mu}_{\mathrm{aipw}}(m,p))$.
Since the oracle AIPW estimator is a Horvitz--Thompson mean of the
pseudo-values, the usual design variance estimator
\begin{equation}
	\label{V1_OR_hat}
	\widehat{V}_1(m,p)
	=
	\frac{1}{N^2}
	\sum_{k\in S}
	\sum_{\ell\in S}
	\frac{\Delta_{k\ell}}{\pi_{k\ell}}
	\frac{\eta_k}{\pi_k}
	\frac{\eta_\ell}{\pi_\ell}
\end{equation}
is unbiased for this component:
$$
\E_{mpq}[\widehat{V}_1(m,p)]
=
V_1(\widehat{\mu}_{\mathrm{aipw}}(m,p)).
$$

We now turn to the second component,
$V_2(\widehat{\mu}_{\mathrm{aipw}}(m,p))$, which is a population total
of conditional residual variances weighted by the nonresponse odds
$\{1-p(\bx_k)\}/p(\bx_k)$.
A first estimator is obtained by inverse probability weighting.
A respondent's squared oracle residual provides an unbiased estimate
of $\sigma_k^2$. This leads to
\begin{equation}
	\label{V2_OR_IPW_hat}
	\widehat{V}_{2,\mathrm{ipw}}(m,p)
	=
	\frac{1}{N^2}
	\sum_{k\in S_r}
	w_k
	\frac{1-p(\bx_k)}
	{\{p(\bx_k)\}^2}
	\{y_k-m(\bx_k)\}^2.
\end{equation}
Indeed,
$$
\E_{mpq}
[\widehat{V}_{2,\mathrm{ipw}}(m,p)]
=
V_2(\widehat{\mu}_{\mathrm{aipw}}(m,p)).
$$

An alternative estimator uses the known conditional variances
$\sigma_k^2$ to adjust for the realized response pattern.
A nonrespondent contributes zero to \eqref{V2_OR_IPW_hat}, although
its nonresponse error $-\epsilon_k$ has model variance $\sigma_k^2$.
We therefore add $\sigma_k^2$ to each nonrespondent's contribution
and subtract $\{1-p(\bx_k)\}\sigma_k^2/p(\bx_k)$ from each
respondent's contribution. This gives
\begin{equation}
	\label{V2_OR_BC_hat}
	\widehat V_{2,\mathrm{bc}}(m,p)
	=
	\widehat V_{2,\mathrm{ipw}}(m,p)
	+
	\frac{1}{N^2}\sum_{k\in S}w_k
	\left\{1-\frac{r_k}{p(\bx_k)}\right\}\sigma_k^2.
\end{equation}
For a fixed response pattern, the model expectation of each
adjusted contribution equals the model variance of the unit's
nonresponse error $\{r_k/p(\bx_k)-1\}\epsilon_k$:
\begin{equation*}
	\E_m\left[
	r_k\frac{1-p(\bx_k)}{\{p(\bx_k)\}^2}\epsilon_k^2
	+
	\left\{1-\frac{r_k}{p(\bx_k)}\right\}\sigma_k^2
	\right]
	=
	\left\{\frac{r_k}{p(\bx_k)}-1\right\}^2\sigma_k^2.
\end{equation*}
Since the added correction has zero expectation under the response
mechanism, the resulting estimator is unbiased for $V_2$ under
$mpq$:
\begin{equation*}
	\E_{mpq}[\widehat V_{2,\mathrm{bc}}(m,p)]
	=
	V_2(\widehat{\mu}_{\mathrm{aipw}}(m,p)).
\end{equation*}
For either choice
\begin{equation*}
	\widehat V_2(m,p)\in
	\left\{
	\widehat V_{2,\mathrm{ipw}}(m,p),
	\widehat V_{2,\mathrm{bc}}(m,p)
	\right\},
\end{equation*}
the total variance estimator is given by
\begin{equation*}
	\widehat{V}_{\mathrm{aipw}}(m,p)
	=
	\widehat{V}_1(m,p)
	+
	\widehat{V}_{2}(m,p).
\end{equation*}

To use these estimators for studentization, their error must be
negligible relative to the true variance. The following result
gives the required control for either choice.

\begin{theorem}
	\label{theo:oracle_variance}
	Assume \ref{A1}--\ref{A4}, \ref{D1}--\ref{D4}, and \ref{M2}.
	For either choice of $\widehat V_2(m,p)$ above,
	\begin{equation*}
		n_v\left[
		\widehat V_1(m,p)
		-
		V_1(\widehat{\mu}_{\mathrm{aipw}}(m,p))
		\right]
		=
		o_{\P}(1),
		\quad
		n_v\left[
		\widehat V_2(m,p)
		-
		V_2(\widehat{\mu}_{\mathrm{aipw}}(m,p))
		\right]
		=
		o_{\P}(1),
	\end{equation*}
	and consequently
	\begin{equation*}
		n_v\left[
		\widehat V_{\mathrm{aipw}}(m,p)
		-
		V_{\mathrm{tot}}(\widehat{\mu}_{\mathrm{aipw}}(m,p))
		\right]
		=
		o_{\P}(1).
	\end{equation*}
	If, in addition,
	$N_vV_{\mathrm{tot}}(\widehat{\mu}_{\mathrm{aipw}}(m,p))
	\xrightarrow[v\to\infty]{\P}c$ for some $c>0$, then
	\begin{equation*}
		\frac{\widehat V_{\mathrm{aipw}}(m,p)}
		{V_{\mathrm{tot}}(\widehat{\mu}_{\mathrm{aipw}}(m,p))}
		\xrightarrow[v\to\infty]{\P}1.
	\end{equation*}
\end{theorem}

\begin{proof}
	See Appendix~\ref{proof:oracle_variance}.
\end{proof}

\subsection{Central limit theorem}

To obtain a normal approximation, we analyze the sampling and response components separately. The pseudo-values are measurable with respect to $\cB_v=\sigma\{(\bx_k,y_k,r_k):k\in U_v\}$, so we formulate the finite-population design CLT for the sampling term conditionally on $\cB_v$. For the response term, \ref{A2}, \ref{A4}, and \ref{M1} give

\begin{equation}
\label{eq:oracle_response_variance_limit}
N_vV_2(\widehat{\mu}_{\mathrm{aipw}}(m,p))
\xrightarrow[v\to\infty]{\mathrm{a.s.}}
c_2:=\E_P\left[\frac{1-p(\bx_1)}{p(\bx_1)}\sigma^2(\bx_1)\right]<\infty.
\end{equation}
The limit $c_2$ determines whether the response component contributes at the $N_v^{-1/2}$ scale. If $c_2>0$, \ref{M2} yields its CLT; if $c_2=0$, the term is negligible. The sampling CLT requires an additional, design-specific assumption; it does not follow from \ref{D1}--\ref{D4} alone. We impose it below, together with a positive total limiting variance.

\begin{assumptionlist}{CL}
\item \label{CL1} There exists $c_1\geq0$, with $c_1+c_2>0$, such that
\begin{equation*}
N_vV_1(\widehat{\mu}_{\mathrm{aipw}}(m,p))
\xrightarrow[v\to\infty]{}c_1.
\end{equation*}

\item \label{CL2} If $c_1>0$, the sampling component satisfies
\begin{equation*}
\frac{N_v^{-1}\sum_{k\in U_v}(I_k/\pi_k-1)\eta_k}
{\sqrt{V_1(\widehat{\mu}_{\mathrm{aipw}}(m,p))}}
\xrightarrow[v\to\infty]{\cL}\mathcal N(0,1),
\qquad\text{conditionally on }\cB_v.
\end{equation*}
\end{assumptionlist}

The conditional convergence in \ref{CL2} is in probability. Design CLTs usually normalize by the conditional design standard deviation for the realized pseudo-values. Under the assumptions of Theorem~\ref{theo:oracle_variance} and \ref{CL1}, when $c_1>0$, the argument in Appendix~\ref{proof:oracle_variance} gives
$$
\frac{\V_p(\widehat\mu_{\mathrm{aipw}}(m,p)\mid\cB_v)}
{V_1(\widehat\mu_{\mathrm{aipw}}(m,p))}
\xrightarrow[v\to\infty]{\P}1,
$$
so this normalization is equivalent to that in \ref{CL2}. Combining the two components and applying Theorem~\ref{theo:oracle_variance} gives the following studentized limit.

\begin{theorem}
\label{theo1}
Assume \ref{A1}--\ref{A4}, \ref{D1}--\ref{D4}, \ref{M2}, and conditions~\ref{CL1}--\ref{CL2}. Then
\begin{equation*}
\frac{\widehat\mu_{\mathrm{aipw}}(m,p)-\mu}{\sqrt{\widehat V_{\mathrm{aipw}}(m,p)}}\xrightarrow[v\to\infty]{\cL}\mathcal N(0,1).
\end{equation*}
\end{theorem}

Here and below, the statistic is set to zero when the estimated variance is nonpositive, an event whose probability vanishes by ratio consistency.

\begin{proof}
See Appendix~\ref{proof:theo1}.
\end{proof}

The sampling approximation in \ref{CL2} can be checked using design-specific central limit theorems for simple random sampling without replacement \citep{hajek1960}, rejective sampling \citep{hajek1964}, and stratified sampling \citep{bickel1984}. Related results for stratified multistage sampling with replacement are given by \cite{krewski1981}.\\

\subsection{Neyman orthogonality}

To prepare for estimation of the nuisance functions, we examine how errors in them affect the bias of the AIPW estimator.
Let $\varphi(\cdot)$ denote a generic outcome regression function and let
$\gamma(\cdot)$ denote a generic positive response propensity. Consider the
sample estimating equation
\begin{equation*}
	\label{eq:aipw_oracle_score}
	\widehat\Psi_N(\mu,\varphi,\gamma)
	=
	\frac{1}{N}
	\sum_{k\in S}
	w_k
	\left[
	\varphi(\bx_k)
	+
	\frac{r_k}{\gamma(\bx_k)}
	\{y_k-\varphi(\bx_k)\}
	\right]
	-\mu.
\end{equation*}
The solution of $\widehat\Psi_N(\mu,\varphi,\gamma)=0$ is the AIPW estimator evaluated at $(\varphi,\gamma)$. Its bias for fixed working functions is represented by the population moment
$$
\Psi_N(\varphi,\gamma)
=
\E_{mpq}
\left[
\widehat\Psi_N
\{\mu,\varphi,\gamma\}
\right].
$$
For fixed working functions, \ref{A1}--\ref{A3} and design unbiasedness give
\begin{equation}
\label{eq:oracle_product_bias}
\Psi_N(\varphi,\gamma)
=\frac{1}{N}\sum_{k\in U}
\frac{\{\varphi(\bx_k)-m(\bx_k)\}\{\gamma(\bx_k)-p(\bx_k)\}}
{\gamma(\bx_k)}.
\end{equation}
The product vanishes when either working function is correct, giving double robustness. It also shows that a small perturbation of either function has no first-order effect at the truth: for admissible directions $h_m,h_p$ with $p+t h_p>0$ near $t=0$, Neyman orthogonality follows as
\begin{equation*}
\left.\frac{\partial}{\partial t}\Psi_N(m+t h_m,p)\right|_{t=0}=0,
\qquad
\left.\frac{\partial}{\partial t}\Psi_N(m,p+t h_p)\right|_{t=0}=0.
\end{equation*}
The product in \eqref{eq:oracle_product_bias} motivates the rate condition used below to make nuisance-estimation bias negligible at the $n_v^{-1/2}$ scale.

\section{AIPW estimator based on estimated nuisance functions}
\label{sec:aipw2}

\subsection{Point estimation and data reuse}

Replacing the oracle nuisance functions $m(\cdot)$ and $p(\cdot)$ by fitted versions in (\ref{DR_OR}) leads to the feasible AIPW estimator:
\begin{align}
	\label{DR}
	\widehat{\mu}_{\mathrm{aipw}}(\widehat m,\widehat p)
	&=
	\frac{1}{N}
	\left\{
	\sum_{k\in S}w_k\widehat m(\bx_k)
	+
	\sum_{k\in S_r}
	w_k
	\frac{y_k-\widehat m(\bx_k)}
	{\widehat p(\bx_k)}
	\right\} \nonumber\\
	&=
	\frac{1}{N}
	\sum_{k\in S}w_k\widehat\eta_k,
\end{align}
where
\begin{equation*}
	\label{pseudo-value_DR}
	\widehat\eta_k
	=
	\widehat m(\bx_k)
	+
	\frac{r_k}{\widehat p(\bx_k)}
	\{y_k-\widehat m(\bx_k)\},
	\qquad k\in S.
\end{equation*}
Here, $\widehat m$ and $\widehat p$ estimate the outcome regression and response propensity, respectively. We evaluate both functions at every sampled unit, take $0<\widehat p(\bx_k)\leq1$ for $k\in S$, and assume that the expectations below exist.\\

Neyman orthogonality makes the population bias in \eqref{eq:oracle_product_bias}
depend on a product of nuisance errors. With fitted nuisance functions, however,
we must also control the random terms arising from their use in the sample score,
particularly when fitting and evaluation use the same observations. To separate
these effects, define
$$
\Delta_m(\bx)
=
\widehat m(\bx)-m(\bx),
\qquad
\Delta_p(\bx)
=
\widehat p(\bx)-p(\bx).
$$
The difference between the feasible and oracle AIPW estimators admits the exact
decomposition
\begin{equation*}
	\widehat{\mu}_{\mathrm{aipw}}(\widehat m,\widehat p)
	-
	\widehat{\mu}_{\mathrm{aipw}}(m,p)
	=
	T_{\mathrm{reg}}
	+
	T_{\mathrm{prop}}
	+
	T_{\mathrm{prod}},
\end{equation*}
where
$$
T_{\mathrm{reg}}
=
\frac{1}{N}
\sum_{k\in S}
w_k
\left\{
1-\frac{r_k}{p(\bx_k)}
\right\}
\Delta_m(\bx_k),
$$
$$
T_{\mathrm{prop}}
=
-
\frac{1}{N}
\sum_{k\in S}
w_kr_k \{y_k-m(\bx_k)\}
\frac{\Delta_p(\bx_k)}
{p(\bx_k)\widehat p(\bx_k)},
$$
and
$$
T_{\mathrm{prod}}
=
\frac{1}{N}
\sum_{k\in S}
w_kr_k
\frac{
	\Delta_m(\bx_k)\Delta_p(\bx_k)
}{
	p(\bx_k)\widehat p(\bx_k)
}.
$$
The term $T_{\mathrm{reg}}$ is the contribution of the outcome regression error,
whereas $T_{\mathrm{prop}}$ is the contribution of the response-propensity error. If the
corresponding nuisance errors could be treated as external to the $k$-th score
contribution, these two terms would be centered under the response
and outcome mechanisms, respectively. With fitted response probabilities bounded
away from zero, $T_{\mathrm{prod}}$ is controlled by a product of the two prediction errors.
A suitable product-rate condition makes this term $o_{\P}(n_v^{-1/2})$;
consistency of the nuisance estimators alone does not suffice.\\

Without sample splitting, the nuisance errors cannot generally be treated as
external to the score contributions. More explicitly, one may write
$\widehat m(\bx_k)
=\widehat m(\bx_k;\mathbf I,\mathbf r,\mathbf y_{\mathrm{obs}},\mathbf X)$,
where $\mathbf y_{\mathrm{obs}}$ denotes the observed respondent outcomes, and
$\widehat p(\bx_k)
=\widehat p(\bx_k;\mathbf I,\mathbf r,\mathbf X)$. Thus,
$\widehat m(\bx_k)$ may depend on the same response indicator $r_k$ and
outcome $y_k$ that enter the $k$-th score contribution. Similarly,
$\widehat p(\bx_k)$ may depend on the same $r_k$ that appears in the
inverse-propensity correction. This is the data-reuse problem.\\

Since $\E_q[r_k]=p(\bx_k)$, the response-bias contribution of $T_{\mathrm{reg}}$ has the exact covariance representation
\begin{equation*}
	\E_q
	\left[
	T_{\mathrm{reg}}
	\right]
	=
	-
	\frac{1}{N}
	\sum_{k\in S}
	\frac{w_k}{p(\bx_k)}
	\operatorname{Cov}_q
	\left(
	r_k,\Delta_m(\bx_k)
	\right).
\end{equation*}
The covariance compares, in expectation, the fitted prediction error when unit $k$ is included among the training respondents with the error when it is excluded.\\

The propensity term has a different interpretation. Since $\widehat p$ uses only $(\mathbf X,\mathbf I,\mathbf r)$, the coefficient multiplying the residual in $T_{\mathrm{prop}}$ is $\sigma(\mathbf X,\mathbf I,\mathbf r)$-measurable. By Lemma~\ref{lem:conditional_moments}, $\E_m[y_k-m(\bx_k)]=0$, and hence
$$
\E_m
\left[
T_{\mathrm{prop}}
\right]
=
0.
$$
Thus, data reuse in a propensity learner that does not use outcomes does not generate $mpq$ bias through $T_{\mathrm{prop}}$. It can nevertheless affect the product term $T_{\mathrm{prod}}$ and the estimator's variance.\\

Since the oracle AIPW estimator is unbiased under the $mpq$ framework, the
preceding identities yield the following exact representation of the
bias:
\begin{align*}
	\E_{mpq}
	\left[
	\widehat{\mu}_{\mathrm{aipw}}(\widehat m,\widehat p)-\mu
	\right]
	\quad =
	-
	\E_{mp}
	\left[
	\frac{1}{N}
	\sum_{k\in S}
	\frac{w_k}{p(\bx_k)}
	\operatorname{Cov}_q
	\left(
	r_k,\Delta_m(\bx_k)
	\right)
	\right]
	+
	\E_{mpq}[T_{\mathrm{prod}}].
\end{align*}
The first term averages the response-bias contribution over the outcome model and sampling design; the second is the expected product remainder. Even when these bias contributions vanish, fluctuations of the combined remainder must also be negligible at the $n_v^{-1/2}$ scale for oracle equivalence.\\

If the outcome learner interpolates the respondents, so that $\widehat m(\bx_k)=y_k$ for $k\in S_r$, all fitted respondent residuals vanish. The AIPW estimator then reduces to customary imputation regardless of $\widehat p$:
$$
\widehat\mu_{\mathrm{aipw}}(\widehat m,\widehat p)
=\widehat\mu_I(\widehat m),
\qquad
T_{\mathrm{prop}}+T_{\mathrm{prod}}=0.
$$

The same vanishing residuals give $\widehat V_{2,\mathrm{ipw}}(\widehat m,\widehat p)=0$ in \eqref{V2_hat_feasible_ipw} below, even when the oracle nonresponse variance component in \eqref{V2_or} is positive. Thus, perfect fit to the respondents can conceal nonresponse variability.

\subsection{Variance estimation}

A natural plug-in approach is to estimate the oracle variance components using fitted nuisance functions. Replacing the oracle pseudo-values gives
\begin{equation}
	\label{V1_hat_feasible}
	\widehat{V}_{1}(\widehat m,\widehat p)
	=
	\frac{1}{N^2}
	\sum_{k\in S}
	\sum_{\ell\in S}
	\frac{\Delta_{k\ell}}{\pi_{k\ell}}
	\frac{\widehat\eta_k}{\pi_k}
	\frac{\widehat\eta_\ell}{\pi_\ell}.
\end{equation}
Replacing $m$ and $p$ in \eqref{V2_OR_IPW_hat} by their fitted counterparts gives
\begin{equation}
	\label{V2_hat_feasible_ipw}
	\widehat{V}_{2,\mathrm{ipw}}(\widehat m,\widehat p)
	=
	\frac{1}{N^2}
	\sum_{k\in S_r}
	w_k
	\frac{1-\widehat p(\bx_k)}
	{\{\widehat p(\bx_k)\}^2}
	\{y_k-\widehat m(\bx_k)\}^2.
\end{equation}
Combining \eqref{V1_hat_feasible} and \eqref{V2_hat_feasible_ipw} gives the direct plug-in estimator
$$
\widehat{V}_{\mathrm{plug}}
\{\widehat{\mu}_{\mathrm{aipw}}(\widehat m,\widehat p)\}
=
\widehat{V}_{1}(\widehat m,\widehat p)
+
\widehat{V}_{2,\mathrm{ipw}}(\widehat m,\widehat p).
$$

An alternative bias-corrected estimator of the second variance component is obtained by replacing  $m$, $p$, and $\sigma_k^2$ in \eqref{V2_OR_BC_hat} by their fitted counterparts, leading to
\begin{equation}
	\label{V2_hat_feasible_bc}
	\widehat{V}_{2,\mathrm{bc}}(\widehat m,\widehat p,\widehat\sigma^2)
	=
	\frac{1}{N^2}
	\sum_{k\in S}
	w_k
	\left[
	r_k
	\frac{1-\widehat p(\bx_k)}
	{\{\widehat p(\bx_k)\}^2}
	\{y_k-\widehat m(\bx_k)\}^2
	+
	\left\{
	1-\frac{r_k}{\widehat p(\bx_k)}
	\right\}
	\widehat\sigma_k^2
	\right].
\end{equation}
The corresponding plug in variance estimator is then given by
$$
\widehat{V}_{\mathrm{bc}}
\{\widehat{\mu}_{\mathrm{aipw}}(\widehat m,\widehat p)\}
=
\widehat{V}_{1}(\widehat m,\widehat p)
+
\widehat{V}_{2,\mathrm{bc}}(\widehat m,\widehat p,\widehat\sigma^2).
$$

The oracle bias correction does not automatically yield an unbiased estimator after its ingredients are fitted. Appendix~\ref{app:fixed_variance_identities} records the related conditional identities for fixed working functions and distinguishes their variance targets.\\

We end this section with a brief discussion of the effect of data reuse on the plug-in variance estimator.
Without sample splitting, the fitted pseudo-values depend on the sample used to
construct them, so the oracle design-variance argument no longer applies directly.
Treating these pseudo-values as fixed can miss variability associated with fitting the nuisance functions, so
$\widehat V_1(\widehat m,\widehat p)$ need not capture the sampling contribution
to the feasible estimator's error variance.\\

The second component is vulnerable to the same data reuse through the squared
fitted residuals in both \eqref{V2_hat_feasible_ipw} and \eqref{V2_hat_feasible_bc}.
An outcome learner can fit the respondent outcomes closely while still making
substantial prediction errors for nonrespondents. The resulting in-sample
residuals can therefore understate the residual variance that the oracle
formulas use. In the bias-corrected estimator, the additional estimate
$\widehat\sigma_k^2$ may be affected as well if it is obtained from those same
in-sample residuals.\\

Response-model overfitting can also distort both point and variance estimation. In the extreme case where $\widehat p(\bx_k)=1$ for every respondent,
$$
\widehat\eta_k=r_ky_k+(1-r_k)\widehat m(\bx_k),
\qquad
\widehat V_{2,\mathrm{ipw}}=0.
$$
The AIPW estimator again reduces to customary imputation, and the IPW estimate of the second variance component vanishes even if the fitted respondent residuals are nonzero. Here, the inverse-propensity factors equal one for respondents; very small fitted propensities for respondents can cause additional instability.\\

Estimation without sample splitting remains possible under suitable restrictions on the complexity of the nuisance fits \citep{chernozhukov2018Double}. We use cross-fitting to control data reuse with flexible learners. The next section establishes oracle equivalence and consistent variance estimation under the stated rate and regularity conditions.

\section{AIPW estimation based on cross-fitting}\label{sec:aipw3}
\subsection{Cross-fitting}
\label{subsec:crossfitting}

Cross-fitting is used to avoid the data-reuse problem described above. The idea is simple: the nuisance functions are estimated on one part of the sample and evaluated on another part. Thus, for a given unit $k$, the predictions used in its AIPW score are obtained from models that were not trained using that unit.\\

Let $L\geq 2$ be a fixed integer. We randomly partition the sample $S$ into $L$ mutually exclusive, nonempty folds whose sizes are fixed in advance and differ by at most one,
$$
\mathcal F_1,\ldots,\mathcal F_L,
\qquad
S=\bigcup_{t=1}^L \mathcal F_t,
\qquad
\mathcal F_t\cap \mathcal F_{t'}=\varnothing
\quad\text{for } t\neq t'.
$$
We write $N_t=|\mathcal F_t|$ for the size of fold $t$.
For $k\in U$ and $t=1,\ldots,L$, define the fold membership indicators and collect them in the vector
$$
\delta_{kt}=\mathds{1}_{\{k\in\mathcal F_t\}},
\qquad
\boldsymbol{\delta}
=(\delta_{11},\ldots,\delta_{1L},\ldots,\delta_{N1},\ldots,\delta_{NL})^\top.
$$
The vector $\boldsymbol{\delta}$ determines the fold partition; all its entries corresponding to a nonsampled unit are zero.
Conditionally on the selected sample, $\boldsymbol{\delta}$ is generated independently of the outcome values and response indicators. It is included in the conditioning information used below. For each unit $k\in S$, let
$
\nu(k)\in\{1,\ldots,L\}
$
denote the index of the fold containing $k$, so that
$
k\in \mathcal F_{\nu(k)}.
$\\

For each fold $t=1,\ldots,L$, the nuisance functions are fitted using only the training sample
$
S_{-t}:=S\setminus \mathcal F_t.
$
The outcome regression function is estimated from the respondents in the training sample,
$
S_{r,-t}:=S_r\cap S_{-t},
$
leading to a fitted function $\widehat m_{-t}(\cdot)$. The response propensity is estimated using only the covariates and response indicators in $S_{-t}$, leading to a fitted function $\widehat p_{-t}(\cdot)$. All fitted response probabilities are taken in $(0,1]$. In practice, the survey weights may be included in these fitting procedures when appropriate.\\

The two nuisance functions may be estimated by different learners. All tuning
and model selection must use only the training sample $S_{-t}$, for example
through inner cross-validation. Suitable criteria include weighted squared
prediction error for the outcome regression and weighted log-loss or a Brier
score for the response probabilities.\\

For each unit $k\in S$, the cross-fitted nuisance predictions are given by
$
\widehat m_{-\nu(k)}(\bx_k)
$
and
$
\widehat p_{-\nu(k)}(\bx_k).
$
The subscript $-\nu(k)$ emphasizes that the prediction for unit $k$ is obtained from models fitted without using any observation in the fold containing $k$. In particular, if $k$ is a respondent, $y_k$ is not used to train the model that produces $\widehat m_{-\nu(k)}(\bx_k)$. Similarly, $r_k$ is not used to train the model that produces $\widehat p_{-\nu(k)}(\bx_k)$. To analyze the resulting estimator, we next specify the training information and how prediction errors are measured on the excluded folds.\\

The balanced construction ensures that each fold grows proportionally to the full sample:
$$
\frac{N_t}{n_v}\xrightarrow[v \to \infty]{}\frac{1}{L},
\qquad t=1,\ldots,L.
$$
The fitted nuisance functions are measurable with respect to the training sigma-algebra
$$
\mathcal D_{-t}
:=
\sigma\!\left(
\mathbf X,\mathbf I,\boldsymbol{\delta},
\bigl\{\mathds{1}_{\{k\in S_{-t}\}}(y_k,r_k):k\in U_v\bigr\}
\right),
$$
where $\sigma(\cdot)$ denotes the sigma-algebra generated by its arguments.
The indicators retain $(y_k,r_k)$ only for units in the training sample $S_{-t}$.
This conditioning information includes all training outcomes, whether
observed or not; fitting uses only the observed data.\\

For $r\geq1$, we measure prediction error on the excluded fold by
\begin{equation}
\label{eq:prediction_risk}
\mathcal R_r(\widehat m_{-t})
=\E_{mpq}\left[
\frac{1}{N_t}
\sum_{k\in\mathcal F_t}
|\widehat m_{-t}(\bx_k)-m(\bx_k)|^r
\right],
\end{equation}
Risks for $p$ and $\sigma^2$ are defined analogously. For cross-fitted quantities,
joint expectations and probabilities include the independent fold assignment. 

\subsection{Cross-fitted point estimation}

Using the cross-fitted nuisance estimates defined above, the cross-fitted AIPW pseudo-value is given by
\begin{equation}
\label{pseudo-value_Dr_cf}
\widehat\eta_{k,\mathrm{cf}}
=
\widehat m_{-\nu(k)}(\bx_k)
+
\frac{r_k}{\widehat p_{-\nu(k)}(\bx_k)}
\{y_k-\widehat m_{-\nu(k)}(\bx_k)\},
\qquad k\in S.
\end{equation}
The cross-fitted AIPW estimator is then given by
\begin{equation}
\label{dr_cross}
\widehat\mu_{\mathrm{aipw,cf}}(\widehat m,\widehat p)
=\frac{1}{N}\sum_{k\in S}w_k\widehat\eta_{k,\mathrm{cf}}.
\end{equation}

Using the three estimation-error terms introduced in
Section~\ref{sec:aipw2}, we decompose the cross-fitted estimation error as
\begin{equation}
\label{eq:cf_oracle_decomposition}
\begin{aligned}
\widehat\mu_{\mathrm{aipw,cf}}(\widehat m,\widehat p)-\mu
=\{\widehat\mu_{\mathrm{aipw}}(m,p)-\mu\}
+\sum_{t=1}^L
\left(T_{\mathrm{reg},t}^{\mathrm{cf}}
+T_{\mathrm{prop},t}^{\mathrm{cf}}
+T_{\mathrm{prod},t}^{\mathrm{cf}}\right),
\end{aligned}
\end{equation}
where the contributions from fold $t$ are given by
\begin{align*}
T_{\mathrm{reg},t}^{\mathrm{cf}}
&=\frac{1}{N_v}\sum_{k\in\mathcal F_t}
w_k\left(1-\frac{r_k}{p(\bx_k)}\right)
\{\widehat m_{-t}(\bx_k)-m(\bx_k)\},\\
T_{\mathrm{prop},t}^{\mathrm{cf}}
&=-\frac{1}{N_v}\sum_{k\in\mathcal F_t}
\frac{w_kr_k\epsilon_k\{\widehat p_{-t}(\bx_k)-p(\bx_k)\}}
{p(\bx_k)\widehat p_{-t}(\bx_k)},\\
T_{\mathrm{prod},t}^{\mathrm{cf}}
&=\frac{1}{N_v}\sum_{k\in\mathcal F_t}
\frac{w_kr_k\{\widehat m_{-t}(\bx_k)-m(\bx_k)\}
\{\widehat p_{-t}(\bx_k)-p(\bx_k)\}}
{p(\bx_k)\widehat p_{-t}(\bx_k)}.
\end{align*}
The first two terms involve the regression and propensity errors, respectively;
the third involves their product. Since the nuisance functions for fold $t$
are fitted without using its outcomes or responses, the first two terms are
centered conditionally on the training information. By
Lemma~\ref{lem:conditional_moments},
$$
\E_{mpq}\left[T_{\mathrm{reg},t}^{\mathrm{cf}}\mid\mathcal D_{-t}\right]
=\E_{mpq}\left[T_{\mathrm{prop},t}^{\mathrm{cf}}\mid\mathcal D_{-t}\right]
=0.
$$
The two centered terms still fluctuate, so their prediction risks must also
be controlled. The following result distinguishes consistency from oracle
equivalence at the $n_v^{-1/2}$ scale.\\

Both conclusions use the following positivity condition on the fitted
response probabilities: there exists $\rho>0$ such that
\begin{equation}
\label{eq:fitted_propensity_positivity}
\P\left(\min_{1\leq t\leq L}\min_{k\in\mathcal F_t}
\widehat p_{-\nu(k)}(\bx_k)\geq\rho\right)
\xrightarrow[v\to\infty]{}1.
\end{equation}

\begin{theorem}
\label{Theo2}
Assume
\ref{A1}--\ref{A4}, \ref{D2}, and the positivity condition
\eqref{eq:fitted_propensity_positivity}.
\begin{enumerate}[label=(\roman*),ref=\roman*,leftmargin=*]
\item \label{thm:cf_consistency} \textbf{Consistency.}
Assume also \ref{D1}, \ref{D3}, and \ref{M1}. If, for every fold $t$,
$$
\mathcal R_2(\widehat m_{-t})=\mathcal O(1),
\qquad
\mathcal R_2(\widehat m_{-t})\mathcal R_2(\widehat p_{-t})=o(1),
$$
then
$$
\widehat\mu_{\mathrm{aipw,cf}}(\widehat m,\widehat p)-\mu=o_{\P}(1).
$$
\item \label{thm:cf_oracle_equivalence} \textbf{Oracle equivalence.}
Assume also \ref{M3}. If, for every fold $t$,
\begin{equation}
\label{eq:cf_product_rate}
\mathcal R_2(\widehat m_{-t})=o(1),
\qquad \mathcal R_2(\widehat p_{-t})=o(1),
\qquad n_v\mathcal R_2(\widehat m_{-t})\mathcal R_2(\widehat p_{-t})=o(1),
\end{equation}
then
\begin{equation*}
\sqrt{n_v}\left\{
\widehat\mu_{\mathrm{aipw,cf}}(\widehat m,\widehat p)
-\widehat\mu_{\mathrm{aipw}}(m,p)\right\}=o_{\P}(1).
\end{equation*}
\end{enumerate}
\end{theorem}

\begin{proof}
See Appendix~\ref{proof:Theo2}.
\end{proof}

Part~(\ref{thm:cf_consistency}) gives double robustness: consistency holds when
either nuisance function is estimated consistently in prediction risk and the
other has bounded risk. The propensity risk is automatically bounded because
both $p$ and $\widehat p$ take values in $(0,1]$.
Part~(\ref{thm:cf_oracle_equivalence}) requires both risks to vanish and their
product to decrease fast enough that nuisance estimation has no first-order
effect. This stronger conclusion supports the variance estimation and
inference results in the next subsection.

\begin{remark}
\label{rem:cf_dependent_sampling}
Theorem~\ref{Theo2} allows the sampling indicators to be dependent while
using an ordinary random partition of the sampled units. Under
\ref{A1}--\ref{A3}, outcomes and responses remain mutually independent
conditionally on the covariates, sampling indicators, and fold assignment.
Consequently, the training information outside a fold leaves the outcome
and response moments within that fold unchanged, as established in
Lemma~\ref{lem:conditional_moments}. This conditional independence is
sufficient to control the nuisance estimation errors through cross-fitting.
The dependence induced by sampling is accounted for in the oracle limit
distribution and the variance estimator. Under the conditions of
Corollary~\ref{cor:cf_clt}, ordinary cross-fitting therefore yields valid
inference for the finite population mean under the joint distribution induced by the imputation, model, the sampling design, and the response mechanism. 
\end{remark}

\begin{remark}
\label{rem:cf_rates}
The product-rate condition allows the two nuisance functions to be estimated
at different rates. For example, suppose that, for every fold $t$,
$$
\mathcal R_2(\widehat m_{-t})=\mathcal O(n_v^{-2\alpha_m}),
\qquad
\mathcal R_2(\widehat p_{-t})=\mathcal O(n_v^{-2\alpha_p}),
$$
with $\alpha_m,\alpha_p>0$. Then the conditions in
\eqref{eq:cf_product_rate} hold whenever
$\alpha_m+\alpha_p>1/2$. Equal exponents larger than $1/4$ suffice.
If one nuisance estimator has root mean squared prediction error
$\mathcal O(n_v^{-1/2})$, the other only needs to be consistent.
For flexible learners, the available rates depend on the method and the
regression functions, as discussed in Section~\ref{sec:literature}.
Aggregation, considered in Section~\ref{sec:aggregation}, can help select or
combine learners with suitable prediction accuracy.
\end{remark}

\subsection{Variance estimation}
\label{sec:aipw_variance_cf}

We estimate the oracle variance in \eqref{Vtot_OR} by adding estimates of
its sampling and nonresponse components:
\begin{equation}
\label{Vtot_hat_cf}
\widehat V_{\mathrm{tot,cf}}
=\widehat V_{1,\mathrm{cf}}+\widehat V_{2,\mathrm{cf}}.
\end{equation}

The first component is estimated by applying the usual Horvitz--Thompson
variance formula to the cross-fitted pseudo-values in
\eqref{pseudo-value_Dr_cf}:
\begin{equation}
\label{V1_hat_cf}
\widehat V_{1,\mathrm{cf}}
=\frac{1}{N^2}\sum_{k\in S}\sum_{\ell\in S}
\frac{\Delta_{k\ell}}{\pi_{k\ell}}
\frac{\widehat\eta_{k,\mathrm{cf}}}{\pi_k}
\frac{\widehat\eta_{\ell,\mathrm{cf}}}{\pi_\ell}.
\end{equation}

For the second component, we consider three estimators. They differ in
whether they use predictions of the conditional residual variance, and
in how much accuracy they require from these predictions.

\begin{itemize}
\item [(a)] IPW estimator:  The simplest choice uses the cross-fitted respondent residuals
$\widehat e_{k,\mathrm{cf}}=y_k-\widehat m_{-\nu(k)}(\bx_k)$.
The IPW estimator is given by
\begin{equation}
\label{V2_hat_cf_ipw}
\widehat V_{2,\mathrm{ipw,cf}}
=\frac{1}{N^2}\sum_{k\in S_r}w_k
\frac{1-\widehat p_{-\nu(k)}(\bx_k)}
{\{\widehat p_{-\nu(k)}(\bx_k)\}^2}
\widehat e_{k,\mathrm{cf}}^2.
\end{equation}
This is the cross-fitted version of \eqref{V2_OR_IPW_hat}. It is nonnegative
and uses only the outcome and propensity fits already needed for point
estimation. Consistent outcome and propensity predictions suffice for
consistency, with no product-rate condition. Its main drawback is that
squared residuals can be noisy, especially when response propensities
are small.

\item [(b)] Plug-in estimator. Suppose we also have predictions
$\widehat\sigma_k^2=\widehat\sigma^2_{-\nu(k)}(\bx_k)$ of the conditional
residual variance, fitted without the fold containing unit $k$.
The plug-in estimator is
\begin{equation}
\label{V2_hat_cf_plugin}
\widehat V_{2,\mathrm{plug,cf}}
=\frac{1}{N^2}\sum_{k\in S}w_k
\frac{1-\widehat p_{-\nu(k)}(\bx_k)}
{\widehat p_{-\nu(k)}(\bx_k)}\widehat\sigma_k^2.
\end{equation}
With the true nuisance functions, this estimator is the conditional
expectation of the IPW estimator given the covariates and the sample.
It removes the variation due to the outcomes and response indicators.
With fitted functions, this smoothing may still help, but the gain
depends on the accuracy of the variance predictions. This component
is consistent when propensity and variance predictions are consistent.
This result does not require consistent predictions of $m$, although
these are still assumed for $\widehat V_{1,\mathrm{cf}}$. The plug-in
estimator is nonnegative when the variance predictions are nonnegative,
but persistent error in those predictions can nonetheless produce biases.

\item  [(c)] Bias-corrected estimator. The third choice adds the correction in \eqref{V2_OR_BC_hat} to the IPW
estimator:
\begin{equation}
\label{V2_hat_cf_bc}
\widehat V_{2,\mathrm{bc,cf}}
=\widehat V_{2,\mathrm{ipw,cf}}
+\frac{1}{N^2}\sum_{k\in S}w_k
\left\{1-\frac{r_k}{\widehat p_{-\nu(k)}(\bx_k)}\right\}
\widehat\sigma_k^2.
\end{equation}
To see why this correction preserves consistency, consider its conditional expectation.
For $k\in\mathcal F_t$, the variance prediction
$\widehat\sigma^2_{-t}(\bx_k)$ is $\mathcal D_{-t}$-measurable and
$\E_q[r_k\mid\mathcal D_{-t}]=p(\bx_k)$, so
\begin{equation*}
\E_q\left[
\left\{1-\frac{r_k}{p(\bx_k)}\right\}
\widehat\sigma^2_{-t}(\bx_k)
\mid\mathcal D_{-t}\right]=0.
\end{equation*}
This identity holds even when the variance prediction is inaccurate.
As shown below, consistent outcome and propensity predictions suffice
if the variance predictions have bounded second moments. Thus, unlike
the plug-in estimator, the bias-corrected estimator does not require a
consistent variance learner. 
\end{itemize}

Any of the three estimators can be used for $\widehat V_{2,\mathrm{cf}}$
in \eqref{Vtot_hat_cf}. The IPW estimator is the simplest to compute
because it uses only the cross-fitted estimates of $m$ and $p$ already
available from point estimation. The plug-in and bias-corrected
estimators also require predictions of the conditional residual
variance $\sigma^2(\bx_k)$.\\

To obtain these predictions for fold $\mathcal F_t$, we first compute
out-of-sample residuals by inner cross-fitting within $S_{-t}$.
We then regress the squared residuals on the covariates using a method
that gives nonnegative predictions. All fitting and tuning are carried
out within $S_{-t}$. Appendix~\ref{app:variance_learning} gives the
details and describes an alternative based on log-scale fitting and
smearing.

\Needspace{15\baselineskip}
\begin{theorem}[Consistency of the variance estimators]
\label{Theo3}
Assume \ref{A1}--\ref{A4}, \ref{D1}--\ref{D4}, \ref{M2}, and the
positivity condition \eqref{eq:fitted_propensity_positivity}.
Suppose $\mathcal R_2(\widehat p_{-t})=o(1)$ for every fold $t$.
\begin{enumerate}[label=(\roman*)]
\item If $\mathcal R_2(\widehat m_{-t})=o(1)$ for every fold $t$, then
\begin{equation*}
n_v\left[\widehat V_{1,\mathrm{cf}}
-V_1(\widehat\mu_{\mathrm{aipw}}(m,p))\right]=o_{\P}(1),
\qquad
n_v\left[\widehat V_{2,\mathrm{ipw,cf}}
-V_2(\widehat\mu_{\mathrm{aipw}}(m,p))\right]=o_{\P}(1).
\end{equation*}
\item If $\mathcal R_1(\widehat\sigma^2_{-t})=o(1)$ for every fold $t$, then
\begin{equation*}
n_v\left[\widehat V_{2,\mathrm{plug,cf}}
-V_2(\widehat\mu_{\mathrm{aipw}}(m,p))\right]=o_{\P}(1).
\end{equation*}
\item If, for every fold $t$, $\mathcal R_2(\widehat m_{-t})=o(1)$ and
$\mathcal R_2(\widehat\sigma^2_{-t})=\mathcal O(1)$, then
\begin{equation*}
n_v\left[\widehat V_{2,\mathrm{bc,cf}}
-V_2(\widehat\mu_{\mathrm{aipw}}(m,p))\right]=o_{\P}(1).
\end{equation*}
\end{enumerate}
If $\mathcal R_2(\widehat m_{-t})=o(1)$ for every fold $t$ and the
conditions for the chosen second component hold, then
\begin{equation*}
n_v\left[\widehat V_{\mathrm{tot,cf}}
-V_{\mathrm{tot}}(\widehat\mu_{\mathrm{aipw}}(m,p))\right]=o_{\P}(1).
\end{equation*}
If, in addition,
$\lim_{v \to \infty} n_vV_{\mathrm{tot}}(\widehat\mu_{\mathrm{aipw}}(m,p))= c\in(0,\infty)$,
then
\begin{equation*}
\frac{\widehat V_{\mathrm{tot,cf}}}
{V_{\mathrm{tot}}(\widehat\mu_{\mathrm{aipw}}(m,p))}
\xrightarrow[v\to\infty]{\P}1.
\end{equation*}
\end{theorem}

\begin{proof}
See Appendix~\ref{proofTheo3}.
\end{proof}

No product-rate condition is needed for these conclusions. The outcome
and propensity risks may tend to zero at any rate. For the bias-corrected
choice, the variance prediction error may even remain bounded away from
zero, provided its mean square stays bounded. Under \ref{M2}, this last
condition is equivalent to a bounded second moment of the variance
predictions averaged over each fold.\\

These results concern the consistency of feasible variance estimators relative to the oracle variance. Estimating it consistently
does not by itself justify confidence intervals for the fitted point
estimator: its bias and additional fluctuations must also be negligible.
For that step, we also need the product-rate condition in
\eqref{eq:cf_product_rate}.\\

The following corollary brings together the oracle CLT, the oracle
equivalence in Theorem~\ref{Theo2}(\ref{thm:cf_oracle_equivalence}), and
the variance consistency in Theorem~\ref{Theo3}.

\Needspace{9\baselineskip}
\begin{corollary}
\label{cor:cf_clt}
Assume the conditions of Theorem~\ref{Theo3} for the total variance with
the chosen second component, together with \ref{M3} and
\ref{CL1}--\ref{CL2}. If, for every fold $t$,
\begin{equation*}
n_v\mathcal R_2(\widehat m_{-t})\mathcal R_2(\widehat p_{-t})=o(1),
\end{equation*}
then
\begin{equation*}
\frac{\widehat\mu_{\mathrm{aipw,cf}}(\widehat m,\widehat p)-\mu}
{\sqrt{\widehat V_{\mathrm{tot,cf}}}}
\xrightarrow[v\to\infty]{\cL}\mathcal N(0,1).
\end{equation*}
\end{corollary}

\begin{proof}
See Appendix~\ref{proof:cf_clt}.
\end{proof}

For $0<\alpha<1$, let $z_{1-\alpha/2}$ denote the $(1-\alpha/2)$-quantile
of the standard normal distribution. An asymptotic $(1-\alpha)$-level
confidence interval for $\mu$ is given by
\begin{equation*}
\mathcal I_{1-\alpha}
=\left[
\widehat\mu_{\mathrm{aipw,cf}}(\widehat m,\widehat p)
-z_{1-\alpha/2}\sqrt{\widehat V_{\mathrm{tot,cf}}},
\widehat\mu_{\mathrm{aipw,cf}}(\widehat m,\widehat p)
+z_{1-\alpha/2}\sqrt{\widehat V_{\mathrm{tot,cf}}}
\right].
\end{equation*}
Under the conditions of Corollary~\ref{cor:cf_clt}, its coverage probability
satisfies
\begin{equation*}
\P(\mu\in\mathcal I_{1-\alpha})
\xrightarrow[v\to\infty]{}1-\alpha.
\end{equation*}

\subsection{Implementation}
\label{sec:implementation}

The cross-fitted AIPW estimator is a weighted mean of the pseudo-values
$\widehat{\eta}_{k,\mathrm{cf}}$:
$$
\widehat{\mu}_{\mathrm{aipw,cf}}(\widehat m,\widehat p)
=
\frac{1}{N}\sum_{k\in S} w_k\widehat{\eta}_{k,\mathrm{cf}}.
$$
This form is not convenient for constructing a completed survey data
file because, even for a respondent, the pseudo-value
$\widehat{\eta}_{k,\mathrm{cf}}$ need not equal the observed value $y_k$. In a typical data file released by a national statistical office,
each completed survey variable occupies a single column containing
observed values for respondents and imputed values for nonrespondents.
A separate column contains the survey weights. Data users apply these
weights to the completed variable to obtain weighted estimates.\\

Our aim is therefore to retain $\widetilde y_k=y_k$ for $k\in S_r$
and to determine imputed values for $k\in S_m$
such that the weighted mean of the completed variable reproduces
the cross-fitted AIPW estimator exactly:
$$
\frac{1}{N}\sum_{k\in S} w_k\widetilde y_k
=
\widehat{\mu}_{\mathrm{aipw,cf}}(\widehat m,\widehat p).
$$

\subsubsection{A weighted sum of completed values}

The mean obtained from the initial regression imputation is
\begin{equation*}
\widehat\mu_{I,\mathrm{cf}}(\widehat m)
=\frac{1}{N}\left\{
\sum_{k\in S_r}w_ky_k
+\sum_{k\in S_m}w_k\widehat m_{-\nu(k)}(\bx_k)
\right\}.
\end{equation*}
The AIPW estimate adds a correction based on the respondent residuals:
\begin{equation}
\label{eq:imputation_mean_difference}
\widehat\mu_{\mathrm{aipw,cf}}(\widehat m,\widehat p)
-\widehat\mu_{I,\mathrm{cf}}(\widehat m)
=\frac{1}{N}\sum_{k\in S_r}w_k
\frac{1-\widehat p_{-\nu(k)}(\bx_k)}{\widehat p_{-\nu(k)}(\bx_k)}
\widehat e_{k,\mathrm{cf}},
\end{equation}
where $\widehat e_{k,\mathrm{cf}}=y_k-\widehat m_{-\nu(k)}(\bx_k)$,
as in Section~\ref{sec:aipw_variance_cf}.\\

To recover this correction through the imputation, we can add the same
amount to each nonrespondent's prediction. A common change in the imputed
values changes the mean by that amount times
$N^{-1}\sum_{\ell\in S_m}w_\ell$.
Thus, set
\begin{equation}
\label{mod_imp_val}
\widehat y_k^*
=\widehat m_{-\nu(k)}(\bx_k)
+\frac{\widehat\mu_{\mathrm{aipw,cf}}(\widehat m,\widehat p)
-\widehat\mu_{I,\mathrm{cf}}(\widehat m)}
{N^{-1}\sum_{\ell\in S_m}w_\ell},
\qquad k\in S_m.
\end{equation}
\Needspace{7\baselineskip}
For implementation, substitute \eqref{eq:imputation_mean_difference}
into \eqref{mod_imp_val}, giving
\begin{equation*}
\widehat y_k^*
=\widehat m_{-\nu(k)}(\bx_k)
+\frac{1}{\sum_{\ell\in S_m}w_\ell}
\sum_{\ell\in S_r}w_\ell
\frac{1-\widehat p_{-\nu(\ell)}(\bx_\ell)}
{\widehat p_{-\nu(\ell)}(\bx_\ell)}
\{y_\ell-\widehat m_{-\nu(\ell)}(\bx_\ell)\},
\qquad k\in S_m.
\end{equation*}

Keeping the respondent values unchanged then gives
\begin{equation}
\label{calibration_constraint_mean}
\begin{aligned}
\frac{1}{N}\left\{\sum_{k\in S_r}w_ky_k
+\sum_{k\in S_m}w_k\widehat y_k^*\right\}=\widehat\mu_{\mathrm{aipw,cf}}(\widehat m,\widehat p).
\end{aligned}
\end{equation}

\subsubsection{A calibration interpretation}

The common adjustment in \eqref{mod_imp_val} is the solution of the
quadratic calibration problem
\begin{equation*}
\min_{\{\widehat y_k^*:k\in S_m\}}
\sum_{k\in S_m}w_k
\frac{\{\widehat y_k^*-\widehat m_{-\nu(k)}(\bx_k)\}^2}{2c_k},
\end{equation*}
subject to \eqref{calibration_constraint_mean}, with $c_k=1$ for every
nonrespondent. It therefore minimizes the weighted squared changes
from the regression predictions among imputations that reproduce the
AIPW mean. This is a calibrated imputation construction
\citep{ren2003reverse,chambers2004outlier,beaumont2005calibrated}.\\

More generally, we can choose positive factors $c_k$ to allow unequal
adjustments. The first-order conditions now make
$\{\widehat y_k^*-\widehat m_{-\nu(k)}(\bx_k)\}/c_k$ the same for all
nonrespondents, giving
\begin{equation*}
\widehat y_k^*
=\widehat m_{-\nu(k)}(\bx_k)
+c_k\frac{\widehat\mu_{\mathrm{aipw,cf}}(\widehat m,\widehat p)
-\widehat\mu_{I,\mathrm{cf}}(\widehat m)}
{N^{-1}\sum_{\ell\in S_m}w_\ell c_\ell},
\qquad k\in S_m.
\end{equation*}
If we have more confidence in a particular prediction, we can give it
a smaller $c_k$ so that the imputed value stays closer to that prediction.
The other nonrespondents then make up more of the difference between the
two estimates. When nonrespondents account for only a small share of the
total survey weight, even a modest difference can require fairly large
adjustments. The resulting imputed values need not lie within the range
of the outcome.\\

Since \eqref{calibration_constraint_mean} holds exactly, the
completed-data mean inherits all properties established for the
cross-fitted AIPW estimator. Its variance is estimated using the
methods in Section~\ref{sec:aipw_variance_cf}. 

\begin{remark}
	In practice, survey users often need estimates for subpopulations,
	or domains, as well as for the population as a whole. For example,
	they may want a mean for each province in addition to the national
	mean. Calibrating the imputed values to reproduce the national
	estimate does not guarantee that they will also reproduce the
	provincial estimates. Calibration can also make one completed variable reproduce several
domain means. Suppose $U$ is partitioned into disjoint domains
$U_1,\ldots,U_D$, with observed membership and known sizes $N_d=|U_d|$.
For each domain, we impose
\begin{equation*}
\frac{1}{N_d}\left\{
\sum_{k\in S_r\cap U_d}w_ky_k
+\sum_{k\in S_m\cap U_d}w_k\widehat y_k^*
\right\}
=\frac{1}{N_d}\sum_{k\in S\cap U_d}w_k\widehat\eta_{k,\mathrm{cf}},
\qquad d=1,\ldots,D.
\end{equation*}
The right-hand side is the cross-fitted AIPW mean for domain $d$, using
the same fitted functions as for the overall mean. We minimize the
same quadratic criterion subject to these domain constraints.\\

\Needspace{6\baselineskip}
Because the domains are disjoint, the minimization separates into one
problem per domain. Applying the preceding calculation within each
domain gives
\begin{equation*}
\widehat y_k^*
=\widehat m_{-\nu(k)}(\bx_k)
+\frac{c_k}{\sum_{\ell\in S_m\cap U_d}w_\ell c_\ell}
\sum_{\ell\in S_r\cap U_d}w_\ell
\frac{1-\widehat p_{-\nu(\ell)}(\bx_\ell)}
{\widehat p_{-\nu(\ell)}(\bx_\ell)}
\widehat e_{\ell,\mathrm{cf}},
\qquad k\in S_m\cap U_d.
\end{equation*}
This distributes each domain's respondent-residual correction among
its own nonrespondents. 
\end{remark}
\section{Aggregation of nuisance learners}
\label{sec:aggregation}

The oracle-equivalence result in Theorem~\ref{Theo2} relies on the
product-rate condition (\ref{eq:cf_product_rate}). In practice, the convergence rates of
machine-learning estimators are rarely known and depend on the
unknown structure of $m$ and $p$. We therefore consider aggregation
methods \citep{wolpert1992stacked,breiman1996stacked,vanderlaan2007super}
that combine predictions from several candidate learners to adapt
to this unknown structure.\\

We consider two dictionaries of candidate learners, one for the outcome
regression and one for the response propensity:
$$
\mathcal M=\{\widehat m_1,\ldots,\widehat m_{J_{\mathrm{reg}}}\},
\qquad
\mathcal P=\{\widehat p_1,\ldots,\widehat p_{J_{\mathrm{prop}}}\}.
$$
Aggregation combines the predictions within each dictionary into a
single predictor. We estimate the aggregation weights by
cross-validation
\citep{wolpert1992stacked,breiman1996stacked,vanderlaan2007super}.
Under the oracle bounds below, each aggregate inherits consistency
from a consistent candidate,
and the two aggregates inherit the joint rate condition from a suitable
pair. The analyst need not identify these candidates in advance.

\subsection{Construction of the aggregated learners}

The partition $\mathcal F_1,\ldots,\mathcal F_L$ introduced in
Section~\ref{subsec:crossfitting} is used for cross-fitting. For a given
outer fold $t$, the nuisance predictions on $\mathcal F_t$ are constructed
using only the remaining sample,
$$
S_{-t}=S\setminus\mathcal F_t
=\bigcup_{s\neq t}\mathcal F_s.
$$
We describe how aggregation is carried out within this training sample.
Figure~\ref{fig:aggregation_splitting} in Appendix~\ref{app:aggregation_diagram}
illustrates the procedure with three outer folds and two inner folds.\\

Estimating the aggregation weights requires predictions on observations
that were excluded from candidate fitting and tuning. For this purpose,
the entire sample $S_{-t}$ is divided again into $K\geq2$ nonempty parts,
called inner folds:
$$
S_{-t}=\bigcup_{\ell=1}^K\mathcal F_{t,\ell}.
$$
\Needspace{5\baselineskip}
These parts form a new random partition of $S_{-t}$; the outer fold
$\mathcal F_t$ remains set aside. Each inner fold takes a turn as the
validation sample, while the candidates are fitted using the other
$K-1$ inner folds. With $K=2$, the candidates are fitted on one part and
predict the other, then the roles of the two parts are reversed.
This rotation is the inner cross-validation step.\\

\Needspace{8\baselineskip}
For a training set $A$, $\widehat m_h(\cdot;A)$ and
$\widehat p_h(\cdot;A)$ denote the fitted candidates, with all fitting and
tuning restricted to observations in $A$. For $k\in\mathcal F_{t,\ell}$,
the validation predictions are therefore
$$
\begin{aligned}
\widehat m^{\mathrm{cv}}_{h,-t}(\bx_k)
&=\widehat m_h(\bx_k;S_{r,-t}\setminus\mathcal F_{t,\ell}),\\
\widehat p^{\mathrm{cv}}_{h,-t}(\bx_k)
&=\widehat p_h(\bx_k;S_{-t}\setminus\mathcal F_{t,\ell}).
\end{aligned}
$$
Thus, outcome fitting uses respondents outside the validation fold,
whereas propensity fitting uses all units outside it. Throughout the
procedure, the propensity learners and their aggregation weights use
only covariates, response indicators, and survey weights.\\

After all inner folds have served as validation samples, these predictions
are pooled over $S_{-t}$. Every respondent has one validation prediction
from each outcome learner, and every unit has one from each propensity
learner. The pooled predictions are used to estimate one outcome weight
vector and one propensity weight vector for the outer fold $t$.

The candidate predictions are combined with nonnegative weights summing
to one. The set of allowed weights is
$$
\Delta_J=\left\{\boldsymbol\alpha\in[0,1]^J:
\sum_{h=1}^J\alpha_h=1\right\}.
$$
\Needspace{6\baselineskip}
This includes a single learner as the special case where its weight is one.
The outcome weights minimize the survey-weighted squared error of the
combined predictions among respondents:
\begin{equation}
\label{eq:aggregation_loss_m}
\widehat{\boldsymbol\alpha}_{\mathrm{reg},-t}
=\argmin_{\boldsymbol\alpha\in\Delta_{J_{\mathrm{reg}}}}
\sum_{k\in S_{r,-t}}w_k
\left\{y_k-\sum_{h=1}^{J_{\mathrm{reg}}}
\alpha_h\widehat m^{\mathrm{cv}}_{h,-t}(\bx_k)\right\}^2.
\end{equation}
Since response status is observed for every sampled unit, the propensity
weights minimize the following score over $S_{-t}$:
\begin{equation}
\label{eq:aggregation_loss_p_brier}
\widehat{\boldsymbol\alpha}_{\mathrm{prop},-t}
=\argmin_{\boldsymbol\alpha\in\Delta_{J_{\mathrm{prop}}}}
\sum_{k\in S_{-t}}w_k
\left\{r_k-\sum_{h=1}^{J_{\mathrm{prop}}}
\alpha_h\widehat p^{\mathrm{cv}}_{h,-t}(\bx_k)\right\}^2.
\end{equation}
Both criteria use predictions from all inner folds. The weights are
therefore estimated once for the whole outer training sample $S_{-t}$.\\

The inner fits have served to estimate the aggregation weights. For the
predictions on $\mathcal F_t$, the candidates are now refitted using all
available outer training observations, giving
$$
\widehat m_{h,-t}=\widehat m_h(\cdot;S_{r,-t}),
\qquad
\widehat p_{h,-t}=\widehat p_h(\cdot;S_{-t}).
$$
\Needspace{8\baselineskip}
With the aggregation weights held fixed, the final combinations are
\begin{equation}
\label{eq:aggregated_learners}
\begin{aligned}
\widehat m_{\mathrm{agg},-t}(\bx)
&=\sum_{h=1}^{J_{\mathrm{reg}}}
\widehat\alpha_{\mathrm{reg},h,-t}\widehat m_{h,-t}(\bx),\\
\widehat p_{\mathrm{agg},-t}(\bx)
&=\sum_{h=1}^{J_{\mathrm{prop}}}
\widehat\alpha_{\mathrm{prop},h,-t}\widehat p_{h,-t}(\bx).
\end{aligned}
\end{equation}
These functions provide the predictions on $\mathcal F_t$ used in the
cross-fitted AIPW estimator \eqref{dr_cross}. The entire construction is
repeated for $t=1,\ldots,L$, with a separate pair of aggregation weight
vectors for each outer fold.\\

The positivity condition \eqref{eq:fitted_propensity_positivity}
requires the fitted propensities to remain bounded away from zero. Truncating the final predictions to $[\rho,1]$ for a fixed
$0<\rho\leq1$ ensures this condition. If $\rho\leq\delta$, where $\delta$ is the lower bound
in \ref{A4}, truncation cannot increase their squared prediction error.
A common lower bound on the propensity candidates is also preserved by
their convex combination. The risk condition below concerns the
predictions actually used, including any truncation.

\Needspace{10\baselineskip}
\subsection{Prediction rates and inference}

The two validation losses measure prediction accuracy. On an inner
validation fold, for a fixed predictor fitted outside that fold, the
expected excess outcome loss weights each squared regression error by
$w_kp(\bx_k)$. The expected excess Brier loss weights each squared
propensity error by $w_k$. Under \ref{A4} and \ref{D2}, these weights are
bounded above and away from zero, so the corresponding weighted and
unweighted errors are comparable. A guarantee for the final aggregate
also requires control of the error from choosing its weights and
refitting its candidates. We impose the following oracle bounds.

\Needspace{8\baselineskip}
\begin{assumptionlist}{AG}
\item \label{AG1}
The dictionary sizes $J_{\mathrm{reg}}$ and $J_{\mathrm{prop}}$ are fixed.
There is a finite constant $C\geq1$, independent of $v$ and $t$, such that,
for every fold and all sufficiently large $v$,
\begin{equation*}
\begin{aligned}
\mathcal R_2(\widehat m_{\mathrm{agg},-t})
&\leq C\min_{1\leq h\leq J_{\mathrm{reg}}}
\mathcal R_2(\widehat m_{h,-t})+\frac{C}{n_v},\\
\mathcal R_2(\widehat p_{\mathrm{agg},-t})
&\leq C\min_{1\leq h\leq J_{\mathrm{prop}}}
\mathcal R_2(\widehat p_{h,-t})+\frac{C}{n_v}.
\end{aligned}
\end{equation*}
\end{assumptionlist}

Condition~\ref{AG1} compares each aggregate with the best individual
candidate, allowing a constant factor and an additional error of order
$n_v^{-1}$. Bounds of this order hold in a simpler setting. For bounded
outcomes and candidate predictions, \citet[Theorem~A]{lecue2013empirical}
shows that least-squares aggregation over the convex hull of a fixed
dictionary has expected prediction risk at most that of the best convex
combination plus a remainder of order $J/n_v$, when the i.i.d. validation
sample is independent of the candidate fits and has size proportional to
$n_v$. The candidates may be
fitted on a separate training sample and are held fixed during validation.
For fixed $J$, this also gives a bound against the best individual
candidate with remainder of order $n_v^{-1}$.\\

\Needspace{5\baselineskip}
That result averages over covariates and concerns a separate validation
sample. It motivates \ref{AG1}; here, we assume the bounds for the final
refitted predictors, under the sampling design and with risks defined
by \eqref{eq:prediction_risk}.

\begin{proposition}[Consistency and prediction rates]
\label{prop:aggregation_rates}
Assume \ref{AG1}.
\begin{enumerate}[label=(\roman*),ref=\roman*,leftmargin=*]
\item \label{prop:aggregation_consistency}
For each fold $t$, if there exists $\widehat m\in\mathcal M$ with
$\mathcal R_2(\widehat m_{-t})=o(1)$, then
$\mathcal R_2(\widehat m_{\mathrm{agg},-t})=o(1)$.
The same holds for propensity learners, and with $o(1)$ replaced by
$\mathcal O(1)$.
\item \label{prop:aggregation_product_rate}
If, in each fold $t$, there exists a pair
$(\widehat m,\widehat p)\in\mathcal M\times\mathcal P$ such that
\begin{equation}
\label{eq:aggregation_candidate_rates}
\begin{gathered}
\mathcal R_2(\widehat m_{-t})=o(1),
\qquad \mathcal R_2(\widehat p_{-t})=o(1), \qquad
n_v\mathcal R_2(\widehat m_{-t})\mathcal R_2(\widehat p_{-t})=o(1),
\end{gathered}
\end{equation}
then, for every fold $t$,
\begin{equation*}
\begin{gathered}
\mathcal R_2(\widehat m_{\mathrm{agg},-t})=o(1),
\qquad \mathcal R_2(\widehat p_{\mathrm{agg},-t})=o(1),\qquad 
n_v\mathcal R_2(\widehat m_{\mathrm{agg},-t})
\mathcal R_2(\widehat p_{\mathrm{agg},-t})=o(1).
\end{gathered}
\end{equation*}
\end{enumerate}
The candidates may vary with the fold and population index.
\end{proposition}
\begin{proof}
See Appendix~\ref{proof:aggregation_rates}.
\end{proof}

The proposition gives several ways to meet the prediction conditions in
Section~\ref{sec:aipw3}. Here, consistency refers to squared prediction
risk tending to zero, and the other assumptions of each cited result,
including the positivity condition \eqref{eq:fitted_propensity_positivity},
are retained. For consistency of the AIPW estimator
in Theorem~\ref{Theo2}(\ref{thm:cf_consistency}), one consistent outcome
candidate in each fold suffices, since propensity risk is automatically
bounded. Alternatively, a consistent propensity candidate suffices if,
in each fold, the outcome dictionary contains a candidate with bounded
risk.\\

If both dictionaries contain consistent candidates in every fold, the
aggregates meet the nuisance consistency conditions for
$\widehat V_{1,\mathrm{cf}}$ and $\widehat V_{2,\mathrm{ipw,cf}}$ in
Theorem~\ref{Theo3}(i), without a product-rate condition. The plug-in and
bias-corrected choices of $\widehat V_{2,\mathrm{cf}}$ retain their
respective conditions on the variance predictions.\\

Finally, a candidate pair satisfying \eqref{eq:aggregation_candidate_rates}
in each outer fold suffices for oracle equivalence in
Theorem~\ref{Theo2}(\ref{thm:cf_oracle_equivalence}). With the remaining
conditions of Corollary~\ref{cor:cf_clt}, it also gives feasible inference.
The suitable candidates need not be identified in advance.\\

Aggregation can improve prediction, but requires repeated fitting of the
candidates and estimation of the aggregation weights. Those weights may
be unstable in small samples, and the rate guarantee in
Proposition~\ref{prop:aggregation_rates} does not imply that aggregation
improves on every candidate in a given sample.

\section{Simulation study}
\label{sec:simulation}
\begingroup
\setlength{\parskip}{0.5\baselineskip}

The simulation study examines how cross-fitting affects coverage and mean
squared error, how closely the feasible estimators approach the oracle,
and how aggregation compares with individual learners. Four settings
combine simple and nonlinear outcome and response models. Prediction
errors provide a further description of the differences between learners.

\subsection{Simulation design}
\label{sec:simulation_design}

We conducted a simulation study to assess the finite-sample performance
of the AIPW estimator, examine the effect of cross-fitting, and evaluate
the accuracy of the variance estimates and confidence intervals.
We also examined the prediction accuracy of the nuisance estimators.

We considered sample sizes $n\in\{500,1000,2000,4000\}$ and finite
populations of size $N=n/f$, with sampling fraction $f=0.025$.
For each population, the ten covariates were generated independently
from the uniform distribution on $[0,1]$ and then held fixed.
Each of the $B=5000$ replications generated new outcomes, selected
a simple random sample without replacement, and generated response
indicators under a missing-at-random mechanism. Thus, the experiment
follows the conditional $mpq$ framework of the paper, with a new
finite-population mean in each replication.

The two outcome regression functions were
\begin{align*}
	m_1(\bx_k)
	&=
	1+2(x_{k1}-0.5)-1.5(x_{k2}-0.5)+(x_{k3}-0.5),\\
	m_2(\bx_k)
	&=
	1+2\sin(\pi x_{k1}x_{k2})
	+2(x_{k3}-0.5)^2+x_{k4}+0.5x_{k5}.
\end{align*}
Outcomes were generated as $y_{jk}=m_j(\bx_k)+\epsilon_{jk}$,
for $j=1,2$. For each outcome model, conditionally on the covariates,
the errors were independent normal random variables with mean zero
and heteroscedastic variances calibrated to give approximately
$R^2=0.50$. The variance specification is given in
Appendix~\ref{app:simulation_specifications}.

Conditionally on the covariates, response indicators were generated
independently under a MAR mechanism, with probabilities given by
one of the following functions:
\begin{align*}
	p_1(\bx_k)
	&=\operatorname{expit}(a+x_{k2}-x_{k1}),\\
	p_2(\bx_k)
	&=0.50+\kappa(z_k-\overline z_U),
\end{align*}
where $z_k=F_{2,4}\{\min(x_{k1},x_{k2})\}$,
$\overline z_U=N^{-1}\sum_{k\in U}z_k$, and $F_{2,4}$ is the
distribution function of a Beta$(2,4)$ random variable.
The intercept $a$ was chosen so that the population average of
$p_1$ was $0.50$. For $p_2$, we used $\kappa=0.70$, reduced to
the largest admissible value if needed to keep all probabilities
in $[0.05,0.95]$. Both mechanisms therefore gave an expected
response rate of $50\%$.

Combining the outcome models and response mechanisms gives four
scenarios: L--L $(m_1,p_1)$, L--NL $(m_1,p_2)$,
NL--L $(m_2,p_1)$, and NL--NL $(m_2,p_2)$.
L and NL denote linear and nonlinear dependence on the covariates,
with the outcome regression listed first and the logit of the
propensity second. Linear and logistic regression are correctly
specified in L--L; the other scenarios vary whether one or both
nuisance functions require a more flexible fit.

\subsection{Estimators}
\label{sec:simulation_estimators}

We compared the feasible AIPW estimator with five-fold cross-fitting
(CF) and without cross-fitting (No CF). For CF, the sample was
divided into $L=5$ folds. The nuisance models were fitted on four
folds and used to predict for the remaining fold, repeating this
for each fold. For No CF, the nuisance models were fitted on the
same sample used to compute the AIPW estimator.

Two benchmarks were also computed: the oracle AIPW estimator,
which uses the true $m$ and $p$, and the full-response
Horvitz--Thompson estimator $\widehat\mu_F$. Comparison with the
oracle measures the cost of estimating the nuisance functions,
whereas comparison with $\widehat\mu_F$ also reflects the
information lost through nonresponse.

The individual learners were linear or logistic regression,
generalized additive models (GAM), and boosted trees (XGBoost).
Linear and logistic regression used all ten original covariates.
The outcome GAM included a smooth interaction between $x_1$ and
$x_2$ and smooth effects of $x_3,x_4,x_5$; the propensity GAM
included smooth effects of $x_1$, $x_2$, and $\min(x_1,x_2)$.
These specifications provide useful structure for the nonlinear
settings. The comparisons concern these particular learner
specifications, rather than a general ranking of the algorithms.

We also used aggregation as described in
Section~\ref{sec:aggregation}. The outcome dictionary contained
linear regression, GAM, elastic-net regression, and a regression
forest. The propensity dictionary contained logistic regression,
GAM, elastic-net logistic regression, and a probability forest.
Nonnegative weights summing to one were estimated by five-fold
inner cross-validation ($K=5$), using squared prediction error
among respondents for the outcome and the Brier score for the
propensity. All tuning and weight estimation took place within
the corresponding training sample. Fitted response probabilities
were truncated to $[0.05,0.95]$.
Appendix~\ref{app:simulation_specifications} gives the tuning
specifications.

Table~\ref{tab:simulation_learner_pairs} lists the learner pairs
considered in each scenario. In the mixed scenarios L--NL and
NL--L, we retained the correctly specified parametric learner
for the linear component and varied the learner for the nonlinear
component. In the table, each pair lists the outcome learner
first and the propensity learner second. LM and Logit denote
linear and logistic regression, respectively, and Agg.\ denotes
aggregation.

\begin{table}[htbp]
	\centering
	\captionsetup{
		format=plain,
		font=small,
		labelfont=bf,
		labelsep=colon,
		justification=justified,
		singlelinecheck=false
	}
	\caption{Outcome and propensity learner pairs used in each scenario.}
	\label{tab:simulation_learner_pairs}
	\small
	\begin{tabular}{
			@{}ccc >{\raggedright\arraybackslash}p{0.43\textwidth}@{}
		}
		\hline
		Scenario
		& Outcome model
		& Response model
		& Learner pairs $(\widehat m,\widehat p)$\\
		\hline
		L--L
		& $m_1$
		& $p_1$
		& LM--Logit; Agg.--Agg.\\
		L--NL
		& $m_1$
		& $p_2$
		& LM--GAM; LM--XGBoost; LM--Agg.\\
		NL--L
		& $m_2$
		& $p_1$
		& GAM--Logit; XGBoost--Logit; Agg.--Logit\\
		NL--NL
		& $m_2$
		& $p_2$
		& GAM--GAM; XGBoost--XGBoost; Agg.--Agg.\\
		\hline
	\end{tabular}
\end{table}

For the oracle, non-cross-fitted, and cross-fitted estimators,
we used the variance estimators in \eqref{V1_OR_hat},
\eqref{V1_hat_feasible}, and \eqref{V1_hat_cf}, respectively.
These retain only $V_1$. At $f=0.025$, the oracle bound in
\eqref{eq:oracle_variance_ratio_bound} gives $V_2/V_1\leq 1/39$,
which motivates this small-sampling-fraction approximation.
Coverage therefore concerns these $V_1$-based intervals;
the study does not compare the estimators of $V_2$.

\subsection{Evaluation criteria}
\label{sec:simulation_criteria}

For replication $b$, let $\widehat\mu^{(b)}$ denote an estimate
of the generated finite-population mean $\mu^{(b)}$, and write
$$
e_b=\widehat\mu^{(b)}-\mu^{(b)},
\qquad
\overline e=\frac{1}{B}\sum_{b=1}^B e_b.
$$
As measures of bias and accuracy, we used the Monte Carlo
percent relative bias and mean squared error:
\begin{align*}
	\operatorname{RB}_{\mathrm{MC}}(\widehat\mu)
	&=
	100\frac{\sum_{b=1}^B e_b}
	{\sum_{b=1}^B\mu^{(b)}},\\
	\operatorname{MSE}_{\mathrm{MC}}(\widehat\mu)
	&=
	\frac{1}{B}\sum_{b=1}^B e_b^2.
\end{align*}
Relative to a reference estimator $\widehat\mu_{\mathrm{ref}}$,
we defined relative efficiency, expressed as a percentage, by
\begin{equation*}
	\operatorname{RE}_{\mathrm{MC}}
	(\widehat\mu;\widehat\mu_{\mathrm{ref}})
	=
	100\times
	\frac{\operatorname{MSE}_{\mathrm{MC}}(\widehat\mu)}
	{\operatorname{MSE}_{\mathrm{MC}}(\widehat\mu_{\mathrm{ref}})}.
\end{equation*}
Table~\ref{tab:simulation_summary} uses the oracle AIPW estimator
as the reference, whereas the complete tables in
Appendix~\ref{app:simulation_results} use the full-response
Horvitz--Thompson estimator $\widehat\mu_F$.
Equal MSE corresponds to $100\%$; smaller values indicate
greater efficiency.

Variance estimation was assessed using the Monte Carlo variance
of the estimation errors:
\begin{equation*}
	V_{\mathrm{MC}}(\widehat\mu)
	=
	\frac{1}{B-1}\sum_{b=1}^B(e_b-\overline e)^2.
\end{equation*}
We use the variance of the estimation errors because the population
mean changes across replications. For a variance estimator
$\widehat V$, with value $\widehat V^{(b)}$ in replication $b$,
the percent relative bias was
\begin{equation*}
	\operatorname{RB}_{\mathrm{MC}}(\widehat V)
	=
	100\frac{
		\overline{\widehat V}-V_{\mathrm{MC}}(\widehat\mu)
	}{
		V_{\mathrm{MC}}(\widehat\mu)
	},
	\qquad
	\overline{\widehat V}
	=
	\frac{1}{B}\sum_{b=1}^B\widehat V^{(b)}.
\end{equation*}

We also computed the percent coverage of the nominal $95\%$
intervals $\widehat\mu^{(b)}\pm1.96\sqrt{\widehat V^{(b)}}$:
\begin{equation*}
	\operatorname{Coverage}_{\mathrm{MC}}(\widehat\mu,\widehat V)
	=
	\frac{100}{B}\sum_{b=1}^B
	\mathds{1}_{\{|e_b|\leq1.96\sqrt{\widehat V^{(b)}}\}}.
\end{equation*}

Since the true $m$ and $p$ were known, we also assessed nuisance
prediction accuracy. In replication $b$, $\mathcal E_b$ was a
random sample of $2000$ nonsampled units. For the cross-fitted
learners, we averaged the predictions from the $L$ outer training
samples at each evaluation unit, writing $\widehat m^{(b)}$ and
$\widehat p^{(b)}$ for these averages. The root mean squared
prediction errors were
\begin{align*}
	\operatorname{RMSE}^{(b)}(\widehat m)
	&=
	\left[
	\frac{1}{|\mathcal E_b|}
	\sum_{k\in\mathcal E_b}
	\{\widehat m^{(b)}(\bx_k)-m(\bx_k)\}^2
	\right]^{1/2},\\
	\operatorname{RMSE}^{(b)}(\widehat p)
	&=
	\left[
	\frac{1}{|\mathcal E_b|}
	\sum_{k\in\mathcal E_b}
	\{\widehat p^{(b)}(\bx_k)-p(\bx_k)\}^2
	\right]^{1/2}.
\end{align*}

\subsection{Results}
\label{sec:simulation_results}

The main tables compare the smallest and largest sample sizes,
$n=500$ and $n=4000$, across all four scenarios. The complete results are presented in
\hyperref[app:simulation_results]{Appendix~\ref*{app:simulation_results}}.
They include all four sample sizes, the oracle and full-response
benchmarks, and additional prediction-error summaries.

\subsubsection{Point estimation}
\label{sec:simulation_point_estimation}

Table~\ref{tab:simulation_summary} presents point-estimator bias and
relative efficiency alongside outcome and propensity prediction errors.

\begin{table}[H]
\centering
\captionsetup{format=plain,font=small,labelfont=bf,labelsep=colon,
  justification=justified,singlelinecheck=false}
\caption{Point-estimator bias, relative efficiency, and prediction errors at the
smallest and largest sample sizes.}
\label{tab:simulation_summary}
\begingroup
\setstretch{1}
\small
\setlength{\tabcolsep}{8pt}
\renewcommand{\arraystretch}{1.15}
\begin{tabular}{@{}l r rr rr rr@{}}
\toprule
Learner pair & $n$ & \multicolumn{2}{c}{RB (\%)}
& \multicolumn{2}{c}{RE (\%)} & \multicolumn{2}{c}{Prediction RMSE (CF)} \\
\cmidrule(lr){3-4}\cmidrule(lr){5-6}\cmidrule(l){7-8}
& & CF & No CF & CF & No CF & Outcome & Propensity \\
\midrule
\multicolumn{8}{@{}l}{\textit{Linear outcome; linear propensity}} \\
Linear + Logistic & 500 & 0.0 & 0.0 & 105 & 102 & 0.164 & 0.072 \\
 & 4000 & 0.0 & 0.0 & 100 & 100 & 0.057 & 0.025 \\
\addlinespace[3pt]
Agg. + Agg. & 500 & 0.0 & -0.1 & 105 & 103 & 0.150 & 0.064 \\
 & 4000 & 0.0 & 0.0 & 100 & 100 & 0.052 & 0.022 \\
\midrule
\multicolumn{8}{@{}l}{\textit{Linear outcome; nonlinear propensity}} \\
Linear + GAM & 500 & 0.0 & 0.0 & 111 & 104 & 0.167 & 0.053 \\
 & 4000 & 0.0 & 0.0 & 101 & 101 & 0.058 & 0.019 \\
\addlinespace[3pt]
Linear + Agg. & 500 & 0.0 & 0.0 & 110 & 102 & 0.167 & 0.055 \\
 & 4000 & 0.0 & 0.0 & 101 & 100 & 0.058 & 0.020 \\
\addlinespace[3pt]
Linear + XGBoost & 500 & 0.2 & 0.0 & 536 & 94 & 0.167 & 0.161 \\
 & 4000 & 0.0 & 0.0 & 121 & 99 & 0.058 & 0.068 \\
\midrule
\multicolumn{8}{@{}l}{\textit{Nonlinear outcome; linear propensity}} \\
GAM + Logistic & 500 & 0.0 & 0.0 & 108 & 104 & 0.212 & 0.072 \\
 & 4000 & 0.0 & 0.0 & 101 & 100 & 0.083 & 0.025 \\
\addlinespace[3pt]
Agg. + Logistic & 500 & 0.0 & 0.0 & 108 & 104 & 0.216 & 0.072 \\
 & 4000 & 0.0 & 0.0 & 101 & 100 & 0.084 & 0.025 \\
\addlinespace[3pt]
XGBoost + Logistic & 500 & 0.0 & -0.1 & 122 & 115 & 0.420 & 0.072 \\
 & 4000 & 0.0 & 0.0 & 104 & 104 & 0.228 & 0.025 \\
\midrule
\multicolumn{8}{@{}l}{\textit{Nonlinear outcome; nonlinear propensity}} \\
GAM + GAM & 500 & -0.1 & 0.1 & 119 & 108 & 0.233 & 0.053 \\
 & 4000 & 0.0 & 0.0 & 102 & 101 & 0.091 & 0.019 \\
\addlinespace[3pt]
Agg. + Agg. & 500 & -0.1 & 0.2 & 119 & 108 & 0.235 & 0.055 \\
 & 4000 & 0.0 & 0.0 & 102 & 101 & 0.092 & 0.020 \\
\addlinespace[3pt]
XGBoost + XGBoost & 500 & -2.5 & 0.9 & 920 & 133 & 0.436 & 0.161 \\
 & 4000 & 0.0 & 0.2 & 127 & 112 & 0.237 & 0.068 \\
\bottomrule
\end{tabular}
\endgroup
\end{table}

Most point-estimator biases in Table~\ref{tab:simulation_summary} are
close to zero. The main exception is the cross-fitted XGBoost pair
when both nuisance functions are nonlinear: its relative bias is
$-2.5\%$ at $n=500$ and rounds to zero at $n=4000$.

Cross-fitting has an MSE cost that depends strongly on the learners.
With the parametric, GAM, and aggregated fits, cross-fitted MSE is
about $5$--$19\%$ above the oracle at $n=500$ and within about $2\%$
at $n=4000$. The difference between CF and No CF also becomes small.
The cost is larger with the XGBoost propensity specification: in
L--NL, cross-fitted relative efficiency is $536\%$ at $n=500$ and
$121\%$ at $n=4000$; in NL--NL, the corresponding values are $920\%$
and $127\%$. Using XGBoost only for the outcome gives smaller values,
$122\%$ and $104\%$.

The prediction errors help explain these differences. Both outcome
and propensity RMSE decrease with sample size, but remain larger
with the boosting specifications. In NL--NL at $n=4000$, the outcome
and propensity RMSEs are $0.091$ and $0.019$ for GAM, compared with
$0.237$ and $0.068$ for XGBoost. Aggregation gives RMSEs of $0.092$
and $0.020$, close to those of GAM, and similar point-estimator
performance. Its relative efficiency decreases from $119\%$ to
$102\%$, as for GAM. Across the four scenarios, aggregation performs
close to the corresponding parametric or GAM fits. This illustrates
how it can retain good performance without selecting a single
candidate in advance, although it does not improve uniformly on
the individual learners.

\subsubsection{Variance estimation}
\label{sec:simulation_variance_estimation}

\begin{table}[!htb]
\centering
\captionsetup{format=plain,font=small,labelfont=bf,labelsep=colon,
  justification=justified,singlelinecheck=false}
\caption{Variance-estimator bias and coverage at the smallest and largest
sample sizes.}
\label{tab:simulation_inference}
\begingroup
\setstretch{1}
\small
\setlength{\tabcolsep}{8pt}
\renewcommand{\arraystretch}{1.10}
\begin{tabular}{@{}l r rr rr@{}}
\toprule
Learner pair & $n$ & \multicolumn{2}{c}{Variance RB (\%)}
& \multicolumn{2}{c}{Coverage (\%)} \\
\cmidrule(lr){3-4}\cmidrule(l){5-6}
& & CF & No CF & CF & No CF \\
\midrule
\multicolumn{6}{@{}l}{\textit{Linear outcome; linear propensity}} \\
Linear + Logistic & 500 & 4.0 & -2.9 & 95.5 & 94.5 \\
 & 4000 & -2.0 & -2.7 & 94.6 & 94.6 \\
\addlinespace[2pt]
Agg. + Agg. & 500 & 2.3 & -11.8 & 95.0 & 93.2 \\
 & 4000 & -2.2 & -4.2 & 94.6 & 94.4 \\
\midrule
\multicolumn{6}{@{}l}{\textit{Linear outcome; nonlinear propensity}} \\
Linear + GAM & 500 & 2.8 & -5.4 & 95.1 & 94.2 \\
 & 4000 & 0.1 & -0.9 & 95.1 & 95.0 \\
\addlinespace[2pt]
Linear + Agg. & 500 & 2.7 & -10.0 & 95.1 & 93.5 \\
 & 4000 & 0.0 & -2.7 & 95.1 & 94.6 \\
\addlinespace[2pt]
Linear + XGBoost & 500 & 27.7 & -45.8 & 97.6 & 85.3 \\
 & 4000 & 9.7 & -12.7 & 95.9 & 93.2 \\
\midrule
\multicolumn{6}{@{}l}{\textit{Nonlinear outcome; linear propensity}} \\
GAM + Logistic & 500 & 4.2 & -6.5 & 95.2 & 93.9 \\
 & 4000 & -0.4 & -2.0 & 94.8 & 94.5 \\
\addlinespace[2pt]
Agg. + Logistic & 500 & 4.1 & -7.7 & 95.3 & 93.7 \\
 & 4000 & -0.5 & -2.3 & 94.7 & 94.5 \\
\addlinespace[2pt]
XGBoost + Logistic & 500 & 3.5 & -44.6 & 95.2 & 85.2 \\
 & 4000 & -0.2 & -12.4 & 94.8 & 93.1 \\
\midrule
\multicolumn{6}{@{}l}{\textit{Nonlinear outcome; nonlinear propensity}} \\
GAM + GAM & 500 & 8.1 & -10.8 & 95.5 & 93.3 \\
 & 4000 & -0.9 & -3.9 & 94.6 & 94.2 \\
\addlinespace[2pt]
Agg. + Agg. & 500 & 6.8 & -16.6 & 95.4 & 92.3 \\
 & 4000 & -0.8 & -6.8 & 94.6 & 93.8 \\
\addlinespace[2pt]
XGBoost + XGBoost & 500 & 33.1 & -60.5 & 97.4 & 75.1 \\
 & 4000 & 14.5 & -29.2 & 96.2 & 89.1 \\
\bottomrule
\end{tabular}
\endgroup
\end{table}

Table~\ref{tab:simulation_inference} shows that, without cross-fitting,
the variance estimates can be much too small.
At $n=500$, using XGBoost for one nuisance function gives coverage
near $85\%$; using it for both gives $75.1\%$. The corresponding
variance biases range from $-44.6\%$ to $-60.5\%$. These biases and
the undercoverage decrease with sample size, but remain substantial
for the XGBoost pair at $n=4000$: variance bias is $-29.2\%$ and
coverage is $89.1\%$.

Cross-fitting avoids this severe undercoverage. With the parametric,
GAM, and aggregated fits, variance biases are small and coverage is
close to $95\%$. When XGBoost estimates the propensity, the
cross-fitted variance estimates are instead too large and the
intervals are conservative: at $n=500$, coverage is $97.6\%$ in
L--NL and $97.4\%$ in NL--NL. Both the variance bias and the
overcoverage decrease with sample size.

\subsection{Main findings}
\label{sec:simulation_findings}

The simulations show why cross-fitting is useful for the proposed
AIPW estimator. Small point-estimator bias alone does not ensure
reliable inference: without cross-fitting, the variance can be
substantially underestimated even when bias is negligible.
Cross-fitting avoids the severe undercoverage seen with boosted trees
and gives coverage close to $95\%$ with the parametric, GAM, and
aggregated fits.

This improvement comes with an MSE cost that decreases with sample
size. For the parametric, GAM, and aggregated fits, cross-fitted MSE
is within about $2\%$ of the oracle at $n=4000$. The boosting results
also show that cross-fitting can yield conservative intervals and a
substantial MSE cost when nuisance predictions are inaccurate.
Cross-fitting and accurate nuisance prediction therefore play
complementary roles in obtaining reliable and efficient inference.

Aggregation gives an encouraging practical option. In the learner
pairs considered, it performs close to the corresponding parametric
or GAM fits across all four settings, both for nuisance prediction
and for estimation of the population mean. This suggests that
combining candidate learners can reduce the need to choose one in
advance while retaining good performance. Paired with cross-fitting,
aggregation gives small biases and coverage close to nominal.

\endgroup

\section{Final remarks}
\label{sec:finalRemarks}

In this paper, we studied inference for finite population means in the presence of item nonresponse when flexible machine-learning methods are used to estimate nuisance functions. We showed that the customary plug-in imputed estimator may be difficult to use for inference because the imputation error enters at first order. We therefore proposed to change the point estimator and use a cross-fitted AIPW estimator, whose Neyman-orthogonal structure makes nuisance estimation second order under suitable product-rate conditions. We also discussed variance estimation, implementation through completed values, and aggregation methods to reduce the dependence on a single learning algorithm.\\

The results were developed for the estimation of an overall finite population mean. A direct extension concerns vectors of domain totals or domain means. In that case, the proposed approach can be applied componentwise, provided that the domains of interest are known in advance. In practice, this means that domain indicators should be included, when appropriate, among the covariates used to estimate both the outcome regression function and the response propensity. If the domains of interest are not known at the imputation stage, the problem is more delicate. A completed data file that is valid for one set of domains may not necessarily yield valid inference for other domains that were not accounted for in the nuisance models.\\

A broader extension concerns the construction of completed data files that can be used for several purposes after dissemination. In practice, data users may not only estimate an overall mean, but also domain means, distribution functions, quantiles, or other finite population parameters defined as solutions of estimating equations. For such parameters, imputing missing values by predicted conditional means is generally not sufficient, because regression imputation tends to shrink the distribution of the completed variable. This may be harmless for some mean estimators, but it can distort distribution functions and quantiles.\\

More generally, future work should consider finite population parameters defined as solutions of census estimating equations,
$$
\mathbf U_N(\boldsymbol\theta)
=
\frac{1}{N}
\sum_{k\in U}
u(\bx_k,y_k;\boldsymbol\theta)
=
\mathbf 0.
$$
This framework includes means, domain means, distribution functions, and quantiles. When the target domains or parameters are known in advance, they can be incorporated into the imputation and calibration procedure. When they are not known at the imputation stage, the problem is more challenging, because a completed data file that is valid for one set of targets may not automatically be valid for another. Developing cross-fitted, distribution-preserving, and possibly calibrated imputation procedures for such multi-purpose inference is currently under investigation.

\newpage

\bibliographystyle{apalike}
\bibliography{bibliography}
\addcontentsline{toc}{section}{References}
\newpage

\appendix

\numberwithin{equation}{section}

\startcontents[appendices]
\section*{Appendix contents}
\begingroup
\singlespacing
\printcontents[appendices]{}{1}[2]{}
\endgroup
\clearpage

\section{Notation}
\label{app:notation}

We retain the notation of the main text. The following conventions and
shorthand are used in several proofs.

For $k\in\mathcal F_t$, we write
\begin{equation*}
\widehat f_{-k}(\bx_k)=\widehat f_{-t}(\bx_k),
\qquad f\in\{m,p\}.
\end{equation*}
The subscript $-k$ means that the entire fold containing unit $k$ was
excluded from training.

For an array $z=(z_k:k\in A)$ on a nonempty finite set $A$, write
\begin{equation*}
\|z\|_{A,2}
=\left\{\frac{1}{|A|}\sum_{k\in A}|z_k|^2\right\}^{1/2}.
\end{equation*}
For functions, this empirical $L_2$ norm is evaluated at the covariates
of the units in $A$.

For arrays $a$ and $b$ on $S_v$, the bilinear form associated with the
design variance estimator is
\begin{equation*}
\mathcal Q(a,b)=\frac{1}{N_v^2}
\sum_{k\in S_v}\sum_{\ell\in S_v}
\frac{\Delta_{k\ell}}{\pi_{k\ell}}
\frac{a_k}{\pi_k}\frac{b_\ell}{\pi_\ell}.
\end{equation*}

For bounds involving inverse fitted propensities, we use the event
\begin{equation*}
\mathcal A=
\left\{\min_{1\leq t\leq L}\min_{k\in\mathcal F_t}
\widehat p_{-t}(\bx_k)\geq\rho\right\},
\end{equation*}
which has probability tending to one by
\eqref{eq:fitted_propensity_positivity}.

\section{Proofs for Section~\ref{sec:aipw1}}
\label{app:proofs_oracle}

\subsection{Proof of Proposition~\ref{prop:oracle_consistency}}
\label{proof:oracle_consistency}

\begin{proof}
Write the error as the sum of the full-sample Horvitz--Thompson error and the additional response error:
\begin{equation*}
\widehat\mu_{\mathrm{aipw}}(m,p)-\mu
=(\widehat\mu_F-\mu)+R_v,
\end{equation*}
where
\begin{equation*}
R_v:=\frac{1}{N_v}\sum_{k\in S_v}w_k
\left\{\frac{r_k}{p(\bx_k)}-1\right\}\epsilon_k.
\end{equation*}
Expanding the squared error gives
\begin{equation}
\label{eq:oracle_mean_square_expansion}
\begin{aligned}
\E_{mpq}\left[\{\widehat\mu_{\mathrm{aipw}}(m,p)-\mu\}^2\right]
&=\E_m[\V_p(\widehat\mu_F)]
+2\E_{mpq}[(\widehat\mu_F-\mu)R_v]+\E_{mpq}[R_v^2].
\end{aligned}
\end{equation}
For the cross term, $\widehat\mu_F-\mu$ is
$\sigma(\mathbf X,\mathbf y,\mathbf I)$-measurable and the response
factors are centered, so
\begin{equation*}
\E_q[R_v]=0,
\qquad
\E_{mpq}[(\widehat\mu_F-\mu)R_v]=0.
\end{equation*}
For the last term, expanding the response second moment gives
\begin{align*}
\E_q[R_v^2]
&=\frac{1}{N_v^2}\sum_{k,\ell\in U_v}
\frac{I_kI_\ell\epsilon_k\epsilon_\ell}{\pi_k\pi_\ell}
\E_q\left[\left\{\frac{r_k}{p(\bx_k)}-1\right\}
\left\{\frac{r_\ell}{p(\bx_\ell)}-1\right\}
\right].
\end{align*}
Independence and centering make the cross-products vanish for $k\ne\ell$.
For $k=\ell$, the response expectation is
$\{1-p(\bx_k)\}/p(\bx_k)$. Thus
\begin{equation*}
\E_q[R_v^2]
=\frac{1}{N_v^2}\sum_{k\in U_v}
\frac{I_k}{\pi_k^2}\frac{1-p(\bx_k)}{p(\bx_k)}\epsilon_k^2.
\end{equation*}
Averaging and substituting into \eqref{eq:oracle_mean_square_expansion} gives
\begin{equation*}
\E_{mpq}\left[\{\widehat\mu_{\mathrm{aipw}}(m,p)-\mu\}^2\right]
=\E_m[\V_p(\widehat\mu_F)]
+\frac{1}{N_v^2}\sum_{k\in U_v}
\frac{1-p(\bx_k)}{\pi_kp(\bx_k)}\sigma_k^2.
\end{equation*}
For the first term, design unbiasedness gives
\begin{equation*}
\E_m[\V_p(\widehat\mu_F)]
=\E_{mp}\left[\left\{\frac{1}{N_v}\sum_{k\in U_v}
\left(\frac{I_k}{\pi_k}-1\right)y_k\right\}^2\right].
\end{equation*}
This is the second moment of a Horvitz--Thompson mean error, so
Lemma~\ref{lem:ht_mean_bound} applies with $\xi_k=y_k$. Its moment condition holds
because \ref{M1} and the strong law give
$N_v^{-1}\sum_{k\in U_v}\E_m[y_k^2]=\mathcal O(1)$.
Thus $\E_m[\V_p(\widehat\mu_F)]=\mathcal O(n_v^{-1})$.
For $\E_{mpq}[R_v^2]$, the same moment bound gives
\begin{align*}
\frac{1}{N_v^2}\sum_{k\in U_v}
\frac{1-p(\bx_k)}{\pi_kp(\bx_k)}\sigma_k^2
&\leq\frac{1}{\lambda_1\delta N_v^2}\sum_{k\in U_v}\sigma_k^2\\
&\leq\frac{1}{\lambda_1\delta N_v^2}\sum_{k\in U_v}\E_m[y_k^2]
=\mathcal{O}(N_v^{-1}).
\end{align*}
Combining the bounds proves the result.
\end{proof}

\subsection{Proof of Theorem~\ref{theo:oracle_variance}}
\label{proof:oracle_variance}

\begin{proof}
Consider the decomposition
\begin{equation}
\label{eq:oracle_variance_error_decomposition}
\begin{aligned}
n_v\left[\widehat V_{\mathrm{aipw}}(m,p)
-V_{\mathrm{tot}}(\widehat\mu_{\mathrm{aipw}}(m,p))\right]
&=n_v\left[\widehat V_1(m,p)
-V_1(\widehat\mu_{\mathrm{aipw}}(m,p))\right]\\
&\quad+n_v\left[\widehat V_2(m,p)
-V_2(\widehat\mu_{\mathrm{aipw}}(m,p))\right].
\end{aligned}
\end{equation}

\paragraph{Consistency of the first component.}

Write
\begin{equation*}
n_v\left[\widehat V_1(m,p)
-V_1(\widehat\mu_{\mathrm{aipw}}(m,p))\right]
=A_{1,v}+A_{2,v},
\end{equation*}
where
\begin{equation*}
A_{1,v}=n_v\left[\widehat V_1(m,p)-\E_p[\widehat V_1(m,p)]\right],
\qquad
A_{2,v}=n_v\left[\E_p[\widehat V_1(m,p)]
-V_1(\widehat\mu_{\mathrm{aipw}}(m,p))\right].
\end{equation*}

\medskip
\noindent\textit{Convergence of $A_{1,v}$.}
The design expectation of the oracle variance estimator is
\begin{equation*}
\E_p[\widehat V_1(m,p)]
=\frac{1}{N_v^2}\sum_{k\in U_v}\sum_{\ell\in U_v}
\frac{\Delta_{k\ell}}{\pi_k\pi_\ell}\eta_k\eta_\ell.
\end{equation*}
This is the design variance of the Horvitz--Thompson mean of the
pseudo-values. Thus $A_{1,v}$ is the scaled error of its variance estimator,
as in Lemma~\ref{lem:ht_variance_consistency} with $\xi_k=\eta_k$.
The pseudo-values are independent of the sample indicators conditionally
on $\mathbf X$. The remaining fourth-moment condition follows from
Lemma~\ref{lem:oracle_moment_bounds}, which gives
\begin{equation*}
\frac{1}{N_v}\sum_{k\in U_v}\eta_k^4=\mathcal{O}_{\P}(1).
\end{equation*}
Lemma~\ref{lem:ht_variance_consistency} therefore gives $A_{1,v}=o_{\P}(1)$.

\Needspace{7\baselineskip}
\medskip
\noindent\textit{Convergence of $A_{2,v}$.}
Since
\begin{equation*}
\E_{mq}\left[\E_p[\widehat V_1(m,p)]\right]
=V_1(\widehat\mu_{\mathrm{aipw}}(m,p)),
\end{equation*}
we have $\E_{mq}[A_{2,v}]=0$. It therefore suffices to prove
\begin{equation*}
\E_{mq}[A_{2,v}^2]=\V_{mq}(A_{2,v})=o(1).
\end{equation*}

Expanding the variance gives
\begin{equation}
\label{eq:oracle_v1_variance_expansion}
\V_{mq}(A_{2,v})
=\frac{n_v^2}{N_v^4}\sum_{k,\ell,i,j\in U_v}
\frac{\Delta_{k\ell}\Delta_{ij}}{\pi_k\pi_\ell\pi_i\pi_j}
\operatorname{Cov}_{mq}(\eta_k\eta_\ell,\eta_i\eta_j).
\end{equation}
The pseudo-values are independent, so the covariance is zero whenever
$\{k,\ell\}\cap\{i,j\}=\varnothing$.
For the remaining covariances,
\begin{equation*}
|\operatorname{Cov}_{mq}(\eta_k\eta_\ell,\eta_i\eta_j)|
\leq\E_{mq}[|\eta_k\eta_\ell\eta_i\eta_j|]
+\E_{mq}[|\eta_k\eta_\ell|]\E_{mq}[|\eta_i\eta_j|].
\end{equation*}
Cauchy--Schwarz gives
\begin{align*}
\E_{mq}[|\eta_k\eta_\ell\eta_i\eta_j|]
&\leq\{\E_{mq}[\eta_k^2\eta_\ell^2]
\E_{mq}[\eta_i^2\eta_j^2]\}^{1/2}\\
&\leq\{\E_{mq}[\eta_k^4]\E_{mq}[\eta_\ell^4]
\E_{mq}[\eta_i^4]\E_{mq}[\eta_j^4]\}^{1/4}.
\end{align*}
Similarly,
\begin{equation*}
\E_{mq}[|\eta_k\eta_\ell|]
\leq\{\E_{mq}[\eta_k^2\eta_\ell^2]\}^{1/2}
\leq\{\E_{mq}[\eta_k^4]\E_{mq}[\eta_\ell^4]\}^{1/4}.
\end{equation*}
Using the same bound for $(i,j)$ and then the arithmetic--geometric mean inequality,
\begin{align*}
|\operatorname{Cov}_{mq}(\eta_k\eta_\ell,\eta_i\eta_j)|
&\leq2\{\E_{mq}[\eta_k^4]\E_{mq}[\eta_\ell^4]
\E_{mq}[\eta_i^4]\E_{mq}[\eta_j^4]\}^{1/4}\\
&\leq\frac{1}{2}\E_{mq}[\eta_k^4+\eta_\ell^4+\eta_i^4+\eta_j^4].
\end{align*}
Substituting into \eqref{eq:oracle_v1_variance_expansion} and retaining
only overlapping index pairs gives
\begin{equation*}
\V_{mq}(A_{2,v})
\leq\frac{n_v^2}{2N_v^4}
\sum_{\substack{k,\ell,i,j\in U_v\\\{k,\ell\}\cap\{i,j\}\ne\varnothing}}
\frac{|\Delta_{k\ell}\Delta_{ij}|}{\pi_k\pi_\ell\pi_i\pi_j}
\E_{mq}[\eta_k^4+\eta_\ell^4+\eta_i^4+\eta_j^4].
\end{equation*}
Every quadruple with overlapping index pairs satisfies at least one of
$k=i$, $k=j$, $\ell=i$, or $\ell=j$.
That is,
\begin{equation*}
\mathds{1}_{\{\{k,\ell\}\cap\{i,j\}\ne\varnothing\}}
\leq\mathds{1}_{\{k=i\}}+\mathds{1}_{\{k=j\}}
+\mathds{1}_{\{\ell=i\}}+\mathds{1}_{\{\ell=j\}}.
\end{equation*}
Multiplying by the nonnegative summand and summing gives an upper bound
by four restricted sums. By symmetry, these sums are equal.
For $i=k$, the sum of fourth powers is
$\eta_k^4+\eta_\ell^4+\eta_i^4+\eta_j^4
=2\eta_k^4+\eta_\ell^4+\eta_j^4$. Thus
\begin{equation}
\label{eq:oracle_v1_overlap_bound}
\V_{mq}(A_{2,v})
\leq\frac{2n_v^2}{N_v^4}\sum_{k,\ell,j\in U_v}
\frac{|\Delta_{k\ell}\Delta_{kj}|}{\pi_k^2\pi_\ell\pi_j}
\E_{mq}[2\eta_k^4+\eta_\ell^4+\eta_j^4].
\end{equation}
For the contribution from $2\eta_k^4$, 
\begin{equation*}
\begin{aligned}
&\sum_{k,\ell,j\in U_v}
\frac{|\Delta_{k\ell}\Delta_{kj}|}{\pi_k^2\pi_\ell\pi_j}
\E_{mq}[2\eta_k^4]\\
&\quad=2\sum_{k\in U_v}\E_{mq}[\eta_k^4]
\left(\sum_{\ell\in U_v}\frac{|\Delta_{k\ell}|}{\pi_k\pi_\ell}\right)
\left(\sum_{j\in U_v}\frac{|\Delta_{kj}|}{\pi_k\pi_j}\right).
\end{aligned}
\end{equation*}
For every $k\in U_v$, the design assumptions give
\begin{equation*}
\begin{aligned}
\sum_{\ell\in U_v}\frac{|\Delta_{k\ell}|}{\pi_k\pi_\ell}
&=\frac{1-\pi_k}{\pi_k}
+\sum_{\ell\in U_v\setminus\{k\}}
\frac{|\Delta_{k\ell}|}{\pi_k\pi_\ell}\\
&\leq\frac{1}{\lambda_1}
+\frac{N_v-1}{\lambda_1^2}\max_{(i,j)\in D_2}|\Delta_{ij}|
\leq C.
\end{aligned}
\end{equation*}
Here $C$ is finite and independent of $k$ and $v$, for all sufficiently
large $v$. Thus the contribution from $2\eta_k^4$ satisfies
\begin{equation*}
\sum_{k,\ell,j\in U_v}
\frac{|\Delta_{k\ell}\Delta_{kj}|}{\pi_k^2\pi_\ell\pi_j}
\E_{mq}[2\eta_k^4]
\leq2C^2\sum_{k\in U_v}\E_{mq}[\eta_k^4].
\end{equation*}
For the contribution from $\eta_\ell^4$, we have
\begin{align*}
&\sum_{k,\ell,j\in U_v}
\frac{|\Delta_{k\ell}\Delta_{kj}|}{\pi_k^2\pi_\ell\pi_j}
\E_{mq}[\eta_\ell^4]\\
&\quad=\sum_{\ell\in U_v}\E_{mq}[\eta_\ell^4]
\sum_{k\in U_v}\frac{|\Delta_{k\ell}|}{\pi_k\pi_\ell}
\left(\sum_{j\in U_v}\frac{|\Delta_{kj}|}{\pi_k\pi_j}\right)\\
&\quad\leq C^2\sum_{\ell\in U_v}\E_{mq}[\eta_\ell^4].
\end{align*}
By symmetry,
\begin{equation*}
\sum_{k,\ell,j\in U_v}
\frac{|\Delta_{k\ell}\Delta_{kj}|}{\pi_k^2\pi_\ell\pi_j}
\E_{mq}[\eta_j^4]
\leq C^2\sum_{j\in U_v}\E_{mq}[\eta_j^4].
\end{equation*}
Adding the three inequalities yields
\begin{align*}
&\sum_{k,\ell,j\in U_v}
\frac{|\Delta_{k\ell}\Delta_{kj}|}{\pi_k^2\pi_\ell\pi_j}
\E_{mq}[2\eta_k^4+\eta_\ell^4+\eta_j^4]\\
&\quad\leq2C^2\sum_{k\in U_v}\E_{mq}[\eta_k^4]
+C^2\sum_{\ell\in U_v}\E_{mq}[\eta_\ell^4]
+C^2\sum_{j\in U_v}\E_{mq}[\eta_j^4]
=4C^2\sum_{k\in U_v}\E_{mq}[\eta_k^4].
\end{align*}
Substituting into \eqref{eq:oracle_v1_overlap_bound} and using
Lemma~\ref{lem:oracle_moment_bounds} gives
\begin{equation*}
\V_{mq}(A_{2,v})
\leq\frac{8C^2n_v^2}{N_v^4}\sum_{k\in U_v}\E_{mq}[\eta_k^4]
=\mathcal{O}(n_v^2/N_v^3)=o(1).
\end{equation*}
Thus $\E_{mq}[A_{2,v}^2]=o(1)$, so $A_{2,v}=o_{\P}(1)$.
It follows that
\begin{equation}
\label{eq:oracle_V1_consistency_proof}
n_v\left[
\widehat V_1(m,p)-V_1(\widehat{\mu}_{\mathrm{aipw}}(m,p))
\right]=A_{1,v}+A_{2,v}=o_{\P}(1).
\end{equation}

\paragraph{Consistency of the IPW variance estimator.}
By unbiasedness, it suffices to prove
\begin{equation*}
\E_{mpq}\left[\left\{n_v[\widehat V_{2,\mathrm{ipw}}(m,p)
-V_2(\widehat\mu_{\mathrm{aipw}}(m,p))]\right\}^2\right]
=n_v^2\V_{mpq}(\widehat V_{2,\mathrm{ipw}}(m,p))=o(1).
\end{equation*}
Using the sample membership and response indicators,
write \eqref{V2_OR_IPW_hat} as
\begin{equation*}
\widehat V_{2,\mathrm{ipw}}(m,p)
=\frac{1}{N_v^2}\sum_{k\in U_v}
\frac{I_kr_k}{\pi_k}
\frac{1-p(\bx_k)}{\{p(\bx_k)\}^2}\epsilon_k^2.
\end{equation*}
Expanding its variance gives
\begin{equation}
\label{eq:oracle_ipw_variance_expansion}
\begin{aligned}
\V_{mpq}(\widehat V_{2,\mathrm{ipw}}(m,p))
&=\frac{1}{N_v^4}\sum_{k\in U_v}
\frac{\{1-p(\bx_k)\}^2}{p(\bx_k)^4}
\V_{mpq}\left(\frac{I_kr_k\epsilon_k^2}{\pi_k}\right)\\
&\quad+\frac{1}{N_v^4}\sum_{(k,\ell)\in D_2}
\frac{\{1-p(\bx_k)\}\{1-p(\bx_\ell)\}}
{p(\bx_k)^2p(\bx_\ell)^2}
\operatorname{Cov}_{mpq}
\left(\frac{I_kr_k\epsilon_k^2}{\pi_k},
\frac{I_\ell r_\ell\epsilon_\ell^2}{\pi_\ell}\right).
\end{aligned}
\end{equation}
For the diagonal terms, $I_k^2=I_k$ and $r_k^2=r_k$ give
\begin{align*}
\E_{mpq}\left[\frac{I_kr_k\epsilon_k^2}{\pi_k}\right]
&=p(\bx_k)\sigma_k^2,\\
\V_{mpq}\left(\frac{I_kr_k\epsilon_k^2}{\pi_k}\right)
&=\frac{p(\bx_k)}{\pi_k}\E_m[\epsilon_k^4]
-p(\bx_k)^2(\sigma_k^2)^2
\leq\frac{p(\bx_k)}{\pi_k}\E_m[\epsilon_k^4].
\end{align*}
For $k\ne\ell$, independence of the outcomes and responses across units gives
\begin{align*}
\operatorname{Cov}_{mpq}
\left(\frac{I_kr_k\epsilon_k^2}{\pi_k},
\frac{I_\ell r_\ell\epsilon_\ell^2}{\pi_\ell}\right)
&=\left(\frac{\pi_{k\ell}}{\pi_k\pi_\ell}-1\right)
p(\bx_k)p(\bx_\ell)\sigma_k^2\sigma_\ell^2\\
&=\frac{\Delta_{k\ell}}{\pi_k\pi_\ell}
p(\bx_k)p(\bx_\ell)\sigma_k^2\sigma_\ell^2.
\end{align*}
Substituting into \eqref{eq:oracle_ipw_variance_expansion} and taking
absolute values of the covariances gives
\begin{equation*}
\begin{aligned}
\V_{mpq}(\widehat V_{2,\mathrm{ipw}}(m,p))
&\leq\frac{1}{N_v^4}\sum_{k\in U_v}
\frac{\{1-p(\bx_k)\}^2}{\pi_kp(\bx_k)^3}
\E_m[\epsilon_k^4]\\
&\quad+\frac{1}{N_v^4}\sum_{(k,\ell)\in D_2}
\frac{|\Delta_{k\ell}|}{\pi_k\pi_\ell}
\frac{\{1-p(\bx_k)\}\{1-p(\bx_\ell)\}}{p(\bx_k)p(\bx_\ell)}
\sigma_k^2\sigma_\ell^2\\
&\leq\frac{1}{N_v^4}
\left\{\frac{1}{\lambda_1\delta^3}\sum_{k\in U_v}\E_m[\epsilon_k^4]
+\frac{\max_{(k,\ell)\in D_2}|\Delta_{k\ell}|}{\lambda_1^2\delta^2}
\left(\sum_{k\in U_v}\sigma_k^2\right)^2\right\}.
\end{aligned}
\end{equation*}
For the last square, Cauchy--Schwarz and
$(\sigma_k^2)^2=\{\E_m[\epsilon_k^2]\}^2
\leq\E_m[\epsilon_k^4]$ give
\begin{equation*}
\left(\sum_{k\in U_v}\sigma_k^2\right)^2
\leq N_v\sum_{k\in U_v}(\sigma_k^2)^2
\leq N_v\sum_{k\in U_v}\E_m[\epsilon_k^4].
\end{equation*}
Substituting this into the variance bound yields
\begin{equation*}
\V_{mpq}(\widehat V_{2,\mathrm{ipw}}(m,p))
\leq\frac{1}{N_v^4}
\left\{\frac{1}{\lambda_1\delta^3}
+\frac{N_v\max_{(k,\ell)\in D_2}|\Delta_{k\ell}|}{\lambda_1^2\delta^2}\right\}
\sum_{k\in U_v}\E_m[\epsilon_k^4]
=\mathcal{O}(N_v^{-3}).
\end{equation*}
Consequently,
\begin{equation*}
\E_{mpq}\left[\left\{n_v[\widehat V_{2,\mathrm{ipw}}(m,p)
-V_2(\widehat\mu_{\mathrm{aipw}}(m,p))]\right\}^2\right]
=n_v^2\V_{mpq}(\widehat V_{2,\mathrm{ipw}}(m,p))
=\mathcal O(n_v^2/N_v^3)=o(1).
\end{equation*}
Chebyshev's inequality gives
\begin{equation*}
n_v\left[
\widehat V_{2,\mathrm{ipw}}(m,p)
-V_2(\widehat{\mu}_{\mathrm{aipw}}(m,p))
\right]=o_{\P}(1).
\end{equation*}
\Needspace{8\baselineskip}
\paragraph{Consistency of the bias-corrected variance estimator.}
Write
\begin{equation}
\label{eq:oracle_bc_error_decomposition}
\begin{aligned}
n_v\left[\widehat V_{2,\mathrm{bc}}(m,p)
-V_2(\widehat\mu_{\mathrm{aipw}}(m,p))\right]
&=n_v\left[\widehat V_{2,\mathrm{ipw}}(m,p)
-V_2(\widehat\mu_{\mathrm{aipw}}(m,p))\right]+n_vG_v,
\end{aligned}
\end{equation}
where
\begin{equation*}
G_v=\frac{1}{N_v^2}\sum_{k\in U_v}\frac{I_k}{\pi_k}
\left\{1-\frac{r_k}{p(\bx_k)}\right\}\sigma_k^2.
\end{equation*}
The IPW result reduces the proof to $n_vG_v=o_{\P}(1)$. Expanding the second moment gives
\begin{equation*}
\E_{pq}[G_v^2]
=\frac{1}{N_v^4}\sum_{k,\ell\in U_v}
\frac{\sigma_k^2\sigma_\ell^2}{\pi_k\pi_\ell}
\E_{pq}\left[I_kI_\ell
\left\{1-\frac{r_k}{p(\bx_k)}\right\}
\left\{1-\frac{r_\ell}{p(\bx_\ell)}\right\}\right].
\end{equation*}
The response factors are centered and independent across units and of
the sample indicators, so
\begin{equation*}
\E_{pq}\left[I_kI_\ell
\left\{1-\frac{r_k}{p(\bx_k)}\right\}
\left\{1-\frac{r_\ell}{p(\bx_\ell)}\right\}\right]
=\pi_k\frac{1-p(\bx_k)}{p(\bx_k)}
\mathds{1}_{\{k=\ell\}}.
\end{equation*}
Substituting into the expansion of $\E_{pq}[G_v^2]$ gives
\begin{equation*}
\E_{pq}[G_v^2]
=\frac{1}{N_v^4}\sum_{k\in U_v}
\frac{1-p(\bx_k)}{\pi_kp(\bx_k)}(\sigma_k^2)^2
\leq\frac{1}{\lambda_1\delta N_v^4}\sum_{k\in U_v}\E_m[\epsilon_k^4]
=\mathcal{O}(N_v^{-3}).
\end{equation*}
It follows that $\E_{pq}[(n_vG_v)^2]=n_v^2\E_{pq}[G_v^2]
=\mathcal O(n_v^2/N_v^3)=o(1)$, and hence $n_vG_v=o_{\P}(1)$.
Substituting into \eqref{eq:oracle_bc_error_decomposition} proves
consistency of the bias-corrected estimator. Thus, for either choice of $\widehat V_2(m,p)$,
\begin{equation*}
n_v\left[
\widehat V_2(m,p)
-V_2(\widehat{\mu}_{\mathrm{aipw}}(m,p))
\right]=o_{\P}(1).
\end{equation*}

\paragraph{Total variance.}
Substituting this result and \eqref{eq:oracle_V1_consistency_proof}
into \eqref{eq:oracle_variance_error_decomposition} gives
\begin{equation*}
n_v\left[\widehat V_{\mathrm{aipw}}(m,p)
-V_{\mathrm{tot}}(\widehat\mu_{\mathrm{aipw}}(m,p))\right]=o_{\P}(1).
\end{equation*}
The additional variance-limit condition gives
\begin{equation*}
n_vV_{\mathrm{tot}}(\widehat{\mu}_{\mathrm{aipw}}(m,p))
\xrightarrow[v\to\infty]{\P}\pi_*c>0.
\end{equation*}
Thus
\begin{equation*}
\frac{\widehat V_{\mathrm{aipw}}(m,p)}
{V_{\mathrm{tot}}(\widehat\mu_{\mathrm{aipw}}(m,p))}-1
=\frac{n_v[\widehat V_{\mathrm{aipw}}(m,p)
-V_{\mathrm{tot}}(\widehat\mu_{\mathrm{aipw}}(m,p))]}
{n_vV_{\mathrm{tot}}(\widehat\mu_{\mathrm{aipw}}(m,p))}
=o_{\P}(1).
\end{equation*}
\end{proof}

\Needspace{8\baselineskip}
\subsection{Proof of Theorem~\ref{theo1}}
\label{proof:theo1}
\label{proof:asympto}

\begin{proof}
Write the oracle error as
\begin{equation*}
\widehat\mu_{\mathrm{aipw}}(m,p)-\mu=A_{1,v}+A_{2,v},
\end{equation*}
where
\begin{equation*}
A_{1,v}=\frac{1}{N_v}\sum_{k\in U_v}
\left(\frac{I_k}{\pi_k}-1\right)\eta_k,
\qquad
A_{2,v}=\frac{1}{N_v}\sum_{k\in U_v}
\left\{\frac{r_k}{p(\bx_k)}-1\right\}\epsilon_k.
\end{equation*}
Both terms are centered, with variances given by \eqref{V1_or} and \eqref{V2_or}, respectively.
The first term is a sampling error for the pseudo-values, and the second
is a sum of independent response contributions. We establish the limit
of $A_{2,v}$ first, then combine it with the conditional limit of $A_{1,v}$.

\paragraph{Limit of $\sqrt{N_v}A_{2,v}$.}
Recall from \eqref{eq:oracle_response_variance_limit} that
$N_vV_2(\widehat\mu_{\mathrm{aipw}}(m,p))\xrightarrow[v\to\infty]{} c_2\geq0$.
Suppose that $c_2>0$. Write
\begin{equation*}
A_{2,v}=\frac{1}{N_v}\sum_{k\in U_v}z_k,
\qquad z_k=\left\{\frac{r_k}{p(\bx_k)}-1\right\}\epsilon_k.
\end{equation*}
The summands are independent and centered conditionally on $\mathbf X$.
Lyapunov's theorem therefore gives a normal limit once we verify
\begin{equation*}
\frac{\sum_{k\in U_v}\E_{mq}[z_k^4]}
{\left\{\sum_{k\in U_v}\V_{mq}(z_k)\right\}^2}=o(1).
\end{equation*}
For the denominator, $\sum_{k\in U_v}z_k=N_vA_{2,v}$ and independence give
\begin{equation*}
\sum_{k\in U_v}\V_{mq}(z_k)
=\V_{mq}\left(\sum_{k\in U_v}z_k\right)
=N_v^2\V_{mq}(A_{2,v})
=N_v^2V_2(\widehat\mu_{\mathrm{aipw}}(m,p)).
\end{equation*}
For the numerator, positivity gives
\begin{equation*}
\begin{aligned}
\E_{mq}[z_k^4]
&=\left[\frac{\{1-p(\bx_k)\}^4}{\{p(\bx_k)\}^3}
+1-p(\bx_k)\right]\E_m[\epsilon_k^4]\\
&\leq(\delta^{-3}+1)\E_m[\epsilon_k^4].
\end{aligned}
\end{equation*}
Combining the variance identity and the fourth-moment bound gives
\begin{align*}
\frac{\sum_{k\in U_v}\E_{mq}[z_k^4]}
{\left\{\sum_{k\in U_v}\V_{mq}(z_k)\right\}^2}
&\leq(\delta^{-3}+1)
\frac{\sum_{k\in U_v}\E_m[\epsilon_k^4]}
{N_v^4\{V_2(\widehat\mu_{\mathrm{aipw}}(m,p))\}^2}\\
&=\frac{\delta^{-3}+1}{N_v}
\frac{N_v^{-1}\sum_{k\in U_v}\E_m[\epsilon_k^4]}
{\{N_vV_2(\widehat\mu_{\mathrm{aipw}}(m,p))\}^2}
=\mathcal{O}(N_v^{-1}).
\end{align*}
The last equality follows because the average fourth moment is bounded by
Lemma~\ref{lem:oracle_moment_bounds} and
$N_vV_2(\widehat\mu_{\mathrm{aipw}}(m,p))\xrightarrow[v\to\infty]{}c_2>0$.
This verifies the remaining condition of Lyapunov's theorem, which gives
\begin{equation}
\label{eq:oracle_response_clt}
\frac{A_{2,v}}{\sqrt{V_2(\widehat\mu_{\mathrm{aipw}}(m,p))}}
\xrightarrow[v\to\infty]{\cL}\mathcal N(0,1).
\end{equation}

By \eqref{eq:oracle_response_variance_limit}, this also gives
$\sqrt{N_v}A_{2,v}\xrightarrow[v\to\infty]{\cL}\mathcal N(0,c_2)$.

\Needspace{9\baselineskip}
\paragraph{Limit of $\sqrt{N_v}(A_{1,v}+A_{2,v})$.}
Recall from \ref{CL1} that
$N_vV_1(\widehat\mu_{\mathrm{aipw}}(m,p))\xrightarrow[v\to\infty]{} c_1\geq0$, with
$c_1+c_2>0$. Suppose first that $c_1>0$ and $c_2>0$.
The term $A_{2,v}$ is $\cB_v$-measurable and has the normal limit
\eqref{eq:oracle_response_clt}, whereas $A_{1,v}$ has conditional mean
zero and the conditional normal approximation in \ref{CL2}.
Theorem~2 of \citet{chen2007asymptotic} combines these two limits once
its conditional variance and normalization conditions hold.
We use it with $U_n=A_{1,v}$, $V_n=A_{2,v}$, and
$\mathcal B_n=\cB_v$. For the conditional variance, write
\begin{equation*}
\V_p(A_{1,v}\mid\cB_v)
=\frac{1}{N_v^2}\sum_{k,\ell\in U_v}
\frac{\Delta_{k\ell}}{\pi_k\pi_\ell}\eta_k\eta_\ell
=\E_p[\widehat V_1(m,p)].
\end{equation*}
The proof of Theorem~\ref{theo:oracle_variance} gives
\begin{equation*}
n_v\left\{\V_p(A_{1,v}\mid\cB_v)
-V_1(\widehat\mu_{\mathrm{aipw}}(m,p))\right\}=o_{\P}(1).
\end{equation*}
Since $N_v/n_v\xrightarrow[v\to\infty]{}\pi_*^{-1}$,
\begin{equation*}
N_v\V_p(A_{1,v}\mid\cB_v)\xrightarrow[v\to\infty]{\P}c_1,
\qquad
\frac{V_2(\widehat\mu_{\mathrm{aipw}}(m,p))}
{\V_p(A_{1,v}\mid\cB_v)}
\xrightarrow[v\to\infty]{\P}\frac{c_2}{c_1}.
\end{equation*}
For the conditional normal approximation, write
\begin{equation*}
\frac{A_{1,v}}{\sqrt{\V_p(A_{1,v}\mid\cB_v)}}
=\frac{A_{1,v}}{\sqrt{V_1(\widehat\mu_{\mathrm{aipw}}(m,p))}}
\left\{\frac{V_1(\widehat\mu_{\mathrm{aipw}}(m,p))}
{\V_p(A_{1,v}\mid\cB_v)}\right\}^{1/2}.
\end{equation*}
The second factor is $\cB_v$-measurable and tends to one in probability,
so \ref{CL2} also holds with the conditional standard deviation.
This verifies the conditional normal
approximation and variance-ratio conditions of their theorem.
Its conclusion and Slutsky's theorem yield
\begin{equation*}
\sqrt{N_v}(A_{1,v}+A_{2,v})
\xrightarrow[v\to\infty]{\cL}\mathcal N(0,c_1+c_2).
\end{equation*}

If $c_1=0$, $\sqrt{N_v}A_{1,v}=o_{\P}(1)$, since
\begin{equation*}
\E_{mpq}[(\sqrt{N_v}A_{1,v})^2]
=N_vV_1(\widehat\mu_{\mathrm{aipw}}(m,p))\xrightarrow[v\to\infty]{}0.
\end{equation*}
Similarly, if $c_2=0$, then
\begin{equation*}
\E_{mq}[(\sqrt{N_v}A_{2,v})^2]
=N_vV_2(\widehat\mu_{\mathrm{aipw}}(m,p))\xrightarrow[v\to\infty]{}0,
\end{equation*}
so $\sqrt{N_v}A_{2,v}=o_{\P}(1)$. Overall, Slutsky's theorem yields
\begin{equation*}
\sqrt{N_v}(A_{1,v}+A_{2,v})
\xrightarrow[v\to\infty]{\cL}\mathcal N(0,c_1+c_2).
\end{equation*}

\paragraph{Studentization.}
Since $N_vV_{\mathrm{tot}}(\widehat\mu_{\mathrm{aipw}}(m,p))\xrightarrow[v\to\infty]{}c_1+c_2$,
the preceding limit and Slutsky's theorem give
\begin{equation*}
\frac{\widehat\mu_{\mathrm{aipw}}(m,p)-\mu}
{\sqrt{V_{\mathrm{tot}}(\widehat\mu_{\mathrm{aipw}}(m,p))}}
\xrightarrow[v\to\infty]{\cL}\mathcal N(0,1).
\end{equation*}
Write
\begin{equation*}
\frac{\widehat\mu_{\mathrm{aipw}}(m,p)-\mu}
{\sqrt{\widehat V_{\mathrm{aipw}}(m,p)}}
=\frac{\widehat\mu_{\mathrm{aipw}}(m,p)-\mu}
{\sqrt{V_{\mathrm{tot}}(\widehat\mu_{\mathrm{aipw}}(m,p))}}
\left\{\frac{V_{\mathrm{tot}}(\widehat\mu_{\mathrm{aipw}}(m,p))}
{\widehat V_{\mathrm{aipw}}(m,p)}\right\}^{1/2}.
\end{equation*}
The positive limiting variance and Theorem~\ref{theo:oracle_variance} yield
\begin{equation*}
\frac{\widehat V_{\mathrm{aipw}}(m,p)}
{V_{\mathrm{tot}}(\widehat\mu_{\mathrm{aipw}}(m,p))}
\xrightarrow[v\to\infty]{\P}1.
\end{equation*}
Hence, Slutsky's theorem gives the result.
\end{proof}

\section{Proofs for Section~\ref{sec:aipw3}}
\label{app:proofs_crossfitting}

\subsection{Proof of Theorem~\ref{Theo2}}
\label{proof:Theo2}

\begin{proof}[Proof of (\ref{thm:cf_consistency})]
Using \eqref{eq:cf_oracle_decomposition}, we write
\begin{equation*}
\widehat\mu_{\mathrm{aipw,cf}}(\widehat m,\widehat p)-\mu
=\{\widehat\mu_{\mathrm{aipw}}(m,p)-\mu\}
+\sum_{t=1}^L
\left(T_{\mathrm{reg},t}^{\mathrm{cf}}
+T_{\mathrm{prop},t}^{\mathrm{cf}}+T_{\mathrm{prod},t}^{\mathrm{cf}}\right).
\end{equation*}
Proposition~\ref{prop:oracle_consistency} gives
$\widehat\mu_{\mathrm{aipw}}(m,p)-\mu=o_{\P}(1)$.
Since $L$ is fixed, it remains to prove, for each fold $t$,
\begin{equation*}
T_{\mathrm{reg},t}^{\mathrm{cf}}=o_{\P}(1),
\qquad T_{\mathrm{prop},t}^{\mathrm{cf}}=o_{\P}(1),
\qquad T_{\mathrm{prod},t}^{\mathrm{cf}}=o_{\P}(1).
\end{equation*}
We use the event $\mathcal A$ of Appendix~\ref{app:notation}, on which
fitted propensities are at least $\rho$ and $\P(\mathcal A^c)=o(1)$.

\paragraph{Convergence of $T_{\mathrm{reg},t}^{\mathrm{cf}}$.}
Since the fitted regression errors are $\mathcal D_{-t}$-measurable,
expanding the conditional second moment gives
\begin{equation*}
\begin{aligned}
&\E_q[(T_{\mathrm{reg},t}^{\mathrm{cf}})^2\mid\mathcal D_{-t}]\\
&\quad=\frac{1}{N_v^2}\sum_{k,\ell\in\mathcal F_t}
\frac{\{\widehat m_{-k}(\bx_k)-m(\bx_k)\}\{\widehat m_{-\ell}(\bx_\ell)-m(\bx_\ell)\}}
{\pi_k\pi_\ell}
\E_q\left[\left(1-\frac{r_k}{p(\bx_k)}\right)
\left(1-\frac{r_\ell}{p(\bx_\ell)}\right)
\mid\mathcal D_{-t}\right].
\end{aligned}
\end{equation*}
Lemma~\ref{lem:conditional_moments} gives
\begin{equation*}
\E_q\left[\left(1-\frac{r_k}{p(\bx_k)}\right)
\left(1-\frac{r_\ell}{p(\bx_\ell)}\right)
\mid\mathcal D_{-t}\right]
=\frac{1-p(\bx_k)}{p(\bx_k)}\mathds{1}_{\{k=\ell\}}.
\end{equation*}
Substituting into the conditional second-moment expansion leaves
only the diagonal terms:
\begin{equation*}
\E_q[(T_{\mathrm{reg},t}^{\mathrm{cf}})^2\mid\mathcal D_{-t}]
=\frac{1}{N_v^2}\sum_{k\in\mathcal F_t}
\frac{1-p(\bx_k)}{\pi_k^2p(\bx_k)}
\{\widehat m_{-k}(\bx_k)-m(\bx_k)\}^2.
\end{equation*}
Taking expectations and using the definition of prediction risk gives
\begin{equation}
\label{eq:cf_reg_second_moment_bound}
\begin{aligned}
\E_{mpq}\left[(T_{\mathrm{reg},t}^{\mathrm{cf}})^2\right]
&=\frac{1}{N_v^2}\E_{mpq}\left[
\sum_{k\in\mathcal F_t}\frac{1-p(\bx_k)}{\pi_k^2p(\bx_k)}
\{\widehat m_{-t}(\bx_k)-m(\bx_k)\}^2\right]\\
&\leq\frac{(1-\delta)N_t}{\lambda_1^2\delta N_v^2}
\mathcal R_2(\widehat m_{-t})=\mathcal O(N_v^{-1}).
\end{aligned}
\end{equation}
Chebyshev's inequality gives $T_{\mathrm{reg},t}^{\mathrm{cf}}=o_{\P}(1)$.

\paragraph{Convergence of $T_{\mathrm{prop},t}^{\mathrm{cf}}$.}
For any $\varepsilon>0$, Chebyshev's inequality gives
\begin{equation*}
\P(|T_{\mathrm{prop},t}^{\mathrm{cf}}|>\varepsilon)
\leq\P(\mathcal A^c)+\frac{1}{\varepsilon^2}
\E_{mpq}[(T_{\mathrm{prop},t}^{\mathrm{cf}})^2\mathds{1}_{\mathcal A}].
\end{equation*}
\Needspace{7\baselineskip}
Since propensity fitting uses no outcomes, the propensity estimates and
$\mathds{1}_{\mathcal A}$ are
$\sigma(\mathbf X,\mathbf I,\mathbf r,\boldsymbol{\delta})$-measurable.
Hence
\begin{equation*}
\begin{aligned}
&\E_m[(T_{\mathrm{prop},t}^{\mathrm{cf}})^2\mid\mathcal D_{-t}]\\
&\quad=\frac{1}{N_v^2}\sum_{k,\ell\in\mathcal F_t}
\frac{r_kr_\ell\{\widehat p_{-k}(\bx_k)-p(\bx_k)\}
\{\widehat p_{-\ell}(\bx_\ell)-p(\bx_\ell)\}}
{\pi_k\pi_\ell p(\bx_k)p(\bx_\ell)\widehat p_{-k}(\bx_k)\widehat p_{-\ell}(\bx_\ell)}
\E_m[\epsilon_k\epsilon_\ell\mid\mathcal D_{-t}].
\end{aligned}
\end{equation*}
Lemma~\ref{lem:conditional_moments} gives
$\E_m[\epsilon_k\epsilon_\ell\mid\mathcal D_{-t}]
=\sigma_k^2\mathds{1}_{\{k=\ell\}}$.
Substituting into the conditional second-moment expansion and using
$r_k^2=r_k$ yields
\begin{equation}
\label{eq:cf_prop_conditional_second_moment}
\E_m[(T_{\mathrm{prop},t}^{\mathrm{cf}})^2\mid\mathcal D_{-t}]
=\frac{1}{N_v^2}\sum_{k\in\mathcal F_t}
\frac{r_k\sigma_k^2\{\widehat p_{-k}(\bx_k)-p(\bx_k)\}^2}
{\pi_k^2p(\bx_k)^2\widehat p_{-k}(\bx_k)^2}.
\end{equation}
Multiplying by $\mathds{1}_{\mathcal A}$ and taking expectations yields
\begin{equation*}
\E_{mpq}\left[(T_{\mathrm{prop},t}^{\mathrm{cf}})^2\mathds{1}_{\mathcal A}\right]
=\frac{1}{N_v^2}\E_{mpq}\left[
\mathds{1}_{\mathcal A}\sum_{k\in\mathcal F_t}
\frac{r_k\sigma_k^2\{\widehat p_{-t}(\bx_k)-p(\bx_k)\}^2}
{\pi_k^2p(\bx_k)^2\{\widehat p_{-t}(\bx_k)\}^2}
\right].
\end{equation*}
On $\mathcal A$, $r_k\leq1$ and $|\widehat p_{-k}(\bx_k)-p(\bx_k)|\leq1$ give
\begin{equation*}
\frac{r_k\{\widehat p_{-k}(\bx_k)-p(\bx_k)\}^2}
{\pi_k^2p(\bx_k)^2\widehat p_{-k}(\bx_k)^2}
\leq\frac{1}{\lambda_1^2\delta^2\rho^2},
\qquad k\in\mathcal F_t.
\end{equation*}
Consequently,
\begin{equation*}
\E_{mpq}\left[(T_{\mathrm{prop},t}^{\mathrm{cf}})^2\mathds{1}_{\mathcal A}\right]
\leq\frac{1}{\lambda_1^2\delta^2\rho^2 N_v^2}
\sum_{k\in U_v}\sigma_k^2
=\mathcal O(N_v^{-1}).
\end{equation*}
Chebyshev's inequality and $\P(\mathcal A^c)\xrightarrow[v\to\infty]{}0$ give
$T_{\mathrm{prop},t}^{\mathrm{cf}}=o_{\P}(1)$.

\paragraph{Convergence of $T_{\mathrm{prod},t}^{\mathrm{cf}}$.}
On $\mathcal A$,
\begin{samepage}
\begin{align*}
|T_{\mathrm{prod},t}^{\mathrm{cf}}|
&\leq\frac{1}{\lambda_1\delta\rho N_v}
\sum_{k\in\mathcal F_t}
|\widehat m_{-t}(\bx_k)-m(\bx_k)|
|\widehat p_{-t}(\bx_k)-p(\bx_k)|\\
&\leq\frac{N_t}{\lambda_1\delta\rho N_v}
\|\widehat m_{-t}-m\|_{\mathcal F_t,2}
\|\widehat p_{-t}-p\|_{\mathcal F_t,2},
\end{align*}
by Cauchy--Schwarz.
\end{samepage}
Taking expectations and applying Cauchy--Schwarz again,
\begin{equation}
\label{eq:cf_product_first_moment_bound}
\begin{aligned}
\E_{mpq}\left[|T_{\mathrm{prod},t}^{\mathrm{cf}}|\mathds{1}_{\mathcal A}\right]
&\leq\frac{N_t}{\lambda_1\delta\rho N_v}
\E_{mpq}\left[\|\widehat m_{-t}-m\|_{\mathcal F_t,2}
\|\widehat p_{-t}-p\|_{\mathcal F_t,2}\right]\\
&\leq\frac{N_t}{\lambda_1\delta\rho N_v}
\left\{\E_{mpq}[\|\widehat m_{-t}-m\|_{\mathcal F_t,2}^2]
\E_{mpq}[\|\widehat p_{-t}-p\|_{\mathcal F_t,2}^2]\right\}^{1/2}\\
&=\frac{N_t}{\lambda_1\delta\rho N_v}
\left\{\mathcal R_2(\widehat m_{-t})\mathcal R_2(\widehat p_{-t})\right\}^{1/2}
=o(1).
\end{aligned}
\end{equation}
Markov's inequality and $\P(\mathcal A^c)\xrightarrow[v\to\infty]{}0$ therefore give
$T_{\mathrm{prod},t}^{\mathrm{cf}}=o_{\P}(1)$. Summing over the folds and using oracle consistency in
\eqref{eq:cf_oracle_decomposition} proves
part~(\ref{thm:cf_consistency}).

\medskip
\noindent\textit{Proof of (\ref{thm:cf_oracle_equivalence}).}
We prove, for each fold $t$,
\begin{equation*}
\sqrt{n_v}T_{\mathrm{reg},t}^{\mathrm{cf}}=o_{\P}(1),
\qquad \sqrt{n_v}T_{\mathrm{prop},t}^{\mathrm{cf}}=o_{\P}(1),
\qquad \sqrt{n_v}T_{\mathrm{prod},t}^{\mathrm{cf}}=o_{\P}(1).
\end{equation*}
We use the preceding moment bounds with the prediction conditions in
\eqref{eq:cf_product_rate}.\\

\medskip
\noindent\textit{Regression term.}
Multiplying \eqref{eq:cf_reg_second_moment_bound} by $n_v$ and using
$\mathcal R_2(\widehat m_{-t})=o(1)$ gives
\begin{equation*}
\E_{mpq}\left[(\sqrt{n_v}T_{\mathrm{reg},t}^{\mathrm{cf}})^2\right]
\leq\frac{1-\delta}{\lambda_1^2\delta}
\mathcal R_2(\widehat m_{-t})=o(1).
\end{equation*}
Chebyshev's inequality gives
$\sqrt{n_v}T_{\mathrm{reg},t}^{\mathrm{cf}}=o_{\P}(1)$.

\medskip
\noindent\textit{Propensity term.}
By \eqref{eq:cf_prop_conditional_second_moment} and $\sigma_k^2\leq\bar\sigma^2$,
\begin{align*}
\E_{mpq}\left[(\sqrt{n_v}T_{\mathrm{prop},t}^{\mathrm{cf}})^2
\mathds{1}_{\mathcal A}\right]
&\leq\frac{n_v\bar\sigma^2}{\lambda_1^2\delta^2\rho^2N_v^2}
\E_{mpq}\left[\sum_{k\in\mathcal F_t}
\{\widehat p_{-k}(\bx_k)-p(\bx_k)\}^2\right]\\
&=\frac{n_vN_t\bar\sigma^2}{\lambda_1^2\delta^2\rho^2N_v^2}
\mathcal R_2(\widehat p_{-t})
\leq\frac{\bar\sigma^2}{\lambda_1^2\delta^2\rho^2}
\mathcal R_2(\widehat p_{-t})=o(1).
\end{align*}
Chebyshev's inequality and $\P(\mathcal A^c)\xrightarrow[v\to\infty]{}0$ therefore give
$\sqrt{n_v}T_{\mathrm{prop},t}^{\mathrm{cf}}=o_{\P}(1)$.

\medskip
\noindent\textit{Product term.}
Multiplying \eqref{eq:cf_product_first_moment_bound} by $\sqrt{n_v}$
and using the product-rate condition in \eqref{eq:cf_product_rate} gives
\begin{equation*}
\E_{mpq}\left[|\sqrt{n_v}T_{\mathrm{prod},t}^{\mathrm{cf}}|
\mathds{1}_{\mathcal A}\right]
\leq\frac{\{n_v\mathcal R_2(\widehat m_{-t})
\mathcal R_2(\widehat p_{-t})\}^{1/2}}{\lambda_1\delta\rho}
=o(1).
\end{equation*}
Markov's inequality and $\P(\mathcal A^c)\xrightarrow[v\to\infty]{}0$ give
$\sqrt{n_v}T_{\mathrm{prod},t}^{\mathrm{cf}}=o_{\P}(1)$.

Consequently,
\begin{equation*}
\sqrt{n_v}\left\{\widehat\mu_{\mathrm{aipw,cf}}(\widehat m,\widehat p)
-\widehat\mu_{\mathrm{aipw}}(m,p)\right\}
=\sum_{t=1}^L\sqrt{n_v}
\left(T_{\mathrm{reg},t}^{\mathrm{cf}}
+T_{\mathrm{prop},t}^{\mathrm{cf}}+T_{\mathrm{prod},t}^{\mathrm{cf}}\right)
=o_{\P}(1).\qedhere
\end{equation*}
\end{proof}

\Needspace{8\baselineskip}
\subsection{Proof of Theorem~\ref{Theo3}}
\label{proofTheo3}

\begin{proof}[Proof of (i)]
We compare $\widehat V_{1,\mathrm{cf}}$ and
$\widehat V_{2,\mathrm{ipw,cf}}$ with their oracle counterparts.
All bounds involving inverse fitted propensities are established on
the event $\mathcal A$ of
Appendix~\ref{app:notation}, where fitted propensities are at least
$\rho$ and $\P(\mathcal A^c)=o(1)$.

\paragraph{Consistency of the first variance component.}
Assume $\mathcal R_2(\widehat m_{-t})=o(1)$ for every fold $t$. Write
\begin{equation*}
n_v\left[\widehat V_{1,\mathrm{cf}}
-V_1(\widehat\mu_{\mathrm{aipw}}(m,p))\right]
=A_{1,v}+A_{2,v},
\end{equation*}
where
\begin{equation*}
A_{1,v}=n_v\left[\widehat V_{1,\mathrm{cf}}-\widehat V_1(m,p)\right],
\qquad
A_{2,v}=n_v\left[\widehat V_1(m,p)
-V_1(\widehat\mu_{\mathrm{aipw}}(m,p))\right].
\end{equation*}
Theorem~\ref{theo:oracle_variance} gives $A_{2,v}=o_{\P}(1)$,
so it remains to prove $A_{1,v}=o_{\P}(1)$.

\Needspace{7\baselineskip}
\medskip
\noindent\textit{Convergence of $A_{1,v}$.}
Recall from Appendix~\ref{app:notation} that, for arrays $a$ and $b$ on $S_v$,
\begin{equation*}
\mathcal Q(a,b)=\frac{1}{N_v^2}
\sum_{k\in S_v}\sum_{\ell\in S_v}
\frac{\Delta_{k\ell}}{\pi_{k\ell}}
\frac{a_k}{\pi_k}\frac{b_\ell}{\pi_\ell}.
\end{equation*}
Writing $\widehat\eta_{\mathrm{cf}}=\eta+(\widehat\eta_{\mathrm{cf}}-\eta)$,
expanding by bilinearity, and then using symmetry gives
\begin{align*}
A_{1,v}
&=n_v\{\mathcal Q(\widehat\eta_{\mathrm{cf}},\widehat\eta_{\mathrm{cf}})
-\mathcal Q(\eta,\eta)\}\\
&=n_v\mathcal Q(\eta,\widehat\eta_{\mathrm{cf}}-\eta)
+n_v\mathcal Q(\widehat\eta_{\mathrm{cf}}-\eta,\eta)
+n_v\mathcal Q(\widehat\eta_{\mathrm{cf}}-\eta,\widehat\eta_{\mathrm{cf}}-\eta)\\
&=2n_v\mathcal Q(\eta,\widehat\eta_{\mathrm{cf}}-\eta)
+n_v\mathcal Q(\widehat\eta_{\mathrm{cf}}-\eta,\widehat\eta_{\mathrm{cf}}-\eta).
\end{align*}
The triangle inequality and Lemma~\ref{lem:bilinear_design_bound} give
\begin{equation}
\label{eq:cf_v1_difference_bound}
\begin{aligned}
|A_{1,v}|
&\leq2n_v|\mathcal Q(\eta,\widehat\eta_{\mathrm{cf}}-\eta)|
+n_v|\mathcal Q(\widehat\eta_{\mathrm{cf}}-\eta,\widehat\eta_{\mathrm{cf}}-\eta)|\\
&\leq\frac{n_v^2}{N_v^2}
\left\{\frac{1-\lambda_1}{\lambda_1^2}
+\frac{n_v\max_{(k,\ell)\in D_2}|\Delta_{k\ell}|}{\lambda_1^2\lambda_2}\right\}\\
&\quad\times\|\widehat\eta_{\mathrm{cf}}-\eta\|_{S_v,2}
\{2\|\eta\|_{S_v,2}+\|\widehat\eta_{\mathrm{cf}}-\eta\|_{S_v,2}\}\\
&\leq C\|\widehat\eta_{\mathrm{cf}}-\eta\|_{S_v,2}
\{2\|\eta\|_{S_v,2}+\|\widehat\eta_{\mathrm{cf}}-\eta\|_{S_v,2}\}.
\end{aligned}
\end{equation}
The last inequality follows from \ref{D3}, with a constant $C$ independent of $v$.
The oracle norm is bounded by Cauchy--Schwarz and
Lemma~\ref{lem:oracle_moment_bounds}:
\begin{equation*}
\|\eta\|_{S_v,2}^2
=\frac{1}{n_v}\sum_{k\in S_v}\eta_k^2
\leq\left(\frac{1}{n_v}\sum_{k\in S_v}\eta_k^4\right)^{1/2}
=\mathcal O_{\P}(1).
\end{equation*}
It therefore remains to show that
$\|\widehat\eta_{\mathrm{cf}}-\eta\|_{S_v,2}=o_{\P}(1)$.
Subtracting the two pseudo-values gives
\begin{equation*}
\widehat\eta_{k,\mathrm{cf}}-\eta_k
=\left(1-\frac{r_k}{\widehat p_{-k}(\bx_k)}\right)\{\widehat m_{-k}(\bx_k)-m(\bx_k)\}
-\frac{r_k\epsilon_k\{\widehat p_{-k}(\bx_k)-p(\bx_k)\}}{p(\bx_k)\widehat p_{-k}(\bx_k)}.
\end{equation*}
Using $(a+b)^2\leq2a^2+2b^2$ and the propensity lower bounds gives, on $\mathcal A$,
\begin{equation*}
\|\widehat\eta_{\mathrm{cf}}-\eta\|_{S_v,2}^2
\leq\frac{2}{\rho^2n_v}\sum_{k\in S_v}\{\widehat m_{-k}(\bx_k)-m(\bx_k)\}^2
+\frac{2}{\delta^2\rho^2n_v}\sum_{k\in S_v}\epsilon_k^2\{\widehat p_{-k}(\bx_k)-p(\bx_k)\}^2.
\end{equation*}
For $f\in\{m,p\}$,
\begin{equation*}
\E_{mpq}\left[\frac{1}{n_v}\sum_{k\in S_v}\{\widehat f_{-k}(\bx_k)-f(\bx_k)\}^2\right]
=\sum_{t=1}^L\frac{N_t}{n_v}
\mathcal R_2(\widehat f_{-t})=o(1).
\end{equation*}
Markov's inequality therefore yields
\begin{equation}
\label{eq:sample_prediction_errors}
\frac{1}{n_v}\sum_{k\in S_v}\{\widehat f_{-k}(\bx_k)-f(\bx_k)\}^2=o_{\P}(1),
\qquad f\in\{m,p\}.
\end{equation}
Also,
\begin{samepage}
\begin{equation}
\label{eq:residual_propensity_error}
\begin{aligned}
&\frac{1}{n_v}\sum_{k\in S_v}\epsilon_k^2\{\widehat p_{-k}(\bx_k)-p(\bx_k)\}^2\\
&\quad\leq\frac{1}{n_v}\sum_{k\in S_v}\epsilon_k^2|\widehat p_{-k}(\bx_k)-p(\bx_k)|
\qquad\text{(Since $|\widehat p_{-k}(\bx_k)-p(\bx_k)|\leq1$.)}\\
&\quad\leq\left(\frac{1}{n_v}\sum_{k\in S_v}\epsilon_k^4\right)^{1/2}
\left(\frac{1}{n_v}\sum_{k\in S_v}\{\widehat p_{-k}(\bx_k)-p(\bx_k)\}^2\right)^{1/2}
\qquad\text{(By Cauchy--Schwarz.)}\\
&\quad=o_{\P}(1),
\end{aligned}
\end{equation}
by Lemma~\ref{lem:oracle_moment_bounds} and \eqref{eq:sample_prediction_errors}.
\end{samepage}
Consequently, $\|\widehat\eta_{\mathrm{cf}}-\eta\|_{S_v,2}=o_{\P}(1)$.
Equation~\eqref{eq:cf_v1_difference_bound} now gives $A_{1,v}=o_{\P}(1)$, and hence
\begin{equation}
\label{eq:v1_final_consistency}
n_v\left[\widehat V_{1,\mathrm{cf}}
-V_1(\widehat\mu_{\mathrm{aipw}}(m,p))\right]
=A_{1,v}+A_{2,v}=o_{\P}(1).
\end{equation}

\Needspace{6\baselineskip}
\paragraph{Consistency of the IPW variance estimator.}
Assume $\mathcal R_2(\widehat m_{-t})=o(1)$ for every fold $t$. Write
\begin{align*}
n_v\left[\widehat V_{2,\mathrm{ipw,cf}}
-V_2(\widehat\mu_{\mathrm{aipw}}(m,p))\right]
&=n_v\left[\widehat V_{2,\mathrm{ipw,cf}}
-\widehat V_{2,\mathrm{ipw}}(m,p)\right]\\
&\quad+n_v\left[\widehat V_{2,\mathrm{ipw}}(m,p)
-V_2(\widehat\mu_{\mathrm{aipw}}(m,p))\right].
\end{align*}
By Theorem~\ref{theo:oracle_variance}, it remains to prove
\begin{equation*}
n_v|\widehat V_{2,\mathrm{ipw,cf}}-\widehat V_{2,\mathrm{ipw}}(m,p)|=o_{\P}(1).
\end{equation*}
Adding and subtracting the term with fitted weight and oracle residual gives
\begin{equation}
\label{eq:cf_ipw_difference_decomposition}
\begin{aligned}
n_v\{\widehat V_{2,\mathrm{ipw,cf}}-\widehat V_{2,\mathrm{ipw}}(m,p)\}
&=\frac{n_v}{N_v^2}\sum_{k\in S_v}w_kr_k
\left\{\frac{1-\widehat p_{-k}(\bx_k)}{\widehat p_{-k}(\bx_k)^2}\widehat e_{k,\mathrm{cf}}^2
-\frac{1-p(\bx_k)}{p(\bx_k)^2}\epsilon_k^2\right\}\\
&=\frac{n_v}{N_v^2}\sum_{k\in S_v}w_kr_k
\frac{1-\widehat p_{-k}(\bx_k)}{\widehat p_{-k}(\bx_k)^2}
(\widehat e_{k,\mathrm{cf}}^2-\epsilon_k^2)\\
&\quad+\frac{n_v}{N_v^2}\sum_{k\in S_v}w_kr_k
\left\{\frac{1-\widehat p_{-k}(\bx_k)}{\widehat p_{-k}(\bx_k)^2}
-\frac{1-p(\bx_k)}{p(\bx_k)^2}\right\}\epsilon_k^2.
\end{aligned}
\end{equation}

\Needspace{6\baselineskip}
For the first sum, expanding the squared residual gives
\begin{equation*}
\widehat e_{k,\mathrm{cf}}^2-\epsilon_k^2
=-2\epsilon_k\{\widehat m_{-k}(\bx_k)-m(\bx_k)\}
+\{\widehat m_{-k}(\bx_k)-m(\bx_k)\}^2,
\end{equation*}
and thus
\begin{equation*}
|\widehat e_{k,\mathrm{cf}}^2-\epsilon_k^2|
\leq2|\epsilon_k||\widehat m_{-k}(\bx_k)-m(\bx_k)|
+\{\widehat m_{-k}(\bx_k)-m(\bx_k)\}^2.
\end{equation*}
For the second sum, the propensity bounds give, on $\mathcal A$,
\begin{align*}
\left|\frac{1-\widehat p_{-k}(\bx_k)}{\widehat p_{-k}(\bx_k)^2}
-\frac{1-p(\bx_k)}{p(\bx_k)^2}\right|
&=\frac{|\widehat p_{-k}(\bx_k)-p(\bx_k)|
\left[1-\{1-p(\bx_k)\}(1-\widehat p_{-k}(\bx_k))\right]}
{p(\bx_k)^2\widehat p_{-k}(\bx_k)^2}\\
&\leq\frac{|\widehat p_{-k}(\bx_k)-p(\bx_k)|}{\delta^2\rho^2}.
\end{align*}
On $\mathcal A$, we also have
\begin{equation*}
0\leq\frac{1-\widehat p_{-k}(\bx_k)}{\widehat p_{-k}(\bx_k)^2}\leq\frac{1}{\rho^2},
\qquad w_k=\frac{1}{\pi_k}\leq\frac{1}{\lambda_1},
\qquad k\in S_v.
\end{equation*}
Substituting these bounds into \eqref{eq:cf_ipw_difference_decomposition}
and taking absolute values gives
\begin{equation}
\label{eq:cf_ipw_difference_bound}
\begin{aligned}
n_v|\widehat V_{2,\mathrm{ipw,cf}}-\widehat V_{2,\mathrm{ipw}}(m,p)|
&\leq\frac{C}{n_v}\sum_{k\in S_v}
\left[2|\epsilon_k||\widehat m_{-k}(\bx_k)-m(\bx_k)|
+\{\widehat m_{-k}(\bx_k)-m(\bx_k)\}^2\right]\\
&\quad+\frac{C}{n_v}\sum_{k\in S_v}
\epsilon_k^2|\widehat p_{-k}(\bx_k)-p(\bx_k)|.
\end{aligned}
\end{equation}
Equations~\eqref{eq:sample_prediction_errors} and
\eqref{eq:residual_propensity_error} give
\begin{equation*}
\frac{1}{n_v}\sum_{k\in S_v}\{\widehat m_{-k}(\bx_k)-m(\bx_k)\}^2=o_{\P}(1),
\qquad
\frac{1}{n_v}\sum_{k\in S_v}\epsilon_k^2|\widehat p_{-k}(\bx_k)-p(\bx_k)|=o_{\P}(1).
\end{equation*}
Also,
\begin{samepage}
\begin{align*}
&\frac{1}{n_v}\sum_{k\in S_v}|\epsilon_k||\widehat m_{-k}(\bx_k)-m(\bx_k)|\\
&\quad\leq\left(\frac{1}{n_v}\sum_{k\in S_v}\epsilon_k^2\right)^{1/2}
\left(\frac{1}{n_v}\sum_{k\in S_v}\{\widehat m_{-k}(\bx_k)-m(\bx_k)\}^2\right)^{1/2}
\qquad\text{(By Cauchy--Schwarz.)}\\
&\quad=o_{\P}(1),
\end{align*}
by Lemma~\ref{lem:oracle_moment_bounds} and \eqref{eq:sample_prediction_errors}. 
\end{samepage}
Substituting into \eqref{eq:cf_ipw_difference_bound} gives
\begin{equation*}
n_v|\widehat V_{2,\mathrm{ipw,cf}}-\widehat V_{2,\mathrm{ipw}}(m,p)|
=o_{\P}(1).
\end{equation*}

\Needspace{5\baselineskip}
\medskip
\noindent\textit{Proof of (ii): consistency of the plug-in estimator.}
Only the propensity and variance risk
conditions are needed. Its oracle counterpart is
\begin{equation*}
\widehat V_{2,\mathrm{plug}}(p,\sigma^2)
=\frac{1}{N_v^2}\sum_{k\in S_v}w_k
\frac{1-p(\bx_k)}{p(\bx_k)}\sigma_k^2.
\end{equation*}
We decompose the error as
\begin{align*}
n_v\left[\widehat V_{2,\mathrm{plug,cf}}
-V_2(\widehat\mu_{\mathrm{aipw}}(m,p))\right]
&=n_v\left[\widehat V_{2,\mathrm{plug,cf}}
-\widehat V_{2,\mathrm{plug}}(p,\sigma^2)\right]\\
&\quad+n_v\left[\widehat V_{2,\mathrm{plug}}(p,\sigma^2)
-V_2(\widehat\mu_{\mathrm{aipw}}(m,p))\right].
\end{align*}
For the oracle term, write
\begin{equation*}
N_v\{\widehat V_{2,\mathrm{plug}}(p,\sigma^2)
-V_2(\widehat\mu_{\mathrm{aipw}}(m,p))\}
=\frac{1}{N_v}\sum_{k\in U_v}
\left(\frac{I_k}{\pi_k}-1\right)
\frac{1-p(\bx_k)}{p(\bx_k)}\sigma_k^2.
\end{equation*}
\begin{samepage}
This is a Horvitz--Thompson mean error, and the population values satisfy
\begin{equation*}
\frac{1}{N_v}\sum_{k\in U_v}
\left\{\frac{1-p(\bx_k)}{p(\bx_k)}\sigma_k^2\right\}^2
\leq\frac{1}{\delta^2N_v}\sum_{k\in U_v}(\sigma_k^2)^2
=\mathcal O(1),
\end{equation*}
by Lemma~\ref{lem:oracle_moment_bounds}.
\end{samepage}
Lemma~\ref{lem:ht_mean_bound} therefore gives
\begin{equation*}
\E_p\left[\left\{n_v\left[\widehat V_{2,\mathrm{plug}}(p,\sigma^2)
-V_2(\widehat\mu_{\mathrm{aipw}}(m,p))\right]\right\}^2\right]
=\mathcal O(n_v/N_v^2)=o(1).
\end{equation*}
For the fitted-estimator difference, write
\begin{equation*}
\begin{aligned}
n_v\{\widehat V_{2,\mathrm{plug,cf}}-\widehat V_{2,\mathrm{plug}}(p,\sigma^2)\}
&=\frac{n_v}{N_v^2}\sum_{k\in S_v}w_k
\frac{1-\widehat p_{-k}(\bx_k)}{\widehat p_{-k}(\bx_k)}
(\widehat\sigma_k^2-\sigma_k^2)\\
&\quad+\frac{n_v}{N_v^2}\sum_{k\in S_v}w_k
\frac{p(\bx_k)-\widehat p_{-k}(\bx_k)}{p(\bx_k)\widehat p_{-k}(\bx_k)}\sigma_k^2.
\end{aligned}
\end{equation*}
On $\mathcal A$, the two coefficients satisfy
\begin{equation*}
0\leq\frac{1-\widehat p_{-k}(\bx_k)}{\widehat p_{-k}(\bx_k)}\leq\frac{1}{\rho},
\qquad
\left|\frac{p(\bx_k)-\widehat p_{-k}(\bx_k)}{p(\bx_k)\widehat p_{-k}(\bx_k)}\right|
\leq\frac{|\widehat p_{-k}(\bx_k)-p(\bx_k)|}{\delta\rho},
\qquad k\in S_v.
\end{equation*}
Substituting these bounds into the preceding decomposition
and taking absolute values gives
\begin{equation}
\label{eq:cf_plugin_difference_bound}
\begin{aligned}
n_v|\widehat V_{2,\mathrm{plug,cf}}-\widehat V_{2,\mathrm{plug}}(p,\sigma^2)|
&\leq\frac{C}{n_v}\sum_{k\in S_v}
|\widehat\sigma_k^2-\sigma_k^2|\\
&\quad+\frac{C}{n_v}\sum_{k\in S_v}
|\widehat p_{-k}(\bx_k)-p(\bx_k)|\sigma_k^2.
\end{aligned}
\end{equation}
The variance risk assumption gives
\begin{equation*}
\E_{mpq}\left[\frac{1}{n_v}\sum_{k\in S_v}
|\widehat\sigma_k^2-\sigma_k^2|\right]
=\sum_{t=1}^L\frac{N_t}{n_v}
\mathcal R_1(\widehat\sigma^2_{-t})=o(1).
\end{equation*}
Markov's inequality therefore gives
\begin{equation*}
\frac{1}{n_v}\sum_{k\in S_v}|\widehat\sigma_k^2-\sigma_k^2|=o_{\P}(1).
\end{equation*}

\Needspace{5\baselineskip}
The propensity bound in \eqref{eq:sample_prediction_errors} uses no
condition on the outcome risk, so it also applies here. Thus
\begin{samepage}
\begin{align*}
&\frac{1}{n_v}\sum_{k\in S_v}|\widehat p_{-k}(\bx_k)-p(\bx_k)|\sigma_k^2\\
&\quad\leq\left(\frac{1}{n_v}\sum_{k\in S_v}\{\widehat p_{-k}(\bx_k)-p(\bx_k)\}^2\right)^{1/2}
\left(\frac{1}{n_v}\sum_{k\in S_v}(\sigma_k^2)^2\right)^{1/2}
\qquad\text{(By Cauchy--Schwarz.)}\\
&\quad=o_{\P}(1),
\end{align*}
by Lemma~\ref{lem:oracle_moment_bounds} and \eqref{eq:sample_prediction_errors}.
\end{samepage}
Substituting into \eqref{eq:cf_plugin_difference_bound} gives
\begin{equation*}
n_v|\widehat V_{2,\mathrm{plug,cf}}-\widehat V_{2,\mathrm{plug}}(p,\sigma^2)|
=o_{\P}(1),
\end{equation*}
which proves the plug-in conclusion.

\medskip
\noindent\textit{Proof of (iii): consistency of the bias-corrected estimator.}
Write
\begin{equation*}
n_v\left[\widehat V_{2,\mathrm{bc,cf}}
-V_2(\widehat\mu_{\mathrm{aipw}}(m,p))\right]
=n_v\left[\widehat V_{2,\mathrm{ipw,cf}}
-V_2(\widehat\mu_{\mathrm{aipw}}(m,p))\right]
+\sum_{t=1}^L(G_t+H_t),
\end{equation*}
where
\begin{equation*}
G_t=\frac{n_v}{N_v^2}\sum_{k\in\mathcal F_t}
w_k\left\{1-\frac{r_k}{p(\bx_k)}\right\}\widehat\sigma_k^2,
\qquad
H_t=\frac{n_v}{N_v^2}\sum_{k\in\mathcal F_t}
\frac{w_kr_k\{\widehat p_{-k}(\bx_k)-p(\bx_k)\}\widehat\sigma_k^2}{p(\bx_k)\widehat p_{-k}(\bx_k)}.
\end{equation*}
The IPW error is $o_{\P}(1)$ by part~(i), so it remains to prove
$G_t=o_{\P}(1)$ and $H_t=o_{\P}(1)$ for each fold.

\Needspace{7\baselineskip}
\medskip
\noindent\textit{Convergence of $G_t$.}
Since the variance predictions are $\mathcal D_{-t}$-measurable,
expanding the conditional second moment gives
\begin{equation*}
\begin{aligned}
\E_{mpq}[G_t^2\mid\mathcal D_{-t}]
&=\frac{n_v^2}{N_v^4}\sum_{k,\ell\in\mathcal F_t}
\frac{\widehat\sigma_k^2\widehat\sigma_\ell^2}{\pi_k\pi_\ell}
\E_q\left[\left(1-\frac{r_k}{p(\bx_k)}\right)
\left(1-\frac{r_\ell}{p(\bx_\ell)}\right)
\mid\mathcal D_{-t}\right].
\end{aligned}
\end{equation*}
The conditional response moments in
Lemma~\ref{lem:conditional_moments} give
\begin{equation*}
\E_q\left[\left(1-\frac{r_k}{p(\bx_k)}\right)
\left(1-\frac{r_\ell}{p(\bx_\ell)}\right)
\mid\mathcal D_{-t}\right]
=\frac{1-p(\bx_k)}{p(\bx_k)}\mathds{1}_{\{k=\ell\}}.
\end{equation*}
Substituting into the expansion of $\E_{mpq}[G_t^2\mid\mathcal D_{-t}]$ gives
\begin{equation*}
\E_{mpq}[G_t^2\mid\mathcal D_{-t}]
=\frac{n_v^2}{N_v^4}\sum_{k\in\mathcal F_t}
\frac{1-p(\bx_k)}{\pi_k^2p(\bx_k)}(\widehat\sigma_k^2)^2.
\end{equation*}
Taking expectations gives
\begin{equation}
\label{eq:cf_correction_second_moment_bound}
\E_{mpq}[G_t^2]
\leq\frac{n_v^2}{\lambda_1^2\delta N_v^4}
\E_{mpq}\left[\sum_{k\in\mathcal F_t}(\widehat\sigma_k^2)^2\right].
\end{equation}
Since $(a+b)^2\leq2a^2+2b^2$,
\begin{samepage}
\begin{equation}
\label{eq:variance_prediction_moment}
\begin{aligned}
\E_{mpq}\left[\frac{1}{N_t}
\sum_{k\in\mathcal F_t}(\widehat\sigma_k^2)^2\right]
&\leq2\E_{mpq}\left[\frac{1}{N_t}\sum_{k\in\mathcal F_t}
(\widehat\sigma_k^2-\sigma_k^2)^2\right]
+2\E_{mpq}\left[\frac{1}{N_t}\sum_{k\in\mathcal F_t}(\sigma_k^2)^2\right]\\
&\leq2\mathcal R_2(\widehat\sigma^2_{-t})
+\frac{2}{N_t}\sum_{k\in U_v}(\sigma_k^2)^2\\
&=\mathcal O(1),
\end{aligned}
\end{equation}
by Lemma~\ref{lem:oracle_moment_bounds} and the risk assumption.
\end{samepage}
Substituting \eqref{eq:variance_prediction_moment} into
\eqref{eq:cf_correction_second_moment_bound} gives
$\E_{mpq}[G_t^2]=\mathcal O(n_v^3/N_v^4)=o(1)$, so
Chebyshev's inequality yields $G_t=o_{\P}(1)$.

\Needspace{4\baselineskip}
\medskip
\noindent\textit{Convergence of $H_t$.}
The weight and propensity bounds on $\mathcal A$ give
\begin{samepage}
\begin{align*}
&\E_{mpq}[|H_t|\mathds{1}_{\mathcal A}]\\
&\quad\leq\frac{n_v}{\lambda_1\delta\rho N_v^2}
\E_{mpq}\left[\sum_{k\in\mathcal F_t}
|\widehat p_{-k}(\bx_k)-p(\bx_k)||\widehat\sigma_k^2|\right]\\
&\quad\leq\frac{n_vN_t}{\lambda_1\delta\rho N_v^2}
\E_{mpq}\left[
\left(\frac{1}{N_t}\sum_{k\in\mathcal F_t}\{\widehat p_{-k}(\bx_k)-p(\bx_k)\}^2\right)^{1/2}
\left(\frac{1}{N_t}\sum_{k\in\mathcal F_t}(\widehat\sigma_k^2)^2\right)^{1/2}\right]\\
&\quad\leq\frac{n_vN_t}{\lambda_1\delta\rho N_v^2}
\left\{\mathcal R_2(\widehat p_{-t})
\E_{mpq}\left[\frac{1}{N_t}\sum_{k\in\mathcal F_t}(\widehat\sigma_k^2)^2\right]\right\}^{1/2}\\
&\quad=o(1),
\end{align*}
by \eqref{eq:variance_prediction_moment} and $\mathcal R_2(\widehat p_{-t})=o(1)$.
\end{samepage}
The second and third inequalities use Cauchy--Schwarz in the sum and in
expectation, respectively. Markov's inequality and $\P(\mathcal A^c)=o(1)$ give
$H_t=o_{\P}(1)$.
Thus $G_t+H_t=o_{\P}(1)$ for each fold. Summing over the folds proves
the bias-corrected conclusion.

\paragraph{Total variance.}
When $\mathcal R_2(\widehat m_{-t})=o(1)$ for every fold $t$,
\eqref{eq:v1_final_consistency}
and the result for the chosen second component give
\begin{align*}
n_v\left[\widehat V_{\mathrm{tot,cf}}
-V_{\mathrm{tot}}(\widehat\mu_{\mathrm{aipw}}(m,p))\right]
&=n_v\left[\widehat V_{1,\mathrm{cf}}
-V_1(\widehat\mu_{\mathrm{aipw}}(m,p))\right]\\
&\quad+n_v\left[\widehat V_{2,\mathrm{cf}}
-V_2(\widehat\mu_{\mathrm{aipw}}(m,p))\right]
=o_{\P}(1).
\end{align*}
Under the additional variance-limit condition,
\begin{equation*}
\frac{\widehat V_{\mathrm{tot,cf}}}
{V_{\mathrm{tot}}(\widehat\mu_{\mathrm{aipw}}(m,p))}-1
=\frac{n_v[\widehat V_{\mathrm{tot,cf}}
-V_{\mathrm{tot}}(\widehat\mu_{\mathrm{aipw}}(m,p))]}
{n_vV_{\mathrm{tot}}(\widehat\mu_{\mathrm{aipw}}(m,p))}
=o_{\P}(1).\qedhere
\end{equation*}
\end{proof}

\subsection{Proof of Corollary~\ref{cor:cf_clt}}
\label{proof:cf_clt}

\begin{proof}
On the event where the estimated variance is positive, write
\begin{align*}
&\frac{\widehat\mu_{\mathrm{aipw,cf}}(\widehat m,\widehat p)-\mu}
{\sqrt{\widehat V_{\mathrm{tot,cf}}}}\\
&\quad=\left\{
\frac{\widehat\mu_{\mathrm{aipw}}(m,p)-\mu}
{\sqrt{V_{\mathrm{tot}}(\widehat\mu_{\mathrm{aipw}}(m,p))}}
+\frac{\widehat\mu_{\mathrm{aipw,cf}}(\widehat m,\widehat p)
-\widehat\mu_{\mathrm{aipw}}(m,p)}
{\sqrt{V_{\mathrm{tot}}(\widehat\mu_{\mathrm{aipw}}(m,p))}}
\right\}
\left\{\frac{V_{\mathrm{tot}}(\widehat\mu_{\mathrm{aipw}}(m,p))}
{\widehat V_{\mathrm{tot,cf}}}\right\}^{1/2}.
\end{align*}

The oracle term has a standard normal limit by Theorem~\ref{theo1}, and
\begin{samepage}
\begin{equation*}
\frac{\widehat V_{\mathrm{tot,cf}}}
{V_{\mathrm{tot}}(\widehat\mu_{\mathrm{aipw}}(m,p))}
\xrightarrow[v\to\infty]{\P}1,
\end{equation*}
by Theorem~\ref{Theo3}.
\end{samepage}
For the point-estimator difference,
\begin{equation*}
\frac{\widehat\mu_{\mathrm{aipw,cf}}(\widehat m,\widehat p)
-\widehat\mu_{\mathrm{aipw}}(m,p)}
{\sqrt{V_{\mathrm{tot}}(\widehat\mu_{\mathrm{aipw}}(m,p))}}
=\frac{\sqrt{n_v}\{\widehat\mu_{\mathrm{aipw,cf}}(\widehat m,\widehat p)
-\widehat\mu_{\mathrm{aipw}}(m,p)\}}
{\sqrt{n_vV_{\mathrm{tot}}(\widehat\mu_{\mathrm{aipw}}(m,p))}}.
\end{equation*}
The numerator is $o_{\P}(1)$ by the oracle equivalence in
Theorem~\ref{Theo2}(\ref{thm:cf_oracle_equivalence}). For the denominator,
\ref{D1}, \ref{CL1}, and \eqref{eq:oracle_response_variance_limit} give
\begin{equation*}
n_vV_{\mathrm{tot}}(\widehat\mu_{\mathrm{aipw}}(m,p))
\xrightarrow[v\to\infty]{}\pi_*(c_1+c_2)>0.
\end{equation*}
The displayed ratio is therefore $o_{\P}(1)$, and Slutsky's theorem gives
\begin{equation*}
\frac{\widehat\mu_{\mathrm{aipw,cf}}(\widehat m,\widehat p)-\mu}
{\sqrt{\widehat V_{\mathrm{tot,cf}}}}
=\left\{\frac{\widehat\mu_{\mathrm{aipw}}(m,p)-\mu}
{\sqrt{V_{\mathrm{tot}}(\widehat\mu_{\mathrm{aipw}}(m,p))}}
+o_{\P}(1)\right\}\{1+o_{\P}(1)\}
\xrightarrow[v\to\infty]{\cL}\mathcal N(0,1).
\end{equation*}
\end{proof}

\section{Proofs for Section~\ref{sec:aggregation}}
\label{app:proofs_aggregation}

\subsection{Proof of Proposition~\ref{prop:aggregation_rates}}
\label{proof:aggregation_rates}

\begin{proof}[Proof of (\ref{prop:aggregation_consistency})]
Fix a fold $t$ and let $\widehat m_{h,-t}$ be a candidate with vanishing
risk. Comparing the aggregate with this candidate in \ref{AG1} gives
\begin{equation*}
\mathcal R_2(\widehat m_{\mathrm{agg},-t})
\leq C\left\{\mathcal R_2(\widehat m_{h,-t})+n_v^{-1}\right\}=o(1).
\end{equation*}
If a propensity candidate $\widehat p_{j,-t}$ has vanishing risk, the
second bound in \ref{AG1} gives
\begin{equation*}
\mathcal R_2(\widehat p_{\mathrm{agg},-t})
\leq C\left\{\mathcal R_2(\widehat p_{j,-t})+n_v^{-1}\right\}=o(1).
\end{equation*}
If either candidate risk is only bounded, the corresponding right-hand
side is $\mathcal O(1)$, proving the bounded-risk assertion.

\medskip
\noindent\textit{Proof of (\ref{prop:aggregation_product_rate}).}
Fix a fold $t$ and take a candidate pair
$(\widehat m_{h,-t},\widehat p_{j,-t})$ satisfying
\eqref{eq:aggregation_candidate_rates}.
Both aggregate risks tend to zero by part~(\ref{prop:aggregation_consistency}).
By \ref{AG1},
\begin{align*}
n_v\mathcal R_2(\widehat m_{\mathrm{agg},-t})
\mathcal R_2(\widehat p_{\mathrm{agg},-t})
&\leq C^2n_v
\{\mathcal R_2(\widehat m_{h,-t})+n_v^{-1}\}
\{\mathcal R_2(\widehat p_{j,-t})+n_v^{-1}\}\\
&= C^2\Big\{
 n_v\mathcal R_2(\widehat m_{h,-t})\mathcal R_2(\widehat p_{j,-t})\\
&\qquad+\mathcal R_2(\widehat m_{h,-t})
+\mathcal R_2(\widehat p_{j,-t})+n_v^{-1}\Big\}
\xrightarrow[v\to\infty]{}0.
\end{align*}
The argument applies to each fold and proves the product-rate conclusion.
\end{proof}

\Needspace{12\baselineskip}
\section{Auxiliary lemmas}
\label{sec:auxiliary_lemmas}
\label{sec:appendix_design_lemmas}

\setcounter{lemma}{0}
\renewcommand{\thelemma}{\thesection.\arabic{lemma}}
\renewcommand{\theHlemma}{\thesection.\arabic{lemma}}

We collect the auxiliary results used in the proofs.

\begin{lemma}[Conditional moments under ignorability]
\label{lem:conditional_moments}
Assume \ref{A1}--\ref{A3}, and suppose that the outcomes have finite conditional second moments. Given $(\mathbf X,\mathbf I,\boldsymbol{\delta})$, all $y_k$ and $r_k$ are mutually independent, with their original conditional laws given $\mathbf X$. For the training information defined in Section~\ref{subsec:crossfitting} and $k,\ell\in\mathcal F_t$,
$$
\begin{aligned}
\E_m[\epsilon_k\mid\mathcal D_{-t}]&=0,
&\E_m[\epsilon_k\epsilon_\ell\mid\mathcal D_{-t}]&=\sigma_k^2\mathds{1}_{\{k=\ell\}},\\
\E_q[r_k\mid\mathcal D_{-t}]&=p(\bx_k),
&\E_q[(r_k-p(\bx_k))(r_\ell-p(\bx_\ell))\mid\mathcal D_{-t}]&=p(\bx_k)(1-p(\bx_k))\mathds{1}_{\{k=\ell\}}.
\end{aligned}
$$
\end{lemma}

\begin{proof}
By \ref{A2}, the pairs $(y_k,r_k)$ are independent conditionally on
$\mathbf X$, and the conditional law of each pair depends only on
$\bx_k$. Within each pair, \ref{A3} makes $y_k$ and $r_k$ conditionally
independent. Thus
\begin{align*}
\cL(\mathbf y,\mathbf r\mid\mathbf X)
&=\bigotimes_{k\in U_v}\cL(y_k,r_k\mid\bx_k)\\
&=\bigotimes_{k\in U_v}\{\cL(y_k\mid\bx_k)\otimes\cL(r_k\mid\bx_k)\}\\
&=\bigotimes_{k\in U_v}\{\cL(y_k\mid\bx_k)\otimes\operatorname{Bernoulli}(p(\bx_k))\}.
\end{align*}
The fold assignment is independent of $(\mathbf y,\mathbf r)$ given
$(\mathbf X,\mathbf I)$ by construction, so
\begin{align*}
\cL(\mathbf y,\mathbf r\mid\mathbf X,\mathbf I,\boldsymbol{\delta})
&=\cL(\mathbf y,\mathbf r\mid\mathbf X,\mathbf I)\\
&=\cL(\mathbf y,\mathbf r\mid\mathbf X),
\end{align*}
where the last equality follows from \ref{A1}. This proves the first assertion.\\

The training information $\mathcal D_{-t}$ adds only outcomes and responses
from $S_{-t}$ to $(\mathbf X,\mathbf I,\boldsymbol{\delta})$. Since
$S_{-t}\cap\mathcal F_t=\varnothing$, the product law above gives
\begin{align*}
\cL((y_k)_{k\in\mathcal F_t}\mid\mathcal D_{-t},\mathbf r)
&=\bigotimes_{k\in\mathcal F_t}\cL(y_k\mid\bx_k),\\
\cL((r_k)_{k\in\mathcal F_t}\mid\mathcal D_{-t},\mathbf y)
&=\bigotimes_{k\in\mathcal F_t}\operatorname{Bernoulli}(p(\bx_k)).
\end{align*}
These are the conditional laws used by $\E_m$ and $\E_q$, respectively.
In particular, for $k\in\mathcal F_t$,
\begin{equation*}
\E_m[\epsilon_k\mid\mathcal D_{-t}]
=m(\bx_k)-m(\bx_k)=0,
\qquad
\E_q[r_k\mid\mathcal D_{-t}]=p(\bx_k).
\end{equation*}
For distinct $k,\ell\in\mathcal F_t$, independence gives
\begin{align*}
\E_m[\epsilon_k\epsilon_\ell\mid\mathcal D_{-t}]
&=\E_m[\epsilon_k\mid\mathcal D_{-t}]
\E_m[\epsilon_\ell\mid\mathcal D_{-t}]=0,\\
\E_q[(r_k-p(\bx_k))(r_\ell-p(\bx_\ell))\mid\mathcal D_{-t}]
&=\E_q[r_k-p(\bx_k)\mid\mathcal D_{-t}]
\E_q[r_\ell-p(\bx_\ell)\mid\mathcal D_{-t}]=0.
\end{align*}
For $k=\ell$, the outcome variance is unchanged and $r_k^2=r_k$, so
\begin{align*}
\E_m[\epsilon_k^2\mid\mathcal D_{-t}]&=\sigma_k^2,\\
\E_q[(r_k-p(\bx_k))^2\mid\mathcal D_{-t}]
&=p(\bx_k)-2p(\bx_k)^2+p(\bx_k)^2\\
&=p(\bx_k)\{1-p(\bx_k)\}.
\end{align*}
Combining the cases $k\ne\ell$ and $k=\ell$ gives the stated second moments.
\end{proof}

\begin{lemma}[Fourth moments of the oracle quantities]
\label{lem:oracle_moment_bounds}
Assume \ref{A2}--\ref{A4} and \ref{M2}. Then
\begin{equation*}
\frac{1}{N_v}\sum_{k\in U_v}
\{\E_m[\epsilon_k^4]+\E_{mq}[\eta_k^4]+(\sigma_k^2)^2\}
=\mathcal O(1).
\end{equation*}
Moreover,
\begin{equation*}
\frac{1}{N_v}\sum_{k\in U_v}
\{\epsilon_k^4+\eta_k^4+(\sigma_k^2)^2\}
=\mathcal O_{\P}(1).
\end{equation*}
If \ref{D1} also holds, then
\begin{equation*}
\frac{1}{n_v}\sum_{k\in S_v}
\{\epsilon_k^4+\eta_k^4+(\sigma_k^2)^2\}
=\mathcal O_{\P}(1).
\end{equation*}
\end{lemma}

\begin{proof}
\noindent\textit{Population moments.}
\begin{samepage}
For the residual,
\begin{align*}
\E_m[\epsilon_k^4]
&=\E_m[\{y_k-m(\bx_k)\}^4]\\
&\leq8\E_m[y_k^4]+8m^4(\bx_k)
\qquad\text{(By $(a+b)^4\leq8(a^4+b^4)$.)}\\
&=8\E_m[y_k^4]+8\{\E_m[y_k]\}^4\\
&\leq16\E_m[y_k^4],
\end{align*}
by Jensen's inequality.
\end{samepage}
For the pseudo-value,
\begin{align*}
\E_{mq}[\eta_k^4]
&=\E_{mq}\left[\left\{m(\bx_k)+\frac{r_k\epsilon_k}{p(\bx_k)}\right\}^4\right]\\
&\leq8m^4(\bx_k)+\frac{8}{p(\bx_k)^4}\E_{mq}[r_k^4\epsilon_k^4]
\qquad\text{(By $(a+b)^4\leq8(a^4+b^4)$.)}\\
&=8m^4(\bx_k)+\frac{8}{p(\bx_k)^3}\E_m[\epsilon_k^4]
\qquad\text{(Since $r_k^4=r_k$ and $r_k$ and $\epsilon_k$ are independent.)}\\
&\leq8(1+16\delta^{-3})\E_m[y_k^4].
\end{align*}
Finally,
\begin{samepage}
\begin{equation*}
(\sigma_k^2)^2\leq\{\E_m[y_k^2]\}^2\leq\E_m[y_k^4],
\end{equation*}
by Jensen's inequality.
\end{samepage}
By the strong law, the population average of $\E_m[y_k^4]$ converges
almost surely to $\E_P[y_1^4]<\infty$, so these bounds prove the expectation
bound. Markov's inequality then gives the bound for the realized population
average.

\medskip
\noindent\textit{Sample moments.}
The sample sum is bounded by the population sum, so
\begin{align*}
\E_{mpq}\left[\frac{1}{n_v}\sum_{k\in S_v}
\{\epsilon_k^4+\eta_k^4+(\sigma_k^2)^2\}\right]
&\leq\frac{1}{n_v}\sum_{k\in U_v}
\{\E_m[\epsilon_k^4]+\E_{mq}[\eta_k^4]+(\sigma_k^2)^2\}\\
&=\mathcal O(N_v/n_v)=\mathcal O(1).
\end{align*}
Markov's inequality gives the sample bound.
\end{proof}

\begin{lemma}[Horvitz--Thompson error bound]
\label{tlemma1}
\label{lem:ht_mean_bound}
Let $\boldsymbol\xi=(\xi_k:k\in U_v)$ be independent of $\mathbf I$
conditionally on $\mathbf X$. Under \ref{D2},
\begin{equation*}
\E_p\left[\left\{\frac{1}{N_v}\sum_{k\in U_v}
\left(\frac{I_k}{\pi_k}-1\right)\xi_k\right\}^2
\mid\boldsymbol\xi\right]
\leq\left\{\frac{1-\lambda_1}{\lambda_1N_v}
+\frac{\max_{(k,\ell)\in D_2}|\Delta_{k\ell}|}{\lambda_1^2}\right\}
\frac{1}{N_v}\sum_{k\in U_v}\xi_k^2.
\end{equation*}
If \ref{D3} also holds and
$N_v^{-1}\sum_{k\in U_v}\E_{mpq}[\xi_k^2]=\mathcal O(1)$, then
\begin{equation*}
\E_{mpq}\left[\left\{\frac{1}{N_v}\sum_{k\in U_v}
\left(\frac{I_k}{\pi_k}-1\right)\xi_k\right\}^2\right]
=\mathcal O(n_v^{-1}).
\end{equation*}
\end{lemma}

\Needspace{6\baselineskip}
\begin{proof}
Since $\boldsymbol\xi$ and $\mathbf I$ are independent conditionally on
$\mathbf X$, design unbiasedness gives
\begin{equation*}
\begin{aligned}
\E_p\left[\left\{\frac{1}{N_v}\sum_{k\in U_v}
\left(\frac{I_k}{\pi_k}-1\right)\xi_k\right\}^2
\mid\boldsymbol\xi\right]
&=\frac{1}{N_v^2}\sum_{k\in U_v}\frac{1-\pi_k}{\pi_k}\xi_k^2
+\frac{1}{N_v^2}\sum_{(k,\ell)\in D_2}
\frac{\Delta_{k\ell}}{\pi_k\pi_\ell}\xi_k\xi_\ell.
\end{aligned}
\end{equation*}
Since $2|\xi_k\xi_\ell|\leq\xi_k^2+\xi_\ell^2$,
\begin{equation*}
\sum_{(k,\ell)\in D_2}|\xi_k\xi_\ell|
\leq\frac{1}{2}\sum_{(k,\ell)\in D_2}(\xi_k^2+\xi_\ell^2)
=(N_v-1)\sum_{k\in U_v}\xi_k^2.
\end{equation*}
Substituting into the second-moment expansion and bounding
the inclusion probabilities gives
\begin{align*}
&\E_p\left[\left\{\frac{1}{N_v}\sum_{k\in U_v}
\left(\frac{I_k}{\pi_k}-1\right)\xi_k\right\}^2
\mid\boldsymbol\xi\right]\\
&\quad\leq\left\{\frac{1-\lambda_1}{\lambda_1N_v^2}
+\frac{(N_v-1)\max_{(k,\ell)\in D_2}|\Delta_{k\ell}|}
{\lambda_1^2N_v^2}\right\}\sum_{k\in U_v}\xi_k^2\\
&\quad\leq\left\{\frac{1-\lambda_1}{\lambda_1N_v}
+\frac{\max_{(k,\ell)\in D_2}|\Delta_{k\ell}|}{\lambda_1^2}\right\}
\frac{1}{N_v}\sum_{k\in U_v}\xi_k^2.
\end{align*}
This proves the conditional bound. Taking expectations and using
$\max_{(k,\ell)\in D_2}|\Delta_{k\ell}|=\mathcal O(n_v^{-1})$ gives the second
conclusion.
\end{proof}

\begin{lemma}[Consistency of the Horvitz--Thompson variance estimator]
\label{lemma4}
\label{lem:ht_variance_consistency}

Let $\boldsymbol\xi=(\xi_k:k\in U_v)$ be independent of $\mathbf I$
conditionally on $\mathbf X$. Define
\begin{equation*}
V_\pi=\frac{1}{N_v^2}\sum_{k\in U_v}\sum_{\ell\in U_v}
\Delta_{k\ell}\frac{\xi_k}{\pi_k}\frac{\xi_\ell}{\pi_\ell},
\qquad
\widehat V_\pi=\frac{1}{N_v^2}\sum_{k\in S_v}\sum_{\ell\in S_v}
\frac{\Delta_{k\ell}}{\pi_{k\ell}}\frac{\xi_k}{\pi_k}\frac{\xi_\ell}{\pi_\ell}.
\end{equation*}
Assume \ref{D1}--\ref{D4} and
$N_v^{-1}\sum_{k\in U_v}\xi_k^4=\mathcal{O}_{\P}(1)$. Then
\begin{equation*}
n_v(\widehat V_\pi-V_\pi)=o_{\P}(1).
\end{equation*}
If, in addition, $N_vV_\pi\xrightarrow[v\to\infty]{\P}c>0$, then
$\widehat V_\pi/V_\pi\xrightarrow[v\to\infty]{\P}1$.
\end{lemma}

\begin{proof}
For $\varepsilon>0$, conditioning on $\boldsymbol\xi$ and applying
Chebyshev's inequality gives
\begin{equation}
\label{eq:ht_variance_probability_bound}
\begin{aligned}
\P(|n_v(\widehat V_\pi-V_\pi)|>\varepsilon)
&=\E_{mpq}\left[\P_p(|n_v(\widehat V_\pi-V_\pi)|>\varepsilon
\mid\boldsymbol\xi)\right]\\
&\leq\E_{mpq}\left[\min\left\{1,
\frac{\E_p[\{n_v(\widehat V_\pi-V_\pi)\}^2\mid\boldsymbol\xi]}
{\varepsilon^2}\right\}\right].
\end{aligned}
\end{equation}
It therefore suffices to prove
\begin{equation*}
\E_p\left[\{n_v(\widehat V_\pi-V_\pi)\}^2\mid\boldsymbol\xi\right]
=o_{\P}(1).
\end{equation*}
Since $\boldsymbol\xi$ and $\mathbf I$ are independent conditionally on
$\mathbf X$, the second-moment calculation in the first part of the proof of
Theorem~3 of \citet[pp.~1038--1039]{breidt2000local} applies with $\xi_k$
in place of their population residuals. Keeping the population fourth
moment explicit, \ref{D1}--\ref{D4} give
\begin{equation*}
\E_p\left[\{n_v(\widehat V_\pi-V_\pi)\}^2\mid\boldsymbol\xi\right]
\leq o(1)\frac{1}{N_v}\sum_{k\in U_v}\xi_k^4
=o_{\P}(1),
\end{equation*}
where the $o(1)$ factor depends only on the sampling design.
The expectation in \eqref{eq:ht_variance_probability_bound} tends to zero because its
integrand is bounded by one and tends to zero in probability. Thus
$n_v(\widehat V_\pi-V_\pi)=o_{\P}(1)$. Finally,
$n_vV_\pi\xrightarrow[v\to\infty]{\P}\pi_*c>0$ gives
\begin{equation*}
\frac{\widehat V_\pi}{V_\pi}-1
=\frac{n_v(\widehat V_\pi-V_\pi)}{n_vV_\pi}
=o_{\P}(1).\qedhere
\end{equation*}
\end{proof}

\begin{lemma}[A bilinear bound for the design variance form]
\label{lem:bilinear_design_bound}
For arrays $a$ and $b$ on $S_v$, the form $\mathcal Q$ defined in
Appendix~\ref{app:notation} satisfies, under \ref{D2},
\begin{equation*}
n_v|\mathcal Q(a,b)|
\leq\frac{n_v^2}{N_v^2}
\left\{\frac{1-\lambda_1}{\lambda_1^2}
+\frac{n_v\max_{(k,\ell)\in D_2}|\Delta_{k\ell}|}{\lambda_1^2\lambda_2}\right\}
\|a\|_{S_v,2}\|b\|_{S_v,2}.
\end{equation*}
The coefficient is bounded under \ref{D3}.
\end{lemma}
\pagebreak[0]

\Needspace{7\baselineskip}
\begin{proof}
Since $\pi_{kk}=\pi_k$,
\begin{equation*}
\begin{aligned}
n_v|\mathcal Q(a,b)|
&\leq\frac{n_v}{N_v^2}\sum_{k\in S_v}
\frac{1-\pi_k}{\pi_k^2}|a_kb_k|
+\frac{n_v}{N_v^2}\sum_{\substack{k,\ell\in S_v\\k\ne\ell}}
\frac{|\Delta_{k\ell}|}{\pi_{k\ell}\pi_k\pi_\ell}|a_kb_\ell|\\
&\leq\frac{n_v(1-\lambda_1)}{N_v^2\lambda_1^2}
\sum_{k\in S_v}|a_kb_k|
+\frac{n_v\max_{(k,\ell)\in D_2}|\Delta_{k\ell}|}{N_v^2\lambda_1^2\lambda_2}
\left(\sum_{k\in S_v}|a_k|\right)
\left(\sum_{\ell\in S_v}|b_\ell|\right).
\end{aligned}
\end{equation*}
By Cauchy--Schwarz,
\begin{align*}
\sum_{k\in S_v}|a_kb_k|
&\leq\left(\sum_{k\in S_v}a_k^2\right)^{1/2}
\left(\sum_{k\in S_v}b_k^2\right)^{1/2}
=n_v\|a\|_{S_v,2}\|b\|_{S_v,2},
\end{align*}
and 
\begin{align*}
	\left(\sum_{k\in S_v}|a_k|\right)\left(\sum_{\ell\in S_v}|b_\ell|\right)
&\leq\left(n_v\sum_{k\in S_v}a_k^2\right)^{1/2}
\left(n_v\sum_{\ell\in S_v}b_\ell^2\right)^{1/2}
=n_v^2\|a\|_{S_v,2}\|b\|_{S_v,2}.
\end{align*}
Substituting these two inequalities into the bound for $n_v|\mathcal Q(a,b)|$
proves the result.
\end{proof}

\Needspace{8\baselineskip}
\section{Additional methodological details}
\label{app:methodological_details}

\subsection{Variance identities for fixed working functions}
\label{app:fixed_variance_identities}

We give the identities underlying the variance correction discussed in Section~\ref{sec:aipw2}. Assume \ref{A1}--\ref{A4} and \ref{M1}. For fixed working functions $\widetilde m,\widetilde p$ with $0<\widetilde p(\bx_k)\leq1$, let $\widehat V_{2,\mathrm{bc}}(\widetilde m,\widetilde p)$ denote \eqref{V2_OR_BC_hat} evaluated at these functions with the true $\sigma_k^2$.
We verify each identity by averaging the contribution of one unit, then
using design weighting to recover the population sums.

With the true outcome regression, the squared residual has model
expectation $\sigma_k^2$. Since its coefficient is
$\sigma(\mathbf X,\mathbf r)$-measurable, the residual term and correction
give the following coefficient of $\sigma_k^2$:
\begin{equation*}
r_k\frac{1-\widetilde p(\bx_k)}{\{\widetilde p(\bx_k)\}^2}
+1-\frac{r_k}{\widetilde p(\bx_k)}
=1-\frac{2r_k}{\widetilde p(\bx_k)}
+\frac{r_k}{\{\widetilde p(\bx_k)\}^2}
=\left\{\frac{r_k}{\widetilde p(\bx_k)}-1\right\}^2,
\end{equation*}
where we used $r_k^2=r_k$. Design unbiasedness therefore gives
\begin{equation*}
\E_{pm}\left[\widehat V_{2,\mathrm{bc}}(m,\widetilde p)\right]
=\frac{1}{N^2}\sum_{k\in U}
\left\{\frac{r_k}{\widetilde p(\bx_k)}-1\right\}^2\sigma_k^2.
\end{equation*}

With the true propensity, the squared residual is
$\sigma(\mathbf X,\mathbf y)$-measurable and $\E_q[r_k]=p(\bx_k)$, so
taking response expectations gives
\begin{align*}
\E_q\left[r_k\frac{1-p(\bx_k)}{\{p(\bx_k)\}^2}
\{y_k-\widetilde m(\bx_k)\}^2\right]
&=\frac{1-p(\bx_k)}{p(\bx_k)}\{y_k-\widetilde m(\bx_k)\}^2,\\
\E_q\left[\left\{1-\frac{r_k}{p(\bx_k)}\right\}\sigma_k^2\right]&=0.
\end{align*}
Consequently,
\begin{equation*}
\E_{pq}\left[\widehat V_{2,\mathrm{bc}}(\widetilde m,p)\right]
=\frac{1}{N^2}\sum_{k\in U}
\frac{1-p(\bx_k)}{p(\bx_k)}\{y_k-\widetilde m(\bx_k)\}^2.
\end{equation*}
The two identities concern different targets and do not imply
unbiasedness for \eqref{V2_or} when only one working model is correct.

\Needspace{8\baselineskip}
\subsection{Estimating the conditional residual variance}
\label{app:variance_learning}

The plug-in and bias-corrected variance estimators in
Section~\ref{sec:aipw_variance_cf} require predictions of $\sigma^2(\bx_k)$.
We describe two ways to construct them. For each outer fold $\mathcal F_t$,
the entire procedure, including tuning and any inner validation, uses only
the training sample $S_{-t}$. The resulting function
$\widehat\sigma^2_{-t}$ is then evaluated on $\mathcal F_t$.

\subsubsection{Regression on squared out-of-sample residuals}

A partition of $S_{-t}$ into inner folds provides out-of-sample residuals.
For each inner fold, an outcome regression fitted on respondents in the
remaining folds predicts the outcomes of its respondents. Subtracting
these predictions from the observed outcomes gives residuals
$\widehat e^{\mathrm{in}}_{\ell,-t}$ for $\ell\in S_{r,-t}$.
Thus, each residual uses an outcome prediction fitted without that respondent.

A regression of $(\widehat e^{\mathrm{in}}_{\ell,-t})^2$ on $\bx_\ell$,
constrained to give nonnegative predictions, estimates the conditional
residual variance. Survey weights may be included in the fit.
The prediction at $\bx_k$ is denoted by
$\widehat\sigma^2_{-t}(\bx_k)$ for $k\in\mathcal F_t$.
Squared out-of-sample residuals contain outcome prediction error as well
as residual variation, so the accuracy of the outcome fits remains relevant
to this construction.

\subsubsection{Log-scale fitting and smearing}
\enlargethispage{\baselineskip}

An alternative is to fit the squared residuals on the log scale. Define
\begin{equation*}
Z_\ell=\log\{(\widehat e^{\mathrm{in}}_{\ell,-t})^2+\tau\},
\qquad \ell\in S_{r,-t},
\end{equation*}
where $\tau>0$ prevents taking the logarithm of zero. A regression of
$Z_\ell$ on $\bx_\ell$ gives $\widehat g_{-t}$. Exponentiating a fitted log-mean
does not generally recover the mean on the original scale; the smearing
idea of \cite{duan1983smearing} supplies a retransformation correction.

To identify the required correction, consider a generic outcome error
$\epsilon=Y-m(\bx)$ and the oracle quantity
$Z=\log(\epsilon^2+\tau)$. Define
\begin{equation*}
g(\bx)=\E_P[Z\mid\bx],
\qquad
c(\bx)=\E_P[\exp\{Z-g(\bx)\}\mid\bx].
\end{equation*}
Since $\exp(Z)=\epsilon^2+\tau$,
\begin{equation*}
\exp\{g(\bx)\}c(\bx)
=\E_P[\exp(Z)\mid\bx]
=\sigma^2(\bx)+\tau.
\end{equation*}
The conditional residual variance can therefore be written as
\begin{equation*}
\sigma^2(\bx)=\exp\{g(\bx)\}c(\bx)-\tau.
\end{equation*}
This identity shows that the smearing factor may depend on the covariates.
A single factor is justified when $c(\bx)$ is constant, for example when
the conditional distribution of $Z-g(\bx)$ does not depend on $\bx$.
Constant conditional variance of $Z-g(\bx)$ alone does not ensure this.

If a constant smearing factor is appropriate, it may be estimated by
\begin{equation*}
\widehat c_{-t}
=\frac{\sum_{\ell\in S_{r,-t}}\omega_\ell \exp\{Z_\ell-\widehat g_{-t}(\bx_\ell)\}}
{\sum_{\ell\in S_{r,-t}}\omega_\ell},
\end{equation*}
where $\omega_\ell$ are positive fitting weights, such as the survey
weights. Otherwise, a regression of $\exp\{Z_\ell-\widehat g_{-t}(\bx_\ell)\}$ on $\bx_\ell$ estimates
$c(\bx)$, again with nonnegative predictions. We denote either fit by
$\widehat c_{-t}(\bx)$.
The conditional variance prediction is then given by
\begin{equation*}
\widehat\sigma^2_{-t}(\bx_k)
=\max\left[\underline\sigma^2,
\widehat c_{-t}(\bx_k)\exp\{\widehat g_{-t}(\bx_k)\}-\tau\right],
\qquad k\in\mathcal F_t.
\end{equation*}
The lower bound may be set to $\underline\sigma^2=0$. For consistent
variance predictions, a positive bound should tend to zero unless the
true conditional variance is known to exceed it. Inner validation can
help limit overfitting in the log-scale and smearing fits; all such
steps must remain within $S_{-t}$.

For the plug-in estimator, either construction must satisfy
$\mathcal R_1(\widehat\sigma^2_{-t})=o(1)$ in Theorem~\ref{Theo3}.
For the bias-corrected estimator, the moment bound
$\mathcal R_2(\widehat\sigma^2_{-t})=\mathcal O(1)$ suffices; the variance
predictions need not be consistent. Nonnegative predictions and retransformation
alone establish neither condition. For example, a single smearing factor
when $c(\bx)$ varies can leave a persistent prediction error. Clipping
the final predictions to a fixed bounded interval ensures the moment
condition for the bias-corrected estimator under \ref{M2}, but can prevent
the consistency required by the plug-in estimator.

\clearpage
\subsection{Illustration of the aggregation procedure}
\label{app:aggregation_diagram}

\begin{figure}[H]
\centering
\begingroup
\setstretch{1}
\begin{tikzpicture}[
  font=\small,
  line width=0.55pt,
  text=black!85,
  box/.style={draw=black!35, rounded corners=2pt, align=center,
    inner xsep=8pt, inner ysep=6pt},
  training/.style={box, draw=blue!35!black!65, fill=blue!5},
  validation/.style={box, draw=blue!35!black!65, fill=white},
  weights/.style={box, draw=teal!50!black, fill=teal!5},
  heldout/.style={box, draw=orange!55!black, fill=orange!10},
  flow/.style={-{Latex[length=2mm]}, draw=black!55, line width=0.7pt},
  fold/.style={minimum width=4.05cm, minimum height=0.58cm,
    inner ysep=3pt},
  heading/.style={anchor=west, font=\small\bfseries}
]

\node[heading] at (0,0) {Outer cross-fitting: three runs};
\node[anchor=west] at (0,-0.72) {$t=1$};
\node[anchor=west] at (0,-1.43) {$t=2$};
\node[anchor=west] at (0,-2.14) {$t=3$};

\node[heldout,fold] at (3.55,-0.72) {Predict on $\mathcal F_1$};
\node[training,fold] at (7.95,-0.72) {Training: $\mathcal F_2$};
\node[training,fold] at (12.35,-0.72) {Training: $\mathcal F_3$};
\node[training,fold] at (3.55,-1.43) {Training: $\mathcal F_1$};
\node[heldout,fold] at (7.95,-1.43) {Predict on $\mathcal F_2$};
\node[training,fold] at (12.35,-1.43) {Training: $\mathcal F_3$};
\node[training,fold] at (3.55,-2.14) {Training: $\mathcal F_1$};
\node[training,fold] at (7.95,-2.14) {Training: $\mathcal F_2$};
\node[heldout,fold] at (12.35,-2.14) {Predict on $\mathcal F_3$};

\draw[black!20] (0,-2.83) -- (14.5,-2.83);
\node[heading] at (0,-3.28) {Within the first run: two inner folds};

\node[training,minimum width=10cm,minimum height=0.75cm]
  (outer) at (7.25,-4.1)
  {Outer training sample $S_{-1}=\mathcal F_2\cup\mathcal F_3$};

\node[training,minimum width=10cm,minimum height=0.75cm]
  (inner) at (7.25,-5.65) {};
\draw[blue!35!black!65] (7.25,-5.275) -- (7.25,-6.025);
\node at (4.75,-5.65) {Inner fold $\mathcal F_{1,1}$};
\node at (9.75,-5.65) {Inner fold $\mathcal F_{1,2}$};
\draw[flow] (outer.south) -- node[right,fill=white,inner sep=4pt]
  {New random partition} (inner.north);

\node[training,text width=5.5cm,minimum height=0.95cm]
  (fit1) at (3.4,-7.2)
  {Candidates fitted on $\mathcal F_{1,1}$\\including tuning};
\node[training,text width=5.5cm,minimum height=0.95cm]
  (fit2) at (11.1,-7.2)
  {Candidates fitted on $\mathcal F_{1,2}$\\including tuning};
\draw[flow] (inner.south) -- (7.25,-6.35) -| (fit1.north);
\draw[flow] (7.25,-6.35) -| (fit2.north);

\node[validation,text width=5.5cm,minimum height=0.95cm]
  (pred1) at (3.4,-8.8)
  {Validation predictions\\on $\mathcal F_{1,2}$};
\node[validation,text width=5.5cm,minimum height=0.95cm]
  (pred2) at (11.1,-8.8)
  {Validation predictions\\on $\mathcal F_{1,1}$};
\draw[flow] (fit1.south) -- (pred1.north);
\draw[flow] (fit2.south) -- (pred2.north);

\node[weights,text width=11.3cm,minimum height=1.25cm]
  (pool) at (7.25,-10.65)
  {\textbf{Aggregation weights}\\
   All validation predictions are pooled to estimate\\
   one outcome weight vector and one propensity weight vector.};
\draw[draw=black!55,line width=0.7pt]
  (pred1.south) -- (3.4,-9.65) -- (11.1,-9.65) -- (pred2.south);
\draw[flow] (7.25,-9.65) -- (pool.north);

\node[training,text width=11.3cm,minimum height=1.05cm]
  (refit) at (7.25,-12.55)
  {Candidates refitted on all of $S_{-1}$\\
   Aggregation weights remain fixed.};
\draw[flow] (pool.south) -- (refit.north);

\node[heldout,text width=11.3cm,minimum height=1.05cm]
  (output) at (7.25,-14.35)
  {\textbf{Predictions on $\mathcal F_1$}\\
   Weighted combination of the refitted candidates};
\draw[flow] (refit.south) -- (output.north);
\end{tikzpicture}
\endgroup
\captionsetup{format=plain,font=small,labelfont=bf,labelsep=colon,
  justification=justified,singlelinecheck=false}
\caption{Aggregation with three outer folds and two inner folds.
The first run is shown in detail; the others use their own training
samples and inner partitions. Outcome fitting and weight estimation use
respondents, whereas propensity fitting and weight estimation use all
units in the corresponding samples.}
\label{fig:aggregation_splitting}
\end{figure}
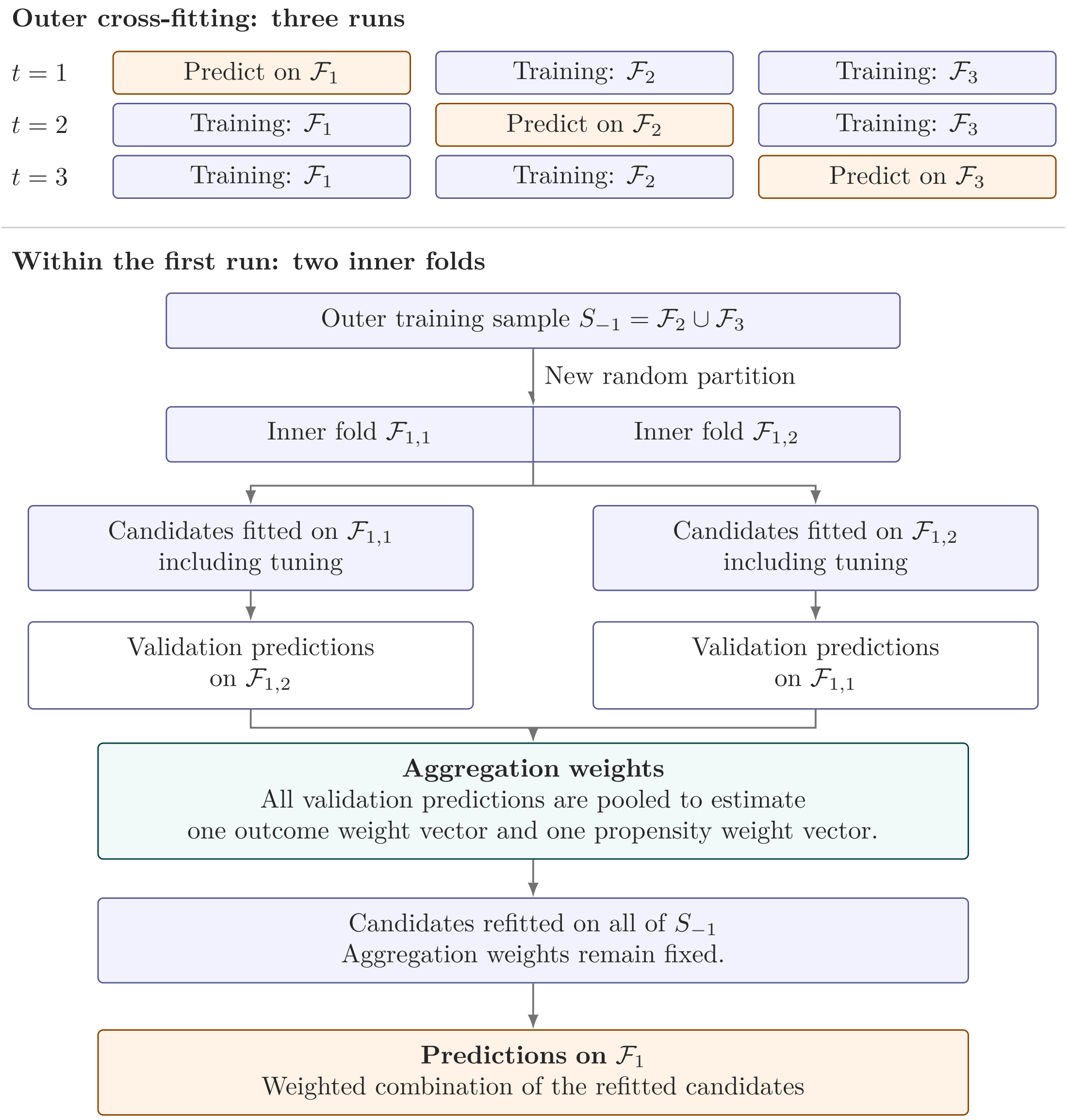

\clearpage
\section{Simulation details and additional results}
\label{app:simulation_results}
\begingroup
\setlength{\parskip}{0.25\baselineskip}

The complete results cover all four sample sizes. Performance tables
compare five-fold cross-fitting (CF) with no cross-fitting (No CF);
the lower panels give the oracle AIPW and full-response
Horvitz--Thompson (HT) benchmarks. RE is relative to HT. The evaluation
criteria are defined in Section~\ref{sec:simulation_criteria}.

\subsection{Variance and learner specifications}
\label{app:simulation_specifications}

For each outcome model, the conditional error variance was
\begin{equation*}
\sigma_k^2=c\frac{h(\bx_k)}{\overline h_U},
\qquad
\overline h_U=\frac{1}{N}\sum_{k\in U}h(\bx_k),
\end{equation*}
where
\begin{equation*}
h(\bx_k)=0.75+0.50x_{k1}+0.25\sin^2(2\pi x_{k2})
+0.25\mathds{1}_{\{x_{k3}>0.5\}}.
\end{equation*}
The constant $c$ was chosen separately for each outcome model so that
the population signal-to-noise ratio corresponded approximately to $R^2=0.50$.

Table~\ref{tab:simulation_tuning} gives the tuning specifications for the
learners described in Section~\ref{sec:simulation_estimators}.

\begin{table}[H]
\centering
\captionsetup{format=plain,font=small,labelfont=bf,labelsep=colon,
  justification=justified,singlelinecheck=false}
\caption{Learner specifications used in the simulation study.}
\label{tab:simulation_tuning}
\begingroup
\setstretch{1}
\small
\renewcommand{\arraystretch}{1.15}
\begin{tabular}{@{}p{0.20\textwidth}p{0.75\textwidth}@{}}
\toprule
Learner & Specification \\
\midrule
GAM & Smoothing parameters selected by restricted maximum likelihood. \\
Elastic net & Mixing parameter $\alpha=0.5$; regularization parameter
selected by five-fold cross-validation, for both linear and logistic
regression. \\
Forests & $500$ trees and sampling fraction $0.80$ per tree. The
regression forest used minimum terminal-node size $10$ and maximum
depth $8$; the probability forest used size $25$ and depth $6$. \\
XGBoost & Maximum depth $2$, learning rate $0.05$, minimum child
weight $10$, and subsampling and column-subsampling fractions $0.80$.
Up to $600$ iterations, selected by three-fold cross-validation and
early stopping, for both nuisance functions. \\
\bottomrule
\end{tabular}
\endgroup
\end{table}

\subsection{Prediction-error summaries}
\label{app:simulation_diagnostics}

The upper panels of the prediction-error tables report Monte Carlo
averages of the RMSEs defined in Section~\ref{sec:simulation_criteria},
their product, and the scaled product
\begin{equation*}
D_n^{(b)}=\sqrt n\operatorname{RMSE}^{(b)}(\widehat m)
\operatorname{RMSE}^{(b)}(\widehat p).
\end{equation*}

The RMSEs use predictions averaged over outer fits and evaluated on
nonsampled units. In the lower panels, $\widehat\alpha_m$ and
$\widehat\alpha_p$ are the negative log--log slopes of the RMSEs against
$n$; the last column gives the slope of the scaled product. These
regressions describe trends across the four sample sizes and do not
verify the prediction conditions in \eqref{eq:cf_product_rate}.

\endgroup

\newcolumntype{d}[1]{D{.}{.}{#1}}
\definecolor{simPanelGray}{gray}{0.94}

\begingroup
\captionsetup[table]{format=plain,justification=justified,singlelinecheck=false,font=small,labelfont=bf,labelsep=colon}
\singlespacing
\setlength{\intextsep}{8pt}

\clearpage
\subsection{Linear outcome, linear propensity}

\begin{table}[H]
    \centering
    \caption{Finite-sample performance with linear outcome and linear propensity.}\label{tab:simulation_performance_ee}
    \small
    \setlength{\tabcolsep}{3.5pt}
    \renewcommand{\arraystretch}{1.00}
    \begin{tabular}{@{}l r *{8}{d{1}}@{}}
        \toprule
        Learner pair & $n$ & \multicolumn{2}{c}{Point RB (\%)} & \multicolumn{2}{c}{RE (\%)} & \multicolumn{2}{c}{Variance RB (\%)} & \multicolumn{2}{c}{Coverage (\%)} \\
        \cmidrule(lr){3-4}\cmidrule(lr){5-6}\cmidrule(lr){7-8}\cmidrule(l){9-10}
        & & \multicolumn{1}{c}{CF} & \multicolumn{1}{c}{No CF} & \multicolumn{1}{c}{CF} & \multicolumn{1}{c}{No CF} & \multicolumn{1}{c}{CF} & \multicolumn{1}{c}{No CF} & \multicolumn{1}{c}{CF} & \multicolumn{1}{c}{No CF} \\
        \midrule
        Linear + Logistic & 500 & -0.0 & -0.0 & 162.7 & 158.5 & 4.0 & -2.9 & 95.5 & 94.5 \\
         & 1000 & -0.0 & -0.0 & 156.4 & 154.5 & 4.3 & 1.1 & 95.3 & 95.0 \\
         & 2000 & -0.0 & -0.0 & 152.9 & 152.3 & 1.9 & 0.2 & 95.3 & 94.9 \\
         & 4000 & -0.0 & -0.0 & 158.1 & 157.7 & -2.0 & -2.7 & 94.6 & 94.6 \\
        \addlinespace[2pt]
        Agg. + Agg. & 500 & 0.0 & -0.1 & 162.9 & 159.4 & 2.3 & -11.8 & 95.0 & 93.2 \\
         & 1000 & -0.0 & -0.1 & 157.5 & 155.4 & 2.8 & -4.4 & 95.2 & 94.3 \\
         & 2000 & -0.0 & -0.0 & 153.4 & 152.7 & 1.2 & -2.7 & 95.2 & 94.6 \\
         & 4000 & -0.0 & -0.0 & 158.1 & 157.9 & -2.2 & -4.2 & 94.6 & 94.4 \\
        \bottomrule
    \end{tabular}
    \par\vspace{6pt}
    \begin{tabular}{@{}r *{8}{d{1}}@{}}
        \rowcolor{simPanelGray}\multicolumn{9}{c}{\textit{Oracle AIPW and full-response Horvitz--Thompson benchmarks}}\\
        \addlinespace[2pt]
        $n$ & \multicolumn{2}{c}{Point RB (\%)} & \multicolumn{2}{c}{RE (\%)} & \multicolumn{2}{c}{Variance RB (\%)} & \multicolumn{2}{c}{Coverage (\%)} \\
        \cmidrule(lr){2-3}\cmidrule(lr){4-5}\cmidrule(lr){6-7}\cmidrule(l){8-9}
        & \multicolumn{1}{c}{Oracle} & \multicolumn{1}{c}{HT} & \multicolumn{1}{c}{Oracle} & \multicolumn{1}{c}{HT} & \multicolumn{1}{c}{Oracle} & \multicolumn{1}{c}{HT} & \multicolumn{1}{c}{Oracle} & \multicolumn{1}{c}{HT} \\
        \midrule
        500 & -0.0 & -0.0 & 155.1 & 100.0 & -1.0 & -0.5 & 94.8 & 95.0 \\
        1000 & -0.0 & 0.0 & 153.5 & 100.0 & 1.6 & 1.1 & 94.9 & 95.1 \\
        2000 & -0.0 & -0.0 & 151.6 & 100.0 & 0.6 & -1.1 & 95.1 & 94.8 \\
        4000 & -0.0 & -0.0 & 157.4 & 100.0 & -2.6 & -0.6 & 94.5 & 95.0 \\
        \bottomrule
    \end{tabular}
\end{table}

\begin{table}[H]
    \centering
    \caption{Prediction-error diagnostics with linear outcome and linear propensity.}\label{tab:simulation_rates_ee}
    \small
    \setlength{\tabcolsep}{6pt}
    \renewcommand{\arraystretch}{1.00}
    \begin{tabular}{@{}l c *{4}{d{3}}@{}}
        \toprule
        Learner pair & {$n$} & \multicolumn{1}{c}{\shortstack{Outcome\\RMSE}} & \multicolumn{1}{c}{\shortstack{Propensity\\RMSE}} & \multicolumn{1}{c}{Product} & \multicolumn{1}{c}{\shortstack{Scaled\\product}} \\
        \midrule
        Linear + Logistic & 500 & 0.164 & 0.072 & 0.012 & 0.263 \\
        & 1000 & 0.115 & 0.051 & 0.006 & 0.184 \\
        & 2000 & 0.080 & 0.036 & 0.003 & 0.127 \\
        & 4000 & 0.057 & 0.025 & 0.001 & 0.090 \\
        \addlinespace[2pt]
        Agg. + Agg. & 500 & 0.150 & 0.064 & 0.010 & 0.215 \\
        & 1000 & 0.105 & 0.045 & 0.005 & 0.150 \\
        & 2000 & 0.073 & 0.031 & 0.002 & 0.103 \\
        & 4000 & 0.052 & 0.022 & 0.001 & 0.072 \\
        \bottomrule
    \end{tabular}
    \par\vspace{6pt}
    \begin{tabular}{@{}l *{4}{d{2}}@{}}
        \rowcolor{simPanelGray}\multicolumn{5}{c}{\textit{Descriptive log--log fits}}\\
        \addlinespace[2pt]
     Learner pair
    & \multicolumn{1}{c}{$\widehat\alpha_m$}
    & \multicolumn{1}{c}{$\widehat\alpha_p$}
    & \multicolumn{1}{c}{$\widehat\alpha_m+\widehat\alpha_p$}
    & \multicolumn{1}{c}{\shortstack{Scaled product\\slope}}
    \\
        \midrule
        Linear + Logistic & 0.51 & 0.50 & 1.02 & -0.52 \\
        Agg. + Agg. & 0.51 & 0.51 & 1.02 & -0.52 \\
        \bottomrule
    \end{tabular}
\end{table}

\clearpage
\subsection{Linear outcome, nonlinear propensity}

\begin{table}[H]
    \centering
    \caption{Finite-sample performance with linear outcome and nonlinear propensity.}\label{tab:simulation_performance_eh}
    \small
    \setlength{\tabcolsep}{3.5pt}
    \renewcommand{\arraystretch}{1.00}
    \begin{tabular}{@{}l r *{8}{d{1}}@{}}
        \toprule
        Learner pair & $n$ & \multicolumn{2}{c}{Point RB (\%)} & \multicolumn{2}{c}{RE (\%)} & \multicolumn{2}{c}{Variance RB (\%)} & \multicolumn{2}{c}{Coverage (\%)} \\
        \cmidrule(lr){3-4}\cmidrule(lr){5-6}\cmidrule(lr){7-8}\cmidrule(l){9-10}
        & & \multicolumn{1}{c}{CF} & \multicolumn{1}{c}{No CF} & \multicolumn{1}{c}{CF} & \multicolumn{1}{c}{No CF} & \multicolumn{1}{c}{CF} & \multicolumn{1}{c}{No CF} & \multicolumn{1}{c}{CF} & \multicolumn{1}{c}{No CF} \\
        \midrule
        Linear + GAM & 500 & -0.0 & -0.0 & 204.9 & 190.8 & 2.8 & -5.4 & 95.1 & 94.2 \\
         & 1000 & 0.0 & 0.0 & 191.7 & 187.1 & 1.7 & -2.6 & 95.2 & 94.5 \\
         & 2000 & -0.0 & -0.0 & 194.3 & 191.8 & -3.0 & -4.9 & 94.7 & 94.6 \\
         & 4000 & 0.0 & 0.0 & 188.2 & 187.2 & 0.1 & -0.9 & 95.1 & 95.0 \\
        \addlinespace[2pt]
        Linear + Agg. & 500 & -0.0 & -0.0 & 203.2 & 188.1 & 2.7 & -10.0 & 95.1 & 93.5 \\
         & 1000 & 0.0 & 0.0 & 192.3 & 184.7 & 1.3 & -5.4 & 95.0 & 94.0 \\
         & 2000 & -0.0 & -0.0 & 194.3 & 189.5 & -3.1 & -7.6 & 94.7 & 94.2 \\
         & 4000 & 0.0 & 0.0 & 188.4 & 185.8 & 0.0 & -2.7 & 95.1 & 94.6 \\
        \addlinespace[2pt]
        Linear + XGBoost & 500 & 0.2 & -0.0 & 988.1 & 174.1 & 27.7 & -45.8 & 97.6 & 85.3 \\
         & 1000 & 0.0 & 0.0 & 462.4 & 176.5 & 26.2 & -33.3 & 97.3 & 88.6 \\
         & 2000 & -0.0 & -0.0 & 294.0 & 185.2 & 12.8 & -24.1 & 96.3 & 91.2 \\
         & 4000 & 0.0 & 0.0 & 224.4 & 183.7 & 9.7 & -12.7 & 95.9 & 93.2 \\
        \bottomrule
    \end{tabular}
    \par\vspace{6pt}
    \begin{tabular}{@{}r *{8}{d{1}}@{}}
        \rowcolor{simPanelGray}\multicolumn{9}{c}{\textit{Oracle AIPW and full-response Horvitz--Thompson benchmarks}}\\
        \addlinespace[2pt]
        $n$ & \multicolumn{2}{c}{Point RB (\%)} & \multicolumn{2}{c}{RE (\%)} & \multicolumn{2}{c}{Variance RB (\%)} & \multicolumn{2}{c}{Coverage (\%)} \\
        \cmidrule(lr){2-3}\cmidrule(lr){4-5}\cmidrule(lr){6-7}\cmidrule(l){8-9}
        & \multicolumn{1}{c}{Oracle} & \multicolumn{1}{c}{HT} & \multicolumn{1}{c}{Oracle} & \multicolumn{1}{c}{HT} & \multicolumn{1}{c}{Oracle} & \multicolumn{1}{c}{HT} & \multicolumn{1}{c}{Oracle} & \multicolumn{1}{c}{HT} \\
        \midrule
        500 & -0.0 & -0.0 & 184.3 & 100.0 & -2.3 & -2.6 & 94.5 & 94.8 \\
        1000 & 0.0 & 0.0 & 184.0 & 100.0 & -1.0 & -1.6 & 94.7 & 94.9 \\
        2000 & -0.0 & 0.0 & 189.8 & 100.0 & -3.8 & -1.4 & 94.7 & 94.8 \\
        4000 & 0.0 & 0.0 & 186.2 & 100.0 & -0.4 & 0.4 & 94.9 & 94.8 \\
        \bottomrule
    \end{tabular}
\end{table}

\begin{table}[H]
    \centering
    \caption{Prediction-error diagnostics with linear outcome and nonlinear propensity.}\label{tab:simulation_rates_eh}
    \small
    \setlength{\tabcolsep}{6pt}
    \renewcommand{\arraystretch}{1.00}
    \begin{tabular}{@{}l c *{4}{d{3}}@{}}
        \toprule
        Learner pair & {$n$} & \multicolumn{1}{c}{\shortstack{Outcome\\RMSE}} & \multicolumn{1}{c}{\shortstack{Propensity\\RMSE}} & \multicolumn{1}{c}{Product} & \multicolumn{1}{c}{\shortstack{Scaled\\product}} \\
        \midrule
        Linear + GAM & 500 & 0.167 & 0.053 & 0.009 & 0.198 \\
        & 1000 & 0.118 & 0.038 & 0.004 & 0.140 \\
        & 2000 & 0.082 & 0.026 & 0.002 & 0.097 \\
        & 4000 & 0.058 & 0.019 & 0.001 & 0.069 \\
        \addlinespace[2pt]
        Linear + Agg. & 500 & 0.167 & 0.055 & 0.009 & 0.206 \\
        & 1000 & 0.118 & 0.039 & 0.005 & 0.145 \\
        & 2000 & 0.082 & 0.028 & 0.002 & 0.101 \\
        & 4000 & 0.058 & 0.020 & 0.001 & 0.073 \\
        \addlinespace[2pt]
        Linear + XGBoost & 500 & 0.167 & 0.161 & 0.027 & 0.600 \\
        & 1000 & 0.118 & 0.124 & 0.015 & 0.461 \\
        & 2000 & 0.082 & 0.093 & 0.008 & 0.342 \\
        & 4000 & 0.058 & 0.068 & 0.004 & 0.252 \\
        \bottomrule
    \end{tabular}
    \par\vspace{6pt}
    \begin{tabular}{@{}l *{4}{d{2}}@{}}
        \rowcolor{simPanelGray}\multicolumn{5}{c}{\textit{Descriptive log--log fits}}\\
        \addlinespace[2pt]
     Learner pair
    & \multicolumn{1}{c}{$\widehat\alpha_m$}
    & \multicolumn{1}{c}{$\widehat\alpha_p$}
    & \multicolumn{1}{c}{$\widehat\alpha_m+\widehat\alpha_p$}
    & \multicolumn{1}{c}{\shortstack{Scaled product\\slope}}
    \\
        \midrule
        Linear + GAM & 0.51 & 0.50 & 1.01 & -0.51 \\
        Linear + Agg. & 0.51 & 0.49 & 1.00 & -0.50 \\
        Linear + XGBoost & 0.51 & 0.41 & 0.92 & -0.42 \\
        \bottomrule
    \end{tabular}
\end{table}

\clearpage
\subsection{Nonlinear outcome, linear propensity}

\begin{table}[H]
    \centering
    \caption{Finite-sample performance with nonlinear outcome and linear propensity.}\label{tab:simulation_performance_he}
    \small
    \setlength{\tabcolsep}{3.5pt}
    \renewcommand{\arraystretch}{1.00}
    \begin{tabular}{@{}l r *{8}{d{1}}@{}}
        \toprule
        Learner pair & $n$ & \multicolumn{2}{c}{Point RB (\%)} & \multicolumn{2}{c}{RE (\%)} & \multicolumn{2}{c}{Variance RB (\%)} & \multicolumn{2}{c}{Coverage (\%)} \\
        \cmidrule(lr){3-4}\cmidrule(lr){5-6}\cmidrule(lr){7-8}\cmidrule(l){9-10}
        & & \multicolumn{1}{c}{CF} & \multicolumn{1}{c}{No CF} & \multicolumn{1}{c}{CF} & \multicolumn{1}{c}{No CF} & \multicolumn{1}{c}{CF} & \multicolumn{1}{c}{No CF} & \multicolumn{1}{c}{CF} & \multicolumn{1}{c}{No CF} \\
        \midrule
        GAM + Logistic & 500 & -0.0 & 0.0 & 165.6 & 160.1 & 4.2 & -6.5 & 95.2 & 93.9 \\
         & 1000 & -0.0 & -0.0 & 159.6 & 156.9 & 1.8 & -3.3 & 95.1 & 94.5 \\
         & 2000 & -0.0 & -0.0 & 156.1 & 156.0 & 1.9 & -1.6 & 95.2 & 94.8 \\
         & 4000 & -0.0 & -0.0 & 156.4 & 156.1 & -0.4 & -2.0 & 94.8 & 94.5 \\
        \addlinespace[2pt]
        Agg. + Logistic & 500 & -0.0 & 0.0 & 166.4 & 160.2 & 4.1 & -7.7 & 95.3 & 93.7 \\
         & 1000 & -0.0 & -0.0 & 159.2 & 157.0 & 2.1 & -3.7 & 95.2 & 94.5 \\
         & 2000 & -0.0 & -0.0 & 156.7 & 156.1 & 1.5 & -2.0 & 95.2 & 94.8 \\
         & 4000 & -0.0 & -0.0 & 156.7 & 156.1 & -0.5 & -2.3 & 94.7 & 94.5 \\
        \addlinespace[2pt]
        XGBoost + Logistic & 500 & -0.0 & -0.1 & 187.1 & 176.2 & 3.5 & -44.6 & 95.2 & 85.2 \\
         & 1000 & -0.0 & -0.1 & 174.9 & 170.6 & 1.4 & -32.0 & 95.1 & 89.3 \\
         & 2000 & -0.0 & -0.0 & 165.9 & 163.5 & 1.6 & -19.5 & 94.9 & 91.9 \\
         & 4000 & -0.0 & -0.0 & 162.2 & 161.1 & -0.2 & -12.4 & 94.8 & 93.1 \\
        \bottomrule
    \end{tabular}
    \par\vspace{6pt}
    \begin{tabular}{@{}r *{8}{d{1}}@{}}
        \rowcolor{simPanelGray}\multicolumn{9}{c}{\textit{Oracle AIPW and full-response Horvitz--Thompson benchmarks}}\\
        \addlinespace[2pt]
        $n$ & \multicolumn{2}{c}{Point RB (\%)} & \multicolumn{2}{c}{RE (\%)} & \multicolumn{2}{c}{Variance RB (\%)} & \multicolumn{2}{c}{Coverage (\%)} \\
        \cmidrule(lr){2-3}\cmidrule(lr){4-5}\cmidrule(lr){6-7}\cmidrule(l){8-9}
        & \multicolumn{1}{c}{Oracle} & \multicolumn{1}{c}{HT} & \multicolumn{1}{c}{Oracle} & \multicolumn{1}{c}{HT} & \multicolumn{1}{c}{Oracle} & \multicolumn{1}{c}{HT} & \multicolumn{1}{c}{Oracle} & \multicolumn{1}{c}{HT} \\
        \midrule
        500 & 0.0 & 0.0 & 153.4 & 100.0 & -0.2 & -0.9 & 95.0 & 94.8 \\
        1000 & -0.0 & -0.0 & 154.3 & 100.0 & -0.3 & -0.3 & 95.0 & 94.9 \\
        2000 & -0.0 & -0.0 & 155.0 & 100.0 & -0.1 & 0.3 & 94.9 & 95.0 \\
        4000 & -0.0 & -0.0 & 155.4 & 100.0 & -1.1 & -0.4 & 94.7 & 94.9 \\
        \bottomrule
    \end{tabular}
\end{table}

\begin{table}[H]
    \centering
    \caption{Prediction-error diagnostics with nonlinear outcome and linear propensity.}\label{tab:simulation_rates_he}
    \small
    \setlength{\tabcolsep}{6pt}
    \renewcommand{\arraystretch}{1.00}
    \begin{tabular}{@{}l c *{4}{d{3}}@{}}
        \toprule
        Learner pair & {$n$} & \multicolumn{1}{c}{\shortstack{Outcome\\RMSE}} & \multicolumn{1}{c}{\shortstack{Propensity\\RMSE}} & \multicolumn{1}{c}{Product} & \multicolumn{1}{c}{\shortstack{Scaled\\product}} \\
        \midrule
        GAM + Logistic & 500 & 0.212 & 0.072 & 0.015 & 0.341 \\
        & 1000 & 0.152 & 0.050 & 0.008 & 0.243 \\
        & 2000 & 0.111 & 0.036 & 0.004 & 0.177 \\
        & 4000 & 0.083 & 0.025 & 0.002 & 0.132 \\
        \addlinespace[2pt]
        Agg. + Logistic & 500 & 0.216 & 0.072 & 0.016 & 0.348 \\
        & 1000 & 0.153 & 0.050 & 0.008 & 0.244 \\
        & 2000 & 0.112 & 0.036 & 0.004 & 0.178 \\
        & 4000 & 0.084 & 0.025 & 0.002 & 0.133 \\
        \addlinespace[2pt]
        XGBoost + Logistic & 500 & 0.420 & 0.072 & 0.030 & 0.675 \\
        & 1000 & 0.350 & 0.050 & 0.018 & 0.559 \\
        & 2000 & 0.282 & 0.036 & 0.010 & 0.449 \\
        & 4000 & 0.228 & 0.025 & 0.006 & 0.362 \\
        \bottomrule
    \end{tabular}
    \par\vspace{6pt}
    \begin{tabular}{@{}l *{4}{d{2}}@{}}
        \rowcolor{simPanelGray}\multicolumn{5}{c}{\textit{Descriptive log--log fits}}\\
        \addlinespace[2pt]
     Learner pair
    & \multicolumn{1}{c}{$\widehat\alpha_m$}
    & \multicolumn{1}{c}{$\widehat\alpha_p$}
    & \multicolumn{1}{c}{$\widehat\alpha_m+\widehat\alpha_p$}
    & \multicolumn{1}{c}{\shortstack{Scaled product\\slope}}
    \\
        \midrule
        GAM + Logistic & 0.45 & 0.51 & 0.96 & -0.46 \\
        Agg. + Logistic & 0.46 & 0.51 & 0.96 & -0.46 \\
        XGBoost + Logistic & 0.29 & 0.51 & 0.80 & -0.30\\
        \bottomrule
    \end{tabular}
\end{table}

\clearpage
\subsection{Nonlinear outcome, nonlinear propensity}

\begin{table}[H]
    \centering
    \caption{Finite-sample performance with nonlinear outcome and nonlinear propensity.}\label{tab:simulation_performance_hh}
    \small
    \setlength{\tabcolsep}{3.5pt}
    \renewcommand{\arraystretch}{1.00}
    \begin{tabular}{@{}l r *{8}{d{1}}@{}}
        \toprule
        Learner pair & $n$ & \multicolumn{2}{c}{Point RB (\%)} & \multicolumn{2}{c}{RE (\%)} & \multicolumn{2}{c}{Variance RB (\%)} & \multicolumn{2}{c}{Coverage (\%)} \\
        \cmidrule(lr){3-4}\cmidrule(lr){5-6}\cmidrule(lr){7-8}\cmidrule(l){9-10}
        & & \multicolumn{1}{c}{CF} & \multicolumn{1}{c}{No CF} & \multicolumn{1}{c}{CF} & \multicolumn{1}{c}{No CF} & \multicolumn{1}{c}{CF} & \multicolumn{1}{c}{No CF} & \multicolumn{1}{c}{CF} & \multicolumn{1}{c}{No CF} \\
        \midrule
        GAM + GAM & 500 & -0.1 & 0.1 & 219.4 & 198.9 & 8.1 & -10.8 & 95.5 & 93.3 \\
         & 1000 & -0.0 & 0.0 & 193.5 & 187.5 & 2.9 & -7.0 & 94.8 & 93.8 \\
         & 2000 & -0.0 & 0.0 & 190.6 & 187.8 & 0.8 & -4.5 & 95.1 & 94.6 \\
         & 4000 & 0.0 & 0.0 & 193.0 & 192.0 & -0.9 & -3.9 & 94.6 & 94.2 \\
        \addlinespace[2pt]
        Agg. + Agg. & 500 & -0.1 & 0.2 & 219.0 & 198.9 & 6.8 & -16.6 & 95.4 & 92.3 \\
         & 1000 & -0.0 & 0.1 & 192.1 & 187.3 & 3.6 & -11.0 & 95.0 & 93.3 \\
         & 2000 & -0.0 & 0.0 & 190.2 & 187.8 & 0.9 & -8.6 & 95.2 & 94.0 \\
         & 4000 & 0.0 & 0.0 & 192.7 & 192.0 & -0.8 & -6.8 & 94.6 & 93.8 \\
        \addlinespace[2pt]
        XGBoost + XGBoost & 500 & -2.5 & 0.9 & 1695.4 & 245.0 & 33.1 & -60.5 & 97.4 & 75.1 \\
         & 1000 & -0.7 & 0.4 & 605.5 & 223.4 & 36.1 & -53.6 & 97.3 & 80.0 \\
         & 2000 & -0.2 & 0.2 & 323.4 & 214.1 & 24.0 & -41.5 & 97.1 & 85.8 \\
         & 4000 & -0.0 & 0.2 & 240.1 & 211.7 & 14.5 & -29.2 & 96.2 & 89.1 \\
        \bottomrule
    \end{tabular}
    \par\vspace{6pt}
    \begin{tabular}{@{}r *{8}{d{1}}@{}}
        \rowcolor{simPanelGray}\multicolumn{9}{c}{\textit{Oracle AIPW and full-response Horvitz--Thompson benchmarks}}\\
        \addlinespace[2pt]
        $n$ & \multicolumn{2}{c}{Point RB (\%)} & \multicolumn{2}{c}{RE (\%)} & \multicolumn{2}{c}{Variance RB (\%)} & \multicolumn{2}{c}{Coverage (\%)} \\
        \cmidrule(lr){2-3}\cmidrule(lr){4-5}\cmidrule(lr){6-7}\cmidrule(l){8-9}
        & \multicolumn{1}{c}{Oracle} & \multicolumn{1}{c}{HT} & \multicolumn{1}{c}{Oracle} & \multicolumn{1}{c}{HT} & \multicolumn{1}{c}{Oracle} & \multicolumn{1}{c}{HT} & \multicolumn{1}{c}{Oracle} & \multicolumn{1}{c}{HT} \\
        \midrule
        500 & -0.0 & 0.0 & 184.3 & 100.0 & 2.2 & 1.5 & 95.3 & 95.0 \\
        1000 & 0.0 & -0.0 & 180.6 & 100.0 & 0.0 & -2.4 & 94.8 & 94.6 \\
        2000 & -0.0 & -0.0 & 183.2 & 100.0 & -0.1 & -0.8 & 95.1 & 95.1 \\
        4000 & 0.0 & 0.0 & 189.5 & 100.0 & -1.5 & 0.8 & 94.6 & 95.2 \\
        \bottomrule
    \end{tabular}
\end{table}

\begin{table}[H]
    \centering
    \caption{Prediction-error diagnostics with nonlinear outcome and nonlinear propensity.}\label{tab:simulation_rates_hh}
    \small
    \setlength{\tabcolsep}{6pt}
    \renewcommand{\arraystretch}{1.00}
    \begin{tabular}{@{}l c *{4}{d{3}}@{}}
        \toprule
        Learner pair & {$n$} & \multicolumn{1}{c}{\shortstack{Outcome\\RMSE}} & \multicolumn{1}{c}{\shortstack{Propensity\\RMSE}} & \multicolumn{1}{c}{Product} & \multicolumn{1}{c}{\shortstack{Scaled\\product}} \\
        \midrule
        GAM + GAM & 500 & 0.233 & 0.053 & 0.012 & 0.278 \\
        & 1000 & 0.167 & 0.038 & 0.006 & 0.198 \\
        & 2000 & 0.122 & 0.027 & 0.003 & 0.145 \\
        & 4000 & 0.091 & 0.019 & 0.002 & 0.109 \\
        \addlinespace[2pt]
        Agg. + Agg. & 500 & 0.235 & 0.055 & 0.013 & 0.290 \\
        & 1000 & 0.167 & 0.039 & 0.006 & 0.205 \\
        & 2000 & 0.123 & 0.028 & 0.003 & 0.152 \\
        & 4000 & 0.092 & 0.020 & 0.002 & 0.115 \\
        \addlinespace[2pt]
        XGBoost + XGBoost & 500 & 0.436 & 0.161 & 0.070 & 1.566 \\
        & 1000 & 0.361 & 0.124 & 0.045 & 1.413 \\
        & 2000 & 0.291 & 0.093 & 0.027 & 1.210 \\
        & 4000 & 0.237 & 0.068 & 0.016 & 1.026 \\
        \bottomrule
    \end{tabular}
    \par\vspace{6pt}
    \begin{tabular}{@{}l *{4}{d{2}}@{}}
        \rowcolor{simPanelGray}\multicolumn{5}{c}{\textit{Descriptive log--log fits}}\\
        \addlinespace[2pt]
     Learner pair
    & \multicolumn{1}{c}{$\widehat\alpha_m$}
    & \multicolumn{1}{c}{$\widehat\alpha_p$}
    & \multicolumn{1}{c}{$\widehat\alpha_m+\widehat\alpha_p$}
    & \multicolumn{1}{c}{\shortstack{Scaled product\\slope}}
    \\
        \midrule
        GAM + GAM & 0.45 & 0.50 & 0.95 & -0.45 \\
        Agg. + Agg. & 0.45 & 0.49 & 0.94 & -0.44 \\
        XGBoost + XGBoost & 0.29 & 0.41 & 0.71 & -0.21 \\
        \bottomrule
    \end{tabular}
\end{table}

\endgroup

\end{document}